\documentclass[preprint,floats,floatfix,aps,epsfig,nofootinbib,amssymb]{revtex4}
\usepackage{subfigure}
\usepackage{graphics,psfrag}
\usepackage{graphicx,psfrag}
\usepackage{amsmath,amssymb}
\usepackage[sort&compress]{natbib}
\usepackage{cancel}
\def\gev{\rm GeV}

\def\ttb{t\bar t}
\def\etmiss{{E\!\!\!\!\slash_{T}}}
\def\ptmiss{p\!\!\!\slash_{T}}
\def\pslash{\not{\hbox{\kern-4pt $p$}}}
\def\qslash{\not{\hbox{\kern-4pt $q$}}}
\def\lv{\not{\hbox{\kern-4pt $L$}}}
\def\lsim{\mathrel{\raise.3ex\hbox{$<$\kern-.75em\lower1ex\hbox{$\sim$}}}}
\def\gsim{\mathrel{\raise.3ex\hbox{$>$\kern-.75em\lower1ex\hbox{$\sim$}}}}
\def\ifmath#1{\relax\ifmmode #1\else $#1$\fi}
\def\beq{\begin{equation}}
\def\eeq{\end{equation}}
\def\bea{\begin{eqnarray}}
\def\eea{\end{eqnarray}}
\def\nn{\nonumber}

\usepackage{graphicx}
\usepackage{bm}
\begin{document}
\draft

\preprint{
   {\vbox {
      \hbox{\bf FERMILAB-PUB-09-597-T}
      \hbox{\bf MADPH-09-1550}	      
      }}}
\vspace*{2cm}

\title{Model-Independent Searches for New Quarks at the LHC}
\vspace*{0.25in}   % makes space between address and abstract
\author{Anupama Atre$^{1, 2}$, Georges Azuelos$^{3,4}$, Marcela
  Carena$^{1,5}$, \\ 
Tao Han$^{6}$, Erkcan Ozcan$^7$, Jos\'e Santiago$^{8}$ and Gokhan Unel$^9$}
%\footnote{avatre@fnal.gov\\ georges.azuelos@umontreal.ca\\ carena@fnal.gov\\ than@hep.wisc.edu\\ eo@hep.ucl.ac.uk\\ santiago@itp.phys.etzh.ch\\ Gokhan.Unel@cern.ch.}}   
\affiliation{\vspace*{0.1in}
$^1$ Fermilab, Batavia U.S.A.\\
$^2$ Michigan State University, East Lansing U.S.A.\\
$^3$ University of Montreal, Montreal CA\\
$^4$ TRIUMF, Vancouver CA\\
$^5$ University of Chicago, Chicago U.S.A.\\
$^6$ University of Wisconsin, Madison U.S.A.\\
$^7$ University College London, U.K. \\
$^8$ University of Granada, Granada Spain\\
$^9$ University of California, Irvine USA.}
%$^1$ Fermi National Accelerator Laboratory, MS106, P.O.Box 500, IL 60510, U.S.A.\\
%$^2$ University of Montreal CA\\
%$^3$ TRIUMF, Vancouver CA\\
%$^4$ Department of Physics and Enrico Fermi Institute, University of Chicago, Chicago, IL 60637, U.S.A.\\
%$^5$ Department of Physics, University of Wisconsin, 1150 University
%Avenue, Madison, WI 53706, U.S.A.\\
%$^6$ University College London\\
%$^7$ CAFPE and Dpto. de F\'{\i}sica Te\'orica y del Cosmos,
%University of Granada, E-18071 Granada, Spain\\
%$^8$ Physics and Astronomy Department, University of California at Irvine, USA.}
\vspace*{0.25 in} % makes space between address and abstract

\begin{abstract}
New vector-like quarks can have sizable couplings to first generation
quarks without conflicting with current experimental constraints. The
coupling with valence quarks and unique kinematics make single
production the optimal discovery process. We perform a
model-independent analysis of the discovery reach at the Large Hadron
Collider for new vector-like quarks considering single production and
subsequent decays via electroweak interactions. An early LHC run with 7 TeV center of mass energy and 1 fb$^{-1}$ of integrated luminosity can probe heavy quark masses up to 1 TeV and can be competitive with the Tevatron reach of 10 fb$^{-1}$. The LHC with 14 TeV center of mass energy and 100 fb$^{-1}$ of integrated luminosity can probe heavy quark masses up to 3.7 TeV for order one couplings.
\end{abstract} 

\maketitle

%%%%%%%%%%%%%%%%%%%%%%%%%%%%%%%%%%%%%%%%
\section{Introduction}
\label{sec:introd}

The Large Hadron Collider (LHC) has finally started its quest to find the origin of electroweak symmetry breaking (EWSB). A natural realization of EWSB is likely to bring along new particles to cancel the ultraviolet sensitivity of the Higgs mass or to unitarize longitudinal gauge boson scattering in the absence of a light Higgs boson. A common occurrence among these new particles is new
vector-like fermions that mix with Standard Model (SM)
particles. Traditionally, vector-like quarks are considered to mix
significantly only with the top sector, as the usual lore states that
large mixings with lighter generations are excluded by electroweak
precision or flavor observables. This statement is however not
necessarily correct, as cancellations among the effects of different
types of new 
quarks can significantly alleviate  the indirect constraints
~\cite{delAguila:2000aa, delAguila:2000rc, delAguila:2008pw}. It was
recently pointed out~\cite{Atre:2008iu} that these cancellations can
be the result of a symmetry ~\cite{Agashe:2006at} and naturally arise
in certain models of extra dimensions or strong EWSB, which also allow
for relatively light 3rd generation partners
~\cite{Carena:2006bn, Cacciapaglia:2006gp, Contino:2006qr,
  Carena:2007ua,carena:2009yt,Albrecht:2009xr,delAguila:2010vg,Casagrande:2010si}. 
Motivated by this type of models, several studies of the LHC reach for top
partners~\cite{carena:2007tn, Contino:2008hi, AguilarSaavedra:2009es,
  Mrazek:2009yu} and tau partners~\cite{delAguila:2010es} 
have appeared recently in the literature. There
are also some studies of  new quarks mixing with first generation
quarks. The discovery prospects of heavy down type vector-like quarks
decaying to light quark generations in the context of the $E_6$ GUT
model \cite{Gursey:1978fu, Gursey:1975ki} were studied in
Refs.~\cite{Sultansoy:2006cw, Mehdiyev:2007pf} for pair production
mode and in Ref.~\cite{delAguila:2008iz} for the single production
mode. Most recently  a more general
analysis of the Tevatron reach for  general new quarks with sizable
couplings to the up or down quark was performed in
Ref. ~\cite{Atre:2008iu}. The goal of this article is   to investigate
the LHC potential in the search for general  new heavy quarks
exploring similar channels as in the previous study for the Tevatron.
  
Heavy quarks can be produced in pairs via strong QCD interactions 
\beq
q\bar q,\ gg \to Q\bar Q,
\label{eq:QQbar}
\eeq
where  $Q$ generically denotes a new heavy quark. This is the most common 
and model-independent  production mechanism for any states with QCD interactions. The heavy quarks can also be produced singly in flavor-changing processes via  electroweak interactions  
\beq
q q'  \stackrel {V^*} {\longrightarrow} q_1 Q,
\label{eq:jQ}
\eeq
where $V = W$ or $Z$ gauge boson. Some of the current authors showed
in Ref.~\cite{Atre:2008iu} that the Tevatron reach to find new
vector-like quarks with sizable  mixings to first generation SM quarks
is significantly better in the single production channel than in pair
production.  

In this article we perform a theory-unbiased study of single
production of new quarks with 
sizable mixing to the up or down quarks at the LHC. The large energy and
luminosity of the LHC allow us to probe  heavy new quark masses,
even with rather moderate mixing to the first generation SM quarks,
thus going beyond the need of exact 
cancellations to ensure compatibility with indirect constraints. We
find quite encouraging results showing that an early run with 1 fb$^{-1}$ of
integrated luminosity at a center of mass energy of 7 TeV can be
competitive with the Tevatron reach of 10 fb$^{-1}$. The mass reach with a
larger integrated luminosity of 100 fb$^{-1}$ at  a center of mass
energy of 14 TeV  can be up to 3.7 TeV  for order one couplings.  

Amongst the many searches for new heavy quarks at collider experiments, one of the popular scenarios searched for is that of a sequential fourth generation quark \cite{Frampton:1999xi, Kribs:2007nz, Erler:2010sk, Martinez:2011ua, 
  Holdom:2009rf}. Besides the limits on their masses from direct
searches at the Tevatron 
%as quoted in section~\ref{sec:results_constraints}, 
there are significant, recently updated bounds
from electroweak (EW) precision data
\cite{Kribs:2007nz, Erler:2010sk, Martinez:2011ua}. The main difference between this model and the models under our consideration is that 
vector-like
 fermions, contrary to chiral fermions, decouple in the limit of large
 masses and therefore their effects can be made arbitrarily small,
 thus avoiding the severe constraints~\cite{delAguila:1982fs}.
Phenomenologically, our theory parameterization and
searching strategy are equally applicable to other heavy quark
searches, as long as there is a sizable mixing with the light
generations. 

The rest of the paper is organized as follows. In Sec.~\ref{sec:simpmod}, we present a general model parameterization for the heavy quarks and discuss the constraints on various models. In Sec.~\ref{sec:hqprod}, we calculate the heavy quark production cross sections at the LHC for the single production processes. In Sec.~\ref{sec:hqdecay}, we calculate the heavy quark decay via the charged and neutral currents and parameterize them by a few model independent parameters. In Sec.~\ref{sec:hqcoll}, we explore the observability of the heavy quarks at the LHC via the single production mechanism with the energy options for 7 and 14 TeV. We present a discussion of our results in Sec.~\ref{sec:results} and conclusions in Sec.~\ref{sec:concl}.  

%%%%%%%%%%%%%%%%%%%%%%%%%%%%%%%%%%%%%%%%
\section{General Parameterization}
\label{sec:simpmod}

In this section we present a general parameterization for models with
new heavy quarks. Let us consider new quarks labeled by $Q=X,U,D,Y$ with electric charges 
\bea
Q_X=\frac{5}{3},\quad Q_U=\frac{2}{3},\quad Q_D=-\frac{1}{3},\quad
{\rm and} \quad Q_Y=-\frac{4}{3},  
\eea
with masses $m_Q$ and arbitrary couplings to the SM gauge bosons and first generation quarks parameterized by 
\bea
\label{Coupling}
&& 
\frac{g}{\sqrt{2}} W_\mu^+ 
\Big[ 
\kappa_{uD} \overline{u}_R \gamma^\mu D_R 
+\kappa_{dY} \overline{d}_R \gamma^\mu Y_R
\Big] 
+
\frac{g}{\sqrt{2}} W_\mu^- 
\Big[ 
\kappa_{uX} \overline{u}_R \gamma^\mu X_R 
+\kappa_{dU} \overline{d}_R \gamma^\mu U_R 
\Big]
\nonumber \\
+&& \frac{g}{2 c_W} Z_\mu
\Big[
\kappa_{uU}\ \overline{u}_R \gamma^\mu U_R
+\kappa_{dD}\ \overline{d}_R \gamma^\mu D_R
\Big]
+ \mathrm{h.c.}~.
\eea
This choice of electric charges exhausts the possibilities of new quarks mixing through gauge couplings with SM quarks. For simplicity we have only considered RH couplings, which is not a restriction since we are not using angular correlations in our analysis. Possible Yukawa couplings between $U$ and $u$ or $D$ and $d$ have not been written explicitly since they will not be used in our analysis as discovery channels. Finally, we have included only one new quark of each charge as multiple new quarks for each charge can be added trivially.   

%%%%%%%%%%%%%%%%%%%%%%%%%%%%%%
\subsection{Generic Constraints on the Couplings}

The couplings in Eq.~(\ref{Coupling}) are subject to various constraints that can be classified as theoretical constraints and indirect experimental constraints.
Theoretical constraints arise from the fact that $\kappa_{qQ}$ originates from the mixing of SM quarks with new vector-like quarks and it is therefore bound by the unitarity of the matrices involved in such mixing. Generically we have 
$$\kappa \leq 1, $$ although this constraint could be stronger in particular
models (see below). Indirect experimental constraints are much more model-dependent. They can arise at tree level through modification of the SM fermion
couplings or at one loop, mainly through contributions to the oblique parameters ~\cite{Peskin:1991sw}. The former crucially depend on whether cancellations among the contributions due to the mixing with different heavy
quarks are present or not. In the absence of cancellations, the couplings $\kappa_{qQ}$ can be quite constrained from low-energy experimental data. There is a generic correlation between the new couplings and corrections to the couplings of SM fermions, namely 
\beq
\frac{\delta g_{SM}}{g_{SM}} \sim \kappa^2,
\label{generic:correction}
\eeq
where $g_{SM}$ represents a generic SM coupling (see for instance Ref.~\cite{delAguila:1982yu}). From EW precision observables, the SM couplings have been measured with a per mille precision.  So, as a rule of thumb, we can expect a generic bound    
\beq
\kappa \lesssim \mbox{ few }\times 10^{-2} \qquad \mbox{(no cancellations)}.
\label{kappa:bound:nocancellations}
\eeq
As we will see below, cancellations among different contributions can however occur and make this bound completely disappear. One loop contributions to the oblique parameters, although also model dependent, are somewhat more robust. The resulting bounds on $\kappa_{qQ}$, which are typically milder than those in Eq.~(\ref{kappa:bound:nocancellations}), can nevertheless be the main
constraint in some models. 

%%%%%%%%%%%%%%%%%%%%%%%%%%%%%%
\subsection{Model Considerations\label{model:considerations}}

As we have emphasized in the previous section, tree level experimental constraints on the values of $\kappa_{qQ}$ are very model dependent and can be, for a given theoretical model, a lot  milder than that of Eq.~(\ref{kappa:bound:nocancellations}). The reason is that there are models in which some of the $\kappa_{qQ}$ can be sizable without modifying in an observable way the SM quark couplings (which are experimentally measured) and without inducing any observable flavor violation. This is the result of cancellations in the SM couplings due to the mixing of SM quarks with several different new quarks. 
These constraints or lack thereof can be better understood with a couple of illustrative examples as follows. 

%%%%%%%%%%%%%%%%%%%%
\subsubsection{Model I: Two degenerate vector-like quark doublets}
\label{sec:degendoublet}

\begin{figure}[htb]
\includegraphics[width=0.5\textwidth,clip=true]{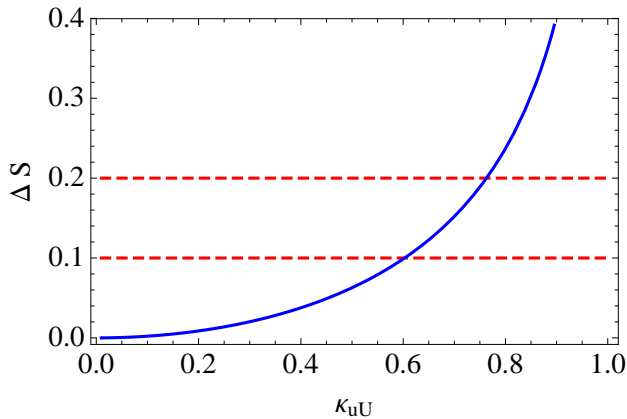}
\caption{ 
One loop contribution to the $S$ parameter, in the model with two degenerate doublets of hypercharge $1/6$ and $7/6$ and mixing only with the up quark, as a function of $\kappa_{uU}$. The result is independent of the mass of the heavy quarks up to corrections $\mathcal{O}(m_u^2/M^2$).}
\label{S:degenerate}
\end{figure}

The first model, discussed in detail in Ref.~\cite{Atre:2008iu}, consists of two new vector-like quark electroweak doublets, with hypercharges $1/6$ and $7/6$
respectively, that mix only with the up quark, in the basis of diagonal Yukawa couplings for the charge $2/3$ SM quarks. The particle content of these two doublets is two charge $2/3$, one charge $-1/3$ and one charge $5/3$ quark. If the two doublets are degenerate, \textit{i.e.} they have the same mass and Yukawa couplings to $u$ before EWSB, their mixing with the up quark does not induce any observable correction to the SM quark couplings. This is due to the
exact cancellation of two large contributions (arising from the mixing
of the up quark with the charge $2/3$ quarks in each doublet) with
opposite signs. These large contributions are functions of
$\kappa_{qQ}$ and since the large contributions cancel, we can have
large values of $\kappa_{qQ}$ without upsetting any SM quark
couplings. The only source of flavour violation in this case turns out
to be suppressed by Cabbibo-Kobayashi-Maskawa (CKM)  angles and by the
ratio of the mass of the up quark and the mass of the new quarks,
which provides enough suppression to keep flavor violation below
experimental limits (details can be found in the Appendix of
Ref.~\cite{Atre:2008iu}). At the loop level, the new quarks do however
contribute to oblique parameters. In the degenerate case, the new
sector is custodially invariant and the $T$ parameter receives no
correction. Interestingly enough, the contribution to the $S$
parameter turns out to be essentially independent of the mass of the
heavy quarks, for fixed values of $\kappa_{uU}$. Note that for a fixed
Yukawa coupling between the new quarks and the up quark, $\kappa_{uU}$
goes to zero in the large mass limit and therefore the vector-like
quarks decouple in the heavy limit as they should.
The dependence on the coupling $\kappa_{uU}$ is displayed in
Fig.~\ref{S:degenerate}. Strict constraints based on the contribution
to the oblique parameters are difficult to impose without knowledge of
other possible sectors of the theory that could also contribute to
them. Nevertheless, assuming $S \leq 0.2$ ($0.1$) gives a pretty
mild bound of  
$\kappa_{uU} \lesssim 0.75~(0.6)$ in the model under
consideration. More details about the calculation of the $S$ parameter
can be found in Appendix~\ref{app:calcS}. 
%%%
%%%

After EWSB one combination of the charge $2/3$ heavy quarks couples to the up quark only through Yukawa couplings whereas the remaining three, of types $U$, $D$ and $X$ in our notation, have the following values of $\kappa$ 
\beq
\kappa_{uU}=\sqrt{2}\kappa_{uD}=\sqrt{2} \kappa_{uX}= s_R,
\eeq
where $s_R$ is the sine of the corresponding mixing angle which depends on the particular values of the model parameters. Thus, we see that in this model, the generic bound from the unitarity of the mixing is saturated for 
$\kappa_{uU}\leq 1$ and is a bit more stringent for  
$\kappa_{uD}=\kappa_{uX} \leq 1/\sqrt{2}$. We refer the reader to Ref.~\cite{Atre:2008iu} for more details. 

Another similar model has two degenerate doublets of hypercharges $1/6$ and $-5/6$, respectively, where the heavy quarks mix only with the down quark. In this case there would be one combination of charge $-1/3$ heavy quarks that couples to the down quark only through Yukawa couplings, while the other three quarks, $U, D$ and $Y$ in our notation, have  large couplings to $d_R$ with the following constraints 
\beq
\kappa_{dD}=\sqrt{2}\kappa_{dU}=\sqrt{2} \kappa_{dY}= s_R, 
\eeq
resulting in a theory upper bound of $\kappa_{dD}\leq 1$, 
$\kappa_{dU}=\kappa_{dY} \leq 1/\sqrt{2}$. As we have stressed above, these cancellations can be due to a symmetry and therefore be natural.  

%%%%%%%%%%%%%%%%%%%%
\subsubsection{Model II: One vector-like quark doublet}

In our second example no cancellations occur and SM couplings are
modified due to the mixing with the heavy quarks. Stringent
experimental constraints arise on the mixing and therefore on the
values of $\kappa_{qQ}$ (see
Eqs.~(\ref{generic:correction})$-$(\ref{kappa:bound:nocancellations})). The
model consists of just one new electroweak doublet with hypercharge
$1/6$ that mixes only with the up quark. The particle content is one
$U$ and one $D$ quark and the mixing with the up quark induces an
anomalous $Z u_R \bar{u}_R$ coupling, proportional to $\kappa_{uD}^2$,
which is strongly constrained experimentally. A detailed analysis,
using an updated version of the code in 
Refs.~\cite{han:2004az, han:2005pr} gives a bound of 
$\kappa_{uD} \lesssim 0.07$.~\footnote{The code in
  Refs.~\cite{han:2004az,han:2005pr} assumes symmetry between the
  first two generations. The reported bound therefore applies to a model
with two new vector-like multiplets that mix equally with the
first two generations. This can be considered a conservative but
indicative bound in our model.} 
The couplings to the heavy quarks can be
computed as discussed in Ref.~\cite{delAguila:1982fs} and Part 13
of Ref.~\cite{Brooijmans:2008se} (see Ref.~\cite{Cacciapaglia:2010vn} for
explicit expressions). 
Denoting the sine and cosine of the
mixing angle by $s_R$ and $c_R$, respectively, we have 
\beq 
\kappa_{uU}=s_Rc_R,\ \ \ \ \kappa_{uD}=s_R. 
\eeq
The indirect
bound on $\kappa_{uD}$, from the constraint on the $Z u_R 
\bar{u}_R$ coupling, is therefore much more restrictive than
the one coming from the unitarity of the mixing  
$\kappa_{uD}\leq 1$. It also induces a very restrictive constraint on
$\kappa_{uU} \lesssim 0.07$.  Incidentally, there are also loop
contributions in this model to the $S$ and $T$ parameters but the tree
level constraints are much more 
restrictive. 

%%%%%%%%%%%%%%%%%%%%
\subsubsection{Model III: Non-degenerate quarks}

An intermediate model between our two examples, with two new doublets
which are not exactly degenerate allows for partial cancellations. For
instance, using the code in Refs.~\cite{han:2004az, han:2005pr} we find that
$\kappa_{uU}\sim 0.25$ is possible if it is within $\sim 10\%$ of 
$\sqrt{2}\kappa_{uX}$. Assuming similar values of the Yukawa couplings, this can translate to a $5-15\%$ difference between the masses of different quarks, which might be experimentally distinguishable. 

Thus we can see that a large number of different possibilities can arise depending on the model but still sizable values of $\kappa_{qQ}$ cannot be excluded on a general basis. 

%%%%%%%%%%%%%%%%%%%%%%%%%%%%%%
\subsection{Generic Parameterization}

In order to consider all these possibilities, we do a model-independent study, 
imposing no \textit{a priori} constraints on the values of $\kappa_{qQ}$. This way our results can be applied to any specific model, for which the relevant theoretical or experimental constraints will have to be taken into account.
 
We can parameterize the coupling $\kappa_{qQ}$ in a model-independent manner as    
\beq
\kappa_{qQ} = ({v}/{m_Q}) \tilde{\kappa}_{qQ},
\eeq
where the dimensionless parameter $\tilde{\kappa}_{qQ}$ encodes the model-dependence and $v\approx 174$ GeV is the Higgs vacuum expectation value. 
This parameterization is useful because we generically have 
\beq
\kappa_{qQ} = 
\frac{\lambda v}{m_Q} \Big( 1+\mathrm{O}(v^2/m_Q^2)\Big), 
\eeq
where $\lambda$ is some Yukawa coupling between SM and heavy quarks. Thus, in the limit that $\kappa_{qQ}$ is relatively small so that the expansion can be well approximated by the first term, $\tilde{\kappa}_{qQ}$ has a direct relation to a model parameter. It corresponds to a Yukawa coupling and it is naturally of order unity in a generic weakly coupled theory, and could therefore be up to $4\pi$ in strongly coupled theories. For larger values of $\kappa_{qQ}$ non-linear effects become important and the dependence of $\tilde{\kappa}_{qQ}$ on the model parameters becomes much more complicated. It should be
noted that all the bounds we have obtained for the couplings, both the
experimental ones and the ones coming from the unitarity of the mixing, apply to $\kappa$ and not $\tilde{\kappa}$. Therefore heavier quarks can have a substantially larger value of $\tilde{\kappa}$ compatible with EW precision observables  and unitarity of the mixing. 

%%%%%%%%%%%%%%%%%%%%%%%%%%%%%%%%%%%%%%%%
\section{Heavy quark production}
\label{sec:hqprod}

In this article, we discuss the single production of heavy quarks via the process shown in Eq.~(\ref{eq:jQ}). In order to keep our study widely applicable, we have factored out the model dependent coupling $\tilde{\kappa}^2$, so that the cross section depends only on the corresponding Parton Distribution Functions (PDFs) and the kinematics (mass of the heavy quark produced). These model-independent cross sections will be denoted as $\sigma^{CC}_Q$ and $\sigma^{NC}_Q$ for single production of quark $Q$ via charged current (CC) and neutral current (NC) interactions respectively. In the rest of the analysis, we adopt the CTEQ6L1 PDFs \cite{Pumplin:2002vw} for the parton distribution functions. 

%%%
\begin{figure}[tb]
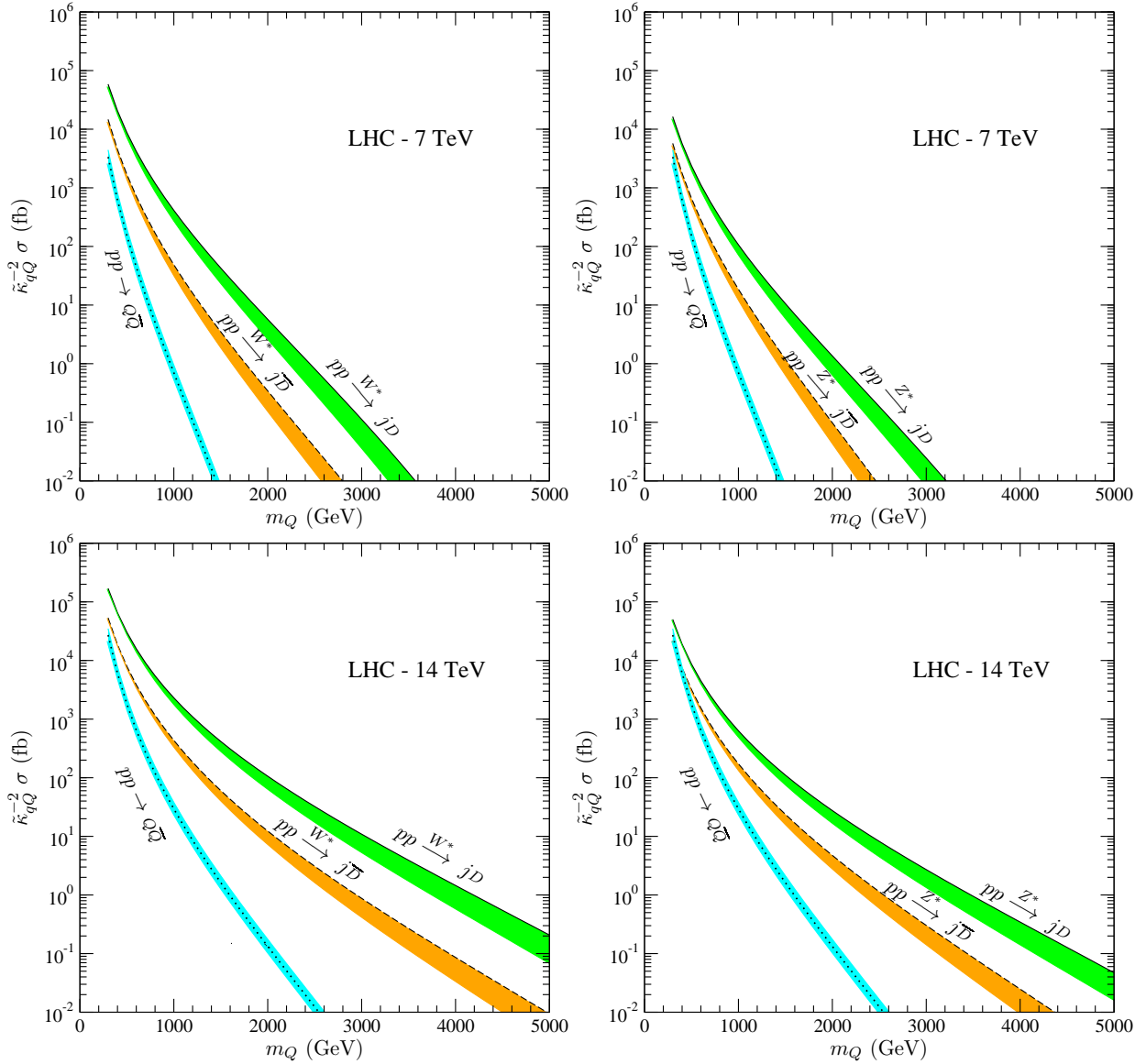

\psfrag{mQ(GeV)}[t][t][0.8]{$m_Q\ \mathrm{(GeV)}$}
\psfrag{k^-2sigma(fb)}[b][b][0.8]{$\tilde \kappa^{-2}_{qQ}\ \sigma\ \mathrm{(fb)}$}  
\psfrag{ppjdccst}[][][0.8]{$pp \stackrel {W^*}
  {\longrightarrow}\  j\scriptstyle {D}$ } 
\psfrag{ppjdbccst}[][][0.8]{$pp \stackrel {W^*}
  {\longrightarrow}\  j\scriptstyle {\overline D}$} 
 \psfrag{ppjdncst}[][][0.8]{$pp \stackrel {Z^*}
  {\longrightarrow}\  j\scriptstyle {D}$ } 
\psfrag{ppjdbncst}[][][0.8]{$pp \stackrel {Z^*}
  {\longrightarrow}\  j\scriptstyle {\overline D}$} 
\psfrag{ppwd}[][][0.8]{$pp \rightarrow\  \scriptstyle {W D}$}
\psfrag{ppwdb}[][][0.8]{$pp \rightarrow\  \scriptstyle {W \overline D}$}
\psfrag{ppzd}[][][0.8]{$pp \rightarrow\  \scriptstyle {Z D}$}
\psfrag{ppzdb}[][][0.8]{$pp \rightarrow\  \scriptstyle {Z \overline D}$}
\psfrag{ppqqb}[][][0.8]{$pp \rightarrow\  \scriptstyle {Q \overline Q}$}
{\includegraphics[width=0.49\textwidth,clip=true]{xsec-d1-cc-7tev.eps}
\includegraphics[width=0.49\textwidth,clip=true]{xsec-d1-nc-7tev.eps}
\includegraphics[width=0.49\textwidth,clip=true]{xsec-d1-cc-14tev.eps}
\includegraphics[width=0.49\textwidth,clip=true]{xsec-d1-nc-14tev.eps}}
\caption{ 
(a) Top left: cross sections (in fb) for singly producing the heavy quark $D$ via CC interactions versus its mass $m_Q^{}$ at the LHC with $\sqrt s=7$ TeV in $pp$ collisions. (b) top right: same as (a) but for NC interactions. (c) bottom left: same as (a) but for $\sqrt s=14$ TeV. (d) bottom right: same as (b) but for 
$\sqrt s=14$ TeV. Solid (dashed) curves represent production of 
$D (\overline D)$ via the process $pp \rightarrow\  jD (\overline D)$. The cross section for pair production (which is independent of $\tilde\kappa$) of heavy quark is shown as dotted (black) curve for comparison. The colored bands represent the variation in the leading order cross section due to the different scale choices as described in the text.  }
\label{fig:xsecd}
\end{figure}
%%%

\begin{figure}[tb]
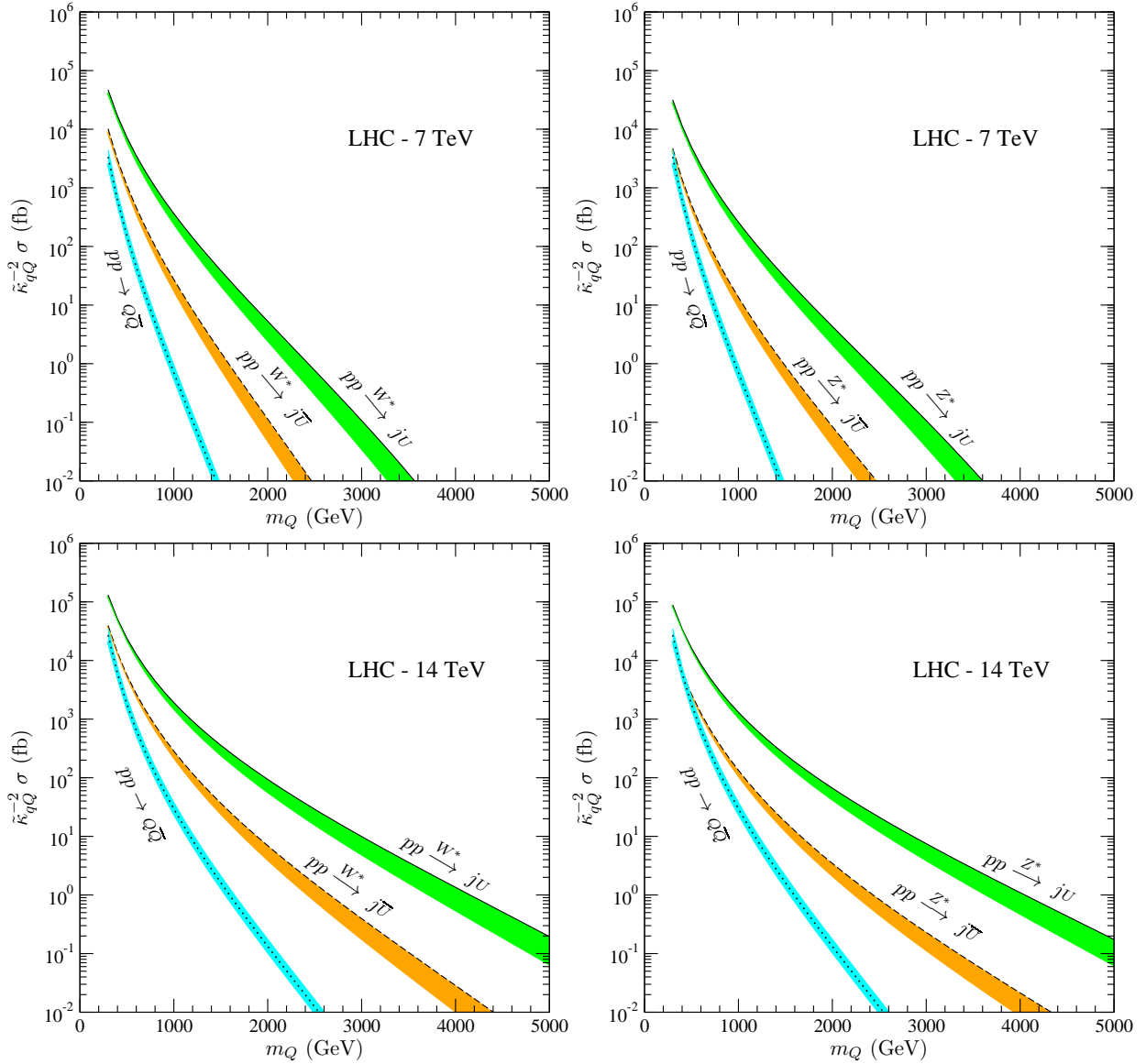

\psfrag{mQ(GeV)}[t][t][0.8]{$m_Q\ \mathrm{(GeV)}$}
\psfrag{k^-2sigma(fb)}[b][b][0.8]{$\tilde \kappa^{-2}_{qQ}\ 
\sigma\ \mathrm{(fb)}$}  
\psfrag{ppjuccst}[][][0.8]{$pp \stackrel {W^*}  {\longrightarrow}\  j\scriptstyle {U}$ } 
\psfrag{ppjubccst}[][][0.8]{$pp \stackrel {W^*}
  {\longrightarrow}\  j\scriptstyle {\overline U}$} 
\psfrag{ppjuncst}[][][0.8]{$pp \stackrel {Z^*}
  {\longrightarrow}\  j\scriptstyle {U}$ } 
\psfrag{ppjubncst}[][][0.8]{$pp \stackrel {Z^*}
  {\longrightarrow}\  j\scriptstyle {\overline U}$} 
\psfrag{ppwu}[][][0.8]{$pp \rightarrow\  \scriptstyle {W U}$}
\psfrag{ppwub}[][][0.8]{$pp \rightarrow\  \scriptstyle {W \overline U}$}
\psfrag{ppzu}[][][0.8]{$pp \rightarrow\  \scriptstyle {Z U}$}
\psfrag{ppzub}[][][0.8]{$pp \rightarrow\  \scriptstyle {Z \overline U}$}
\psfrag{ppqqb}[][][0.8]{$pp \rightarrow\  \scriptstyle {Q \overline Q}$}
{\includegraphics[width=0.49\textwidth,clip=true]{xsec-u1-cc-7tev.eps}
\includegraphics[width=0.49\textwidth,clip=true]{xsec-u1-nc-7tev.eps}
\includegraphics[width=0.49\textwidth,clip=true]{xsec-u1-cc-14tev.eps}
\includegraphics[width=0.49\textwidth,clip=true]{xsec-u1-nc-14tev.eps}}
\caption{ 
Same as Fig.~\ref{fig:xsecd} but for $U$.}
\label{fig:xsecu}
\end{figure}

\begin{figure}[tb]
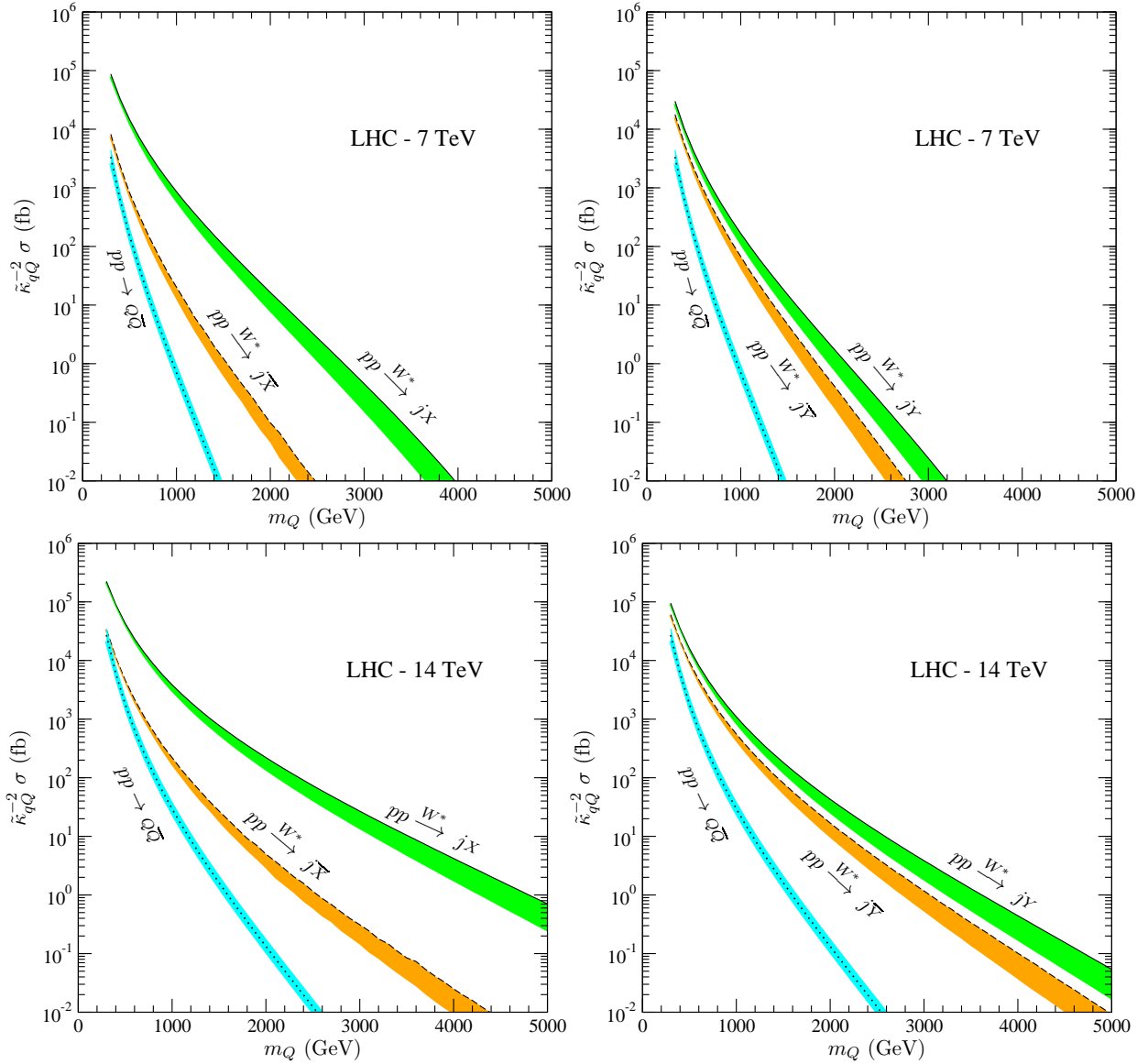

\psfrag{mQ(GeV)}[t][t][0.8]{$m_Q\ \mathrm{(GeV)}$}
\psfrag{k^-2sigma(fb)}[b][b][0.8]{$\tilde \kappa^{-2}_{qQ}\ 
\sigma\ \mathrm{(fb)}$}  
\psfrag{ppjxccst}[][][0.8]{$pp \stackrel {W^*}
  {\longrightarrow}\  j\scriptstyle {X}$ } 
\psfrag{ppjxbccst}[][][0.8]{$pp \stackrel {W^*}
  {\longrightarrow}\  j\scriptstyle {\overline X}$} 
\psfrag{ppjyccst}[][][0.8]{$pp \stackrel {W^*}
  {\longrightarrow}\  j\scriptstyle {Y}$ } 
\psfrag{ppjybccst}[][][0.8]{$pp \stackrel {W^*}
  {\longrightarrow}\  j\scriptstyle {\overline Y}$} 
\psfrag{ppwx}[][][0.8]{$pp \rightarrow\  \scriptstyle {W X}$}
\psfrag{ppwxb}[][][0.8]{$pp \rightarrow\  \scriptstyle {W \overline X}$}
\psfrag{ppwy}[][][0.8]{$pp \rightarrow\  \scriptstyle {W Y}$}
\psfrag{ppwyb}[][][0.8]{$pp \rightarrow\  \scriptstyle {W \overline Y}$}
\psfrag{ppqqb}[][][0.8]{$pp \rightarrow\  \scriptstyle {Q \overline Q}$}
{
\includegraphics[width=0.49\textwidth,clip=true]{xsec-xu1-cc-7tev.eps}
\includegraphics[width=0.49\textwidth,clip=true]{xsec-yd1-cc-7tev.eps}
\includegraphics[width=0.49\textwidth,clip=true]{xsec-xu1-cc-14tev.eps}
\includegraphics[width=0.49\textwidth,clip=true]{xsec-yd1-cc-14tev.eps}
}
\caption{ (a) Top left: cross sections (in fb) for singly producing the heavy quark $X$ via CC interactions versus its mass $m_Q^{}$ at the LHC with $\sqrt s=7$ TeV in $pp$ collisions. (b) top right: same as (a) but for the heavy quark Y. (c) bottom left: same as (a) but for $\sqrt s=14$ TeV. (d) bottom right: same as (b) but for $\sqrt s=14$ TeV. The description of the curves is the same as 
Fig.~\ref{fig:xsecd} but for $X$ and $Y$ with only charged current interactions.}
\label{fig:xsecxy}
\end{figure}

The single production cross section as a function of the mass of the heavy quark is shown in Figs.~\ref{fig:xsecd} $-$ \ref{fig:xsecxy} for various quark species at two different c.m. energies, 7 and 14 TeV. We show separately the production cross section for each quark (solid black curve) and anti-quark (dashed black curve) species through charged ($W^{*}$) or neutral ($Z^{*}$) currents for single production. All the curves for single production correspond to a factorization and renormalization scale $\mu_F=\mu_R=m_W (m_Z)$ for CC (NC) processes. The scale dependence of our LO calculation is represented by colored bands that correspond to a scale variation up to $\mu_F=\mu_R=m_Q$. For comparison, the QCD cross section for pair production (independent of 
$\tilde\kappa$) is also included in the figures and represented by dotted (black) curves. For pair production mode, we choose the central value of the factorization and renormalization scales to be 
$\mu_F=\mu_R=\sqrt{\hat{s}}$ and the scale dependence of our LO calculation is represented by colored bands that correspond to a scale variation up to $\sqrt{\hat{s}}/2 \le \mu_F=\mu_R \le 2\sqrt{\hat{s}}$. 

Fig.~\ref{fig:xsecd} and Fig.~\ref{fig:xsecu} correspond to single production of 
$D$ and $U$, respectively, with the left panels representing charged current channels and the right ones neutral current channels. Although we have included all contributing channels for the production, the leading contributions are from
\bea
\nonumber
&& d u  \stackrel {W^*}  \to jD,\quad q d \stackrel {Z^*}  \to  jD, \\
&& u d \stackrel {W^*}  \to jU,\quad q u \stackrel {Z^*}  \to jU ,
\label{eq:Qprod}
\eea
where $q$ denotes a generic valence quark parton and $j$ a light quark jet. Fig.~\ref{fig:xsecxy} on the other hand provides the production of $X$ on the left and $Y$ on the right. Due to their electric charge, they have only charged current channels and the leading contributions are from
\bea
 u u  \stackrel {W^*}  \to j X,\quad d d \stackrel {W^*}  \to  j Y .
\eea
%

%%%
%%%

%%%
%%%
  
Several interesting features regarding single production at the LHC can be
observed from Figs.~\ref{fig:xsecd} $-$ \ref{fig:xsecxy}. To start with, let us recall the advantages of single production versus pair production as shown in Ref.~\cite{Atre:2008iu}. The cross section for pair production falls off sharply due to phase space suppression and decreasing parton luminosity at large $x$ values. Evidently the LHC with its substantially larger c.m. energy will be able to probe higher quark masses in the pair production channel compared to the Tevatron. However, as at the Tevatron, single heavy quark production has the advantage of less phase-space suppression and longitudinal gauge boson enhancement of order $m^2_Q/M^2_V$ at higher energies compared to pair production. Due to the participation of $u,d$ valence quarks in the initial state with the coupling strength given in Eq.~(\ref{Coupling}), the cross section can be substantial and it falls more slowly for a higher mass. These effects can be easily observed from Figs.~\ref{fig:xsecd} $-$ \ref{fig:xsecxy}. To summarize, the large c.m. energy of the LHC coupled with the above advantages of single production make this an ideal process for discovery of heavy quarks up to very large mass values and very small couplings.

Heavy quarks are produced at a much higher rate compared to anti-quarks with a larger difference for higher masses at the LHC. This difference in the rates of production of $Q$ and $\bar Q$ is due to the difference in the PDFs of valence and sea quarks in the initial states - heavy quarks couple to valence quarks whereas heavy anti-quarks couple to sea quarks. As we will see in later sections this feature will help us in constraining the electromagnetic charge of the heavy quarks. For $jQ$ production valence quarks participate in the initial state and as the quark PDFs dominate at higher energies, the cross section falls slowly. Also note that, the relative sizes of the single production cross sections between different types of quarks are determined by the corresponding valence quark density in the initial state. To illustrate the features of single production explicitly, the cross sections for EW single production computed with MadGraph/MadEvent \cite{Alwall:2007st} are listed in Table~\ref{tab:singl:alt} for all $Q$ with $m_Q = 900\ (1800)$ GeV and $\tilde \kappa = 1$ at the LHC with $\sqrt{s} = 7\ (14)$ TeV. These results reiterate the previous observations that (i) valence quarks dominate; (ii) $u$ quark contributes
more than $d$; and (iii) NC is weaker than CC.
%%%%
\renewcommand{\arraystretch}{1.4}
\begin{table}\footnotesize
\begin{center}
\begin{tabular}{|l|c|c|l|c|c|l|c|c|} 
\hline
\ \ Channel & 7 TeV & 14 TeV & \ \ Channel & 7 TeV & 14 TeV & \ \ Channel & 7 TeV & 14 TeV \\ 
\ \ \ \  {  ($m_Q$)} &   {  (0.9 TeV)} &  {  (1.8 TeV)} & \ \  &  {  (0.9 TeV)} &  {  (1.8 TeV)} & \ \  &  {  (0.9 TeV)} &  {  (1.8 TeV)} \\ 
\hline
\ $pp  \stackrel {W^*} {\longrightarrow} jD$\ \  & 0.69 &  0.18 & 
\ $pp  \stackrel {W^*} {\longrightarrow} jU$\ \  & 0.61 &  0.16 &
\ $pp  \stackrel {W^*} {\longrightarrow} jX$\ \  & 1.4 &  0.36 \\
\ $pp  \stackrel {W^*} {\longrightarrow} j\overline{D}$\ \  & 0.089 & 0.022 & 
\ $pp  \stackrel {W^*} {\longrightarrow} j\overline{U}$\ \  & 0.052 & 0.013 & 
\ $pp  \stackrel {W^*} {\longrightarrow} j\overline{X}$\ \  & 0.037 & 0.0092 \\
\ $pp  \stackrel {Z^*} {\longrightarrow} jD$\ \  & 0.18 &  0.047 & 
\ $pp  \stackrel {Z^*} {\longrightarrow} jU$\ \  & 0.43 &  0.11 &
\ $pp  \stackrel {W^*} {\longrightarrow} jY$\ \  & 0.29 &  0.074 \\
\ $pp  \stackrel {Z^*} {\longrightarrow} j\overline{D}$\ \  & 0.034 & 0.0088 & 
\ $pp  \stackrel {Z^*} {\longrightarrow} j\overline{U}$\ \  & 0.025 & 0.0064 & 
\ $pp  \stackrel {W^*} {\longrightarrow} j\overline{Y}$\ \  & 0.12 & 0.031 \\
\hline
\end{tabular}
\caption{Cross sections in pb for EW single production of $D$, $U$, $X$ and $Y$ with $\tilde \kappa = 1$ and $m_Q = 900 (1800)$ GeV at the LHC with $\sqrt{s} = 7 (14)$ TeV.} 
\label{tab:singl:alt}
\end{center}
\end{table}
%%%%

%%%%%%%%%%%%%%%%%%%%%%%%%%%%%%%%%%%%%%%%
\section{Heavy quark decay}
\label{sec:hqdecay}

The singly produced heavy quarks will decay into jets and gauge or Higgs bosons through gauge and Yukawa interactions. In this article we are not considering the Higgs related channels so the relevant decay channels are
\bea
X&\to& W^+ u,\phantom{, Z u} \quad Y\to W^- d,
\\
U&\to& W^+ d,\ Z u,
\quad D\to W^- u,\  Z d,
\eea
and the charge conjugated ones.

To perform a model-independent study, we classify processes according to their final states. In order to overcome the QCD background, we consider leptonic decays of the corresponding gauge boson. Therefore, the four final state configurations that could be experimentally distinguished from the SM backgrounds are:
\beq
jj (W^-\to \ell^- \bar{\nu}), 
\quad jj (W^+\to \ell^+ \nu), 
\quad jj (Z \to \ell^+ \ell^-), 
\quad jj (Z \to \nu \bar{\nu}).
\eeq
Following this classification, we parameterize the corresponding cross section, under the narrow width approximation as
\beq
\sigma (pp \to jj f f^\prime) = S^{V} Br(V \to ff^\prime),
\eeq
where $V=W^+,W^-$ and $Z$ with $ff^\prime=\ell^+ \nu, \ell^- \bar{\nu}$ and $\ell^+\ell^-$ plus $\nu \bar \nu$, respectively. 

All the model dependence is encoded in the $S^V$ parameters which are nothing but the production cross sections times the corresponding branching ratios summed over all possible channels. These can be written as:
\bea
S^{W^-} &\equiv& \Big[ \tilde \kappa^2_{dD} \sigma_{D}^{NC}
+\tilde \kappa^2_{uD} \sigma_{D}^{CC} \Big] Br(D\to W^- u)  
+\tilde \kappa^2_{dY} \sigma_Y^{CC} Br(Y \to W^- d)
\nonumber \\
&+& \Big[\tilde \kappa^2_{uU} \sigma_{\overline U}^{NC}
+\tilde \kappa^2_{dU} \sigma_{\overline U}^{CC} \Big] 
Br(\overline U \to W^- \bar d)
+ \tilde \kappa^2_{uX} \sigma_{\overline X}^{CC} Br(\overline X \to W^- \bar u), \\
S^{W^+} &\equiv& 
\Big[ \tilde \kappa^2_{dD} \sigma_{\overline D}^{NC}
+\tilde\kappa^2_{uD} \sigma_{\overline D}^{CC}\Big] Br(\overline D \to W^+ \bar u)
+\tilde \kappa^2_{dY} \sigma_{\overline Y}^{CC} Br(\overline Y \to W^+ \bar d)
\nonumber \\
&+& \Big[\tilde \kappa^2_{uU} \sigma_{U}^{NC} +\tilde \kappa^2_{dU} \sigma_{U}^{CC} \Big] Br(U \to W^+ d) 
+ \tilde \kappa^2_{uX} \sigma_{X}^{CC} Br(X \to W^+ u), 
\\
S^{Z} &\equiv & \Big[ \tilde \kappa^2_{dD} \big( \sigma_{D}^{NC} + \sigma_{\overline D}^{NC} \big) + \tilde \kappa^2_{uD} \big( \sigma_{D}^{CC} + \sigma_{\overline D}^{CC} \big) \Big] Br(D \to Z d)
\nonumber \\
&+& \Big[ \tilde \kappa^2_{uU} \big( \sigma_{U}^{NC} + \sigma_{\overline U}^{NC} \big) + \tilde \kappa^2_{dU} \big( \sigma_{U}^{CC} + \sigma_{\overline U}^{CC} \big) \Big] Br(U \to Z u),
\eea
where $\sigma^{CC,NC}_Q$ are defined in Sec.~\ref{sec:hqprod} and displayed in Figs.~\ref{fig:xsecd} $-$ \ref{fig:xsecxy}.

These $S^V$ parameters are what we are sensitive to from the experimental point of view. Their apparently complicated form is due to the high degree of model independence but can simplify greatly in specific models. For instance, assuming two degenerate doublets of hypercharges $1/6$ and $7/6$ that only mix with the up quark~\cite{Atre:2008iu}, we have three quarks of charges 
$5/3$, $2/3$ and $-1/3$, all with the same mass $m_Q$, each one decaying with $100\%$ branching ratio to $W^+ u$, $Z u$ and $W^- u$, respectively. The fourth quark with charge $2/3$, decays via the $H u$ channel and we do not consider that here. In that case we have 
\beq
S^{W^-} = \tilde \kappa^2_{uD} \sigma^{CC}_{D} 
+\tilde \kappa^2_{uX} \sigma^{CC}_{\overline{X}},\ \ \ \ 
S^{W^+} = \tilde \kappa^2_{uX} \sigma^{CC}_{X}
+\tilde \kappa^2_{uD} \sigma^{CC}_{\overline{D}},\ \ \ \ 
S^{Z} = \tilde \kappa^2_{uU} \big(
\sigma^{NC}_{U}  +\sigma^{NC}_{\overline{U}} \big),
\eeq
with $\tilde{\kappa}_{uU}=\sqrt{2}\tilde{\kappa}_{uD}
=\sqrt{2}\tilde{\kappa}_{uX}\leq m_Q/v$, where the inequality is the theoretical bound as discussed in Sec.~\ref{sec:degendoublet}. In this specific case with two degenerate doublets it then implies the following upper bound on the corresponding $S^V$, 
\bea
\nonumber
S^{W^-} &\leq& \frac{1}{2}\left(\frac{m_Q}{v}\right)^2(\sigma^{CC}_D+\sigma^{CC}_{\overline{X}}),\ \ \ \
S^{W^+} \leq \frac{1}{2}\left(\frac{m_Q}{v}\right)^2(\sigma^{CC}_X+\sigma^{CC}_{\overline{D}}),\ \ \ \  \\
S^{Z} &\leq& \left(\frac{m_Q}{v}\right)^2(\sigma^{NC}_U+\sigma^{NC}_{\overline{U}}).
\eea
The expressions for other models can be easily computed using the
general expressions for the partial widths given in
Appendix~\ref{partialwidths}. 
In the next section we describe the prospects for the observability of the new heavy quarks according to the signatures defined in this section.

%%%%%%%%%%%%%%%%%%%%%%%%%%%%%%%%%%%%%%%%
\section{Observability of the heavy quark signal}
\label{sec:hqcoll}

The classification of the possible final states that we have performed in the previous section allows us to perform a model-independent analysis of the LHC reach of new quarks which have sizable mixing with first generation quarks. We report on the results of such an analysis in the next few sections for single production. After a general discussion of the main kinematical features of the signal and background, we discuss the results of such an analysis for two different scenarios, one early run with 1 fb$^{-1}$ integrated luminosity at 
$\sqrt{s}=7$ TeV and one longer run with 100 fb$^{-1}$ of integrated luminosity at $\sqrt{s}=14$ TeV. 

The final state channels of our current interest are
\beq
jQ\  \to\  \ell^\pm \etmiss\ 2j,\quad \ell^+\ell^-\ 2j,\quad \etmiss\ 2j,
\eeq
from $Q$ decaying to a light-quark jet plus a $W^\pm(\to \ell^\pm \nu)$ or 
$Z (\rightarrow \ell^+ \ell^-,\  \nu \bar{\nu})$, respectively. For the leptonic decay modes of the gauge bosons we consider $\ell = e, \mu$ and 
$\nu = \nu_e, \nu_\mu, \nu_\tau$. Although the inclusion of the $\tau$ lepton in the final state could increase the signal statistics, for simplicity we ignore this experimentally more challenging channel. The relevant backgrounds to these processes are
\begin{itemize}
\item $W+$ jets, $Z+$ jets with $W, Z$ leptonic decays;
\item $W^+W^-, W^\pm Z$ and $ZZ$ with semi-leptonic decays;
\item single top production leading to $W^\pm b\ q$;
\item top pair production with fully leptonic and semi-leptonic decays.
\end{itemize}
%

%%%
\begin{figure}[tb]
\includegraphics[width=0.5\textwidth,clip=true]{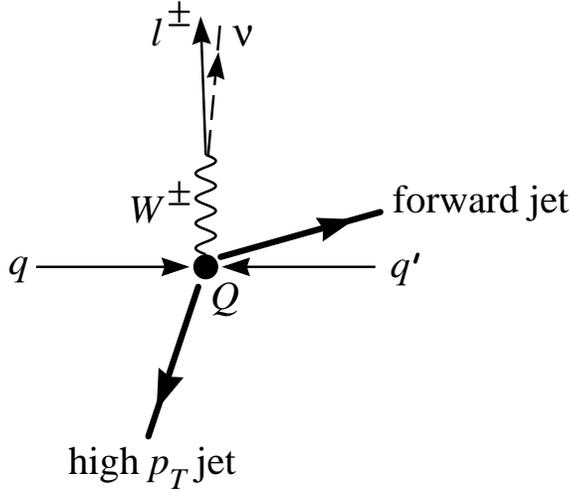}
\caption{ 
Illustration of the signal kinematics for a single heavy quark production.}
\label{fig:Q}
\end{figure}
%%%

An important aspect for our proposed search is the unique kinematics of single production and we discuss it first and present the detailed analysis next. A simple cartoon representation of the kinematic features is shown in Fig.~\ref{fig:Q}. First, one of the two jets is from the heavy quark decay that makes it very energetic with a Jacobian peak in the transverse momentum spectrum near 
\bea
p_T(j)\approx  {1\over 2} m_Q (1-M_V^2/m_Q^2)^{1/2}. 
\eea
Using the $p_T$ of the jets as a discriminant gives very good accuracy in identifying the jet coming from the decay of the heavy quark correctly, especially for high masses. Hence we identify the hardest jet ($j_1$) in the event as the one from heavy quark decay. The second jet  is associated with the $t$-channel exchange of a gauge boson. The collinear behavior of the radiation implies a forward outgoing jet with its transverse momentum governed by the propagator's mass, typically around $M_W/2$, and with a high rapidity. The pseudo-rapidity of the jet associated with the $t$-channel exchange of the gauge boson peaks at large values (typically $|\eta| \gsim  2$). The soft nature of the radiated gauge boson makes the accompanying jet energetic. These features can be used for forward jet tagging to enhance the signal over the backgrounds.  The forward tagged jet candidate is selected as the one with the highest deposited total energy and a minimum threshold in the pseudo-rapidity separation between the tag jet candidate and the decay products of the heavy quark, namely the reconstructed gauge boson and the highest $p_T$ jet, in addition to the basic cuts for jets during generation. These selection criteria accurately identify the forward tagged jet ($j_t$) as the one associated with the $t$-channel exchange of a gauge boson. 
Note that there is also an $s$-channel diagram where the jet is not necessarily forward but the contribution of this channel is negligible. 

Similar to the hard jet, the $W/Z$ from the heavy quark decay is also very energetic. This leads to rather collimated final state leptons in the signal while those leptons in the background still tend to be back-to-back.  The central hard jet and the energetic gauge boson $W/Z$ are both from the heavy quark decay and are thus nearly back-to-back in the transverse plane. The separation ($\Delta R$) of the central hard jet and the tagged jet is typically large for the signal as well, while that for the QCD background tends to present a collinear singularity. 

Of most importance is the reconstruction of the mass peak for the resonant particles. For a heavy quark decay with $Z\to \ell^+\ell^-$ in the final state, it is straightforward to form the invariant mass of the heavy quark $m_Q=M(j_1, Z)$ for the signal and backgrounds. For the final state with $W\to \ell\nu$ and $Z\to \nu\bar{\nu}$, one can define a  cluster transverse mass variable to be
\bea
\nn
M_T^2 = \left(\sqrt{p_{TW,Z}^2+M_{W,Z}^2} + p_{Tj_1}\right)^2 -
\left(\vec p_{TW,Z} + \vec p_{Tj_1} \right)^2.
\eea
However, even for the $W\to \ell\nu$ channel a full invariant mass can be constructed by including the $W$ rest mass approximation (details are presented in Sec.~\ref{sec:hqcoll-cc}). Next, we present some details about the simulation.    

We have performed a detailed simulation of both the signal and all the
relevant backgrounds using Madgraph/Madevent \cite{Alwall:2007st} for
event generation at the partonic level and PYTHIA \cite{Sjostrand:2006za} for parton showering with initial and final state radiation, as well as hadronization. 
Fast detector simulation was performed using {\sc Delphes}~\cite{Ovyn:2009tx}, version 1.8b, with the particle flow option turned off and the jets reconstructed using the anti-$K_{\rm T}$
algorithm with a radius of 0.7 and a minimum transverse momentum of 20 GeV.  All other parameters were set to default values that correspond to the ATLAS Detector.  Missing energy was manually corrected for the muons in the event.  

The CTEQ6L1 PDFs ~\cite{Pumplin:2002vw} were used for both signal and
background samples. For the signal we set the factorization and renormalization scales to be  $\mu_F=\mu_R=m_W (m_Z)$ for CC (NC) processes. Due to the nature of the vector boson fusion, it is known that the natural choice of the factorization scale is around the mass of the propagating particle  \cite{Han:1992hr, Figy:2004ec}. Varying the scale within some reasonable range could result in an uncertainty of about $10\%$. For the background processes the renormalization and factorization scales are set to  
$\mu_F = \mu_R = Q^2 = \sum (m^2 + p_T^2(j))$. To avoid double counting of the phase space overlap between the parton shower and matrix element parton generation, the modified MLM matching procedure ~\cite{Alwall:2007fs} with the $showerkt$ method~\cite{Alwall:2008pm} was applied for the background processes with multijets in the final state. The value of the parameter {\it xqcut}, which sets the minimum $K_T$ measure of partons in the matrix element generation, was set to 30 GeV. We summarize below the minimum cuts imposed in reconstructing the physics objects for our analysis. 
\bea
\nn
&& p_T(j_{1}) > 40\ {\gev},\ \ \ \ \ \ \ \ \  p_T^j > 20\ {\gev},\ \ \ \ p_T^\ell > 15\ {\gev},  \\
&& |\eta_j | < 5,\ \ \ \ |\eta_\ell| < 2.5,\ \ \ \Delta R(j\ell) > 0.4,\ \ \ \ \Delta R(\ell\ell) > 0.2.
\label{eq:basiccuts}
\eea

Various background cross sections are given in Table~\ref{tab:backgrounds}. Finally QCD di-jet samples at various hard-scattering momenta 
($\hat{p}_{\perp}$) were generated with {\sc Pythia} to be used as backgrounds to the $Z_{\nu\nu}+2$\,jets analysis.  The cross-section for the 
280\,GeV$<\hat{p}_{\perp}<$560\,GeV range, which is found to be the most relevant in terms of mimicking the signal, is 1.82\,nb at $7$ TeV collision energy.  

%%%%
\begin{table}\footnotesize
\caption{Cross section (in pb) for the various background processes with the acceptance cuts in
Eq.~(\ref{eq:basiccuts}). 
A requirement that $p_T(j_1)> 40$\,GeV was applied in all cases, except where indicated. }
\begin{centering}
\begin{tabular}{|c|c|c|}
\hline 
 Background  & $\sqrt{s}=$7\,TeV  & $\sqrt{s}=$14\,TeV  \\ \hline
$(W\to \ell \nu)$ + 1,2,3 jets  &   1070 &   2680     \\ \hline
$(W \to \ell \nu)$ + 2 jets &  &            \\
  $p_T(j_1)> 400$ GeV &      -   &  2.76       \\ \hline
$(Z\to \nu \nu)$ + 1,2,3 jets  &   402  &   1206       \\ \hline
$(Z\to \ell \ell)$ + 1,2,3 jets  &   97.8  &  247       \\ \hline 
$W^\pm_{\ell \nu}\ W^\mp_{jj}$ + 0,1,2 jets &  7.83  &      19.0     \\ \hline
$W^\pm_{\ell\nu}\  Z_{jj}$ + 0,1 jets  &   1.64 & 4.18            \\ 
$W^\pm_{jj}\  Z_{\ell\ell}$ + 0,1 jets  &   0.429 & 1.05            \\ \hline
$Z_{\ell\ell}\ Z_{jj}$ + 0,1,2 jets  & 0.291 & 0.689            \\ \hline
single top ($t \to \ell^\pm \nu b$) + 1,2 jets  &  7.24  & 28.8        \\ \hline
%  & 7.24  &  28.8        \\ \hline
$t \bar t$ fully leptonic  &   3.73  & 21.1        \\ \hline
$t \bar t$ semi leptonic   &   20.4  &  121         \\ \hline
\end{tabular}
\par\end{centering}
\label{tab:backgrounds}
\end{table}
%%%%

It is evident from Table~\ref{tab:backgrounds} that $W/Z+$jets is the leading background. However there is also a sizable contribution to the total background from $t \bar t$ production. While the cuts we describe later are sufficient to extract the signal from the large backgrounds, we list below some further improvements that can be made in order to reduce the large $t \bar t$ background. For the case of CC decay modes of the signal, we can impose a veto on a second isolated lepton to reduce the $\ttb$ background from the fully leptonic decay mode. For the NC decay mode of the signal with leptons in the final state, we can veto events with large missing energy. To further reduce $\ttb$ events, we can veto events with $b$-tags (this has virtually no effect on our signal). The semi-leptonic decay mode can be reduced further by vetoing events where any two jets reconstruct a $W$-boson or if any three jets reconstruct the top quark. These simple vetoes, together with the rest of the cuts allow us to reduce the $t \bar t$ background to negligible levels. In the next three subsections we present the CC and NC analyses making use of the unique kinematic features discussed above.

For the rest of the analyses, we consider a variety of heavy quark masses and study the reach at two 
LHC energies, as given below
\bea
\nonumber
&& m_{Q}= 300,\ 600,\  900\ {\rm GeV},\quad \sqrt s = 7\ {\rm TeV}; \\
&& m_{Q}= 900,\ 1800,\  2400\ ({\rm or}\ 3000)\ {\rm GeV},\quad \sqrt s = 14 \ {\rm TeV}.
\label{eqn:masses}
\eea
%

%%%%%%%%%%%%%%%%%%%%%%%%%%%%%%
\subsection{Neutral Current Channel with $Z$-Decays to Charged Leptons }
\label{sec:hqcoll-nclep}

In this section we study the case where the heavy quark decays into a jet and a $Z$ boson, with the  $Z$  boson decaying leptonically into electrons and muons  $(\ell^{\pm}=e^{\pm},\ \mu^{\pm})$ and  investigate the prospects of the LHC in this channel. Since the $Z$-boson's invariant mass can most easily be reconstructed using the charged leptons, the analysis starts with the selection of those. The basic acceptance cuts are listed in Eq.~(\ref{eq:basiccuts}). The invariant mass reconstructed using the two leptons is required to be within 25 GeV of the $Z$-boson mass peak to remove any possible fakes and misidentifications. Additionally, to select the boosted $Z$ bosons from the signal, the azimuthal angle between the two leptons is required to be smaller than a maximum, the value of which is optimized for the best signal significance. 

The event is also required to contain at least two jets isolated from leptons and other jets, the criteria for which are listed in Eq.~(\ref{eq:basiccuts}). The jet with the highest $p_T$ is selected as the one coming from the heavy quark decay. The tag jet candidate is selected as the jet with the highest deposited energy and a minimum threshold on the pseudorapidity separation between the tag jet candidate  and the decay products of the heavy quark, namely, the reconstructed $Z$ boson and the jet with the highest $p_T$. To take advantage of the forward nature of the tagged jet, thresholds optimized for best signal significance are applied for the pseudorapidity of the tagged jet. The efficiencies of these selection criteria in the order they are applied for various signal and background samples are shown in Table~\ref{tab:selection_Z_ll} for the 7 TeV and 14 TeV c.m. energy scenarios.

%%%%
\begin{table}\footnotesize
\caption{Selection criteria and the efficiencies (in percent) for the
  signal ($\epsilon_{m_Q}$) and background ($\epsilon_{BG}$) for the
  case of the neutral current channel with the $Z$ boson decaying to
  electrons or muons. Note that in evaluating $\epsilon_{BG}$ all the
  relevant backgrounds %listed in Sec.~\ref{sec:hqcoll} 
have been included.}
\begin{centering}
\begin{tabular}{|r||l|c|c||c|c|c||c|c|c|}
\hline
\multicolumn{1}{|c||}{} &\multicolumn{6}{|c||}{LHC 7 TeV} & \multicolumn{3}{|c|}{LHC 14 TeV}\tabularnewline
\hline 
 selection &  limit & $\epsilon_{600}$ & $\epsilon_{BG}$ & limit &  $\epsilon_{900}$ & $\epsilon_{BG}$ & limit & $\epsilon_{2400}$ & $\epsilon_{BG}$\tabularnewline
\hline
\hline 
min \# of $e/\mu$ &   2 & 92 & 78 & 2 & 95 & 78 & 2 & 98 & 54 \tabularnewline
\hline 
$|M_{\ell\ell}^{rec}-M_{Z}|<$ & 25 & 95 & 87 & 25 & 95 &  87 & 25 & 54 & 40 \tabularnewline
\hline 
$|\Delta\phi_{\ell,\ell}|<$ &  0.9 & 78 & 26 & 0.7 & 86 & 18 & 0.3 & 82 & 6 \tabularnewline
\hline 
min \# of jets &  2 & 92 & 67 & 2 & 92 & 66 & 2 & 93 & 90 \tabularnewline
\hline 
$|\Delta\eta_{j_1, j_t}|$, $|\Delta\eta_{Z,j_t}|>$ & 1.4 & 80 & 40 & 1.2 & 86 & 47 & 1.5 & 87 & 48 \tabularnewline
\hline 
$p^{j_1}_{T}>$ &  180 & 89 & 37 & 250 & 91 & 16 & 610 & 84 & 13 \tabularnewline
\hline 
$E_{j_t}>$ &  180 & 83 & 54 & 220 &83 & 46 & 380 & 86 & 51 \tabularnewline
\hline 
$\eta_{j_t}>$ &  1.3 & 97 & 96 & 1.8 & 90 & 87 & 0.8 & 100 & 99  \tabularnewline
\hline 
overall &   & 36 & 0.92 &  & 41 & 0.25 &  & 26 & 0.034 \tabularnewline
\hline
\end{tabular}
\par\end{centering}
\label{tab:selection_Z_ll}
\end{table}
%%%%

\begin{figure}[tb]
{\includegraphics[width=8.55cm,clip=true]{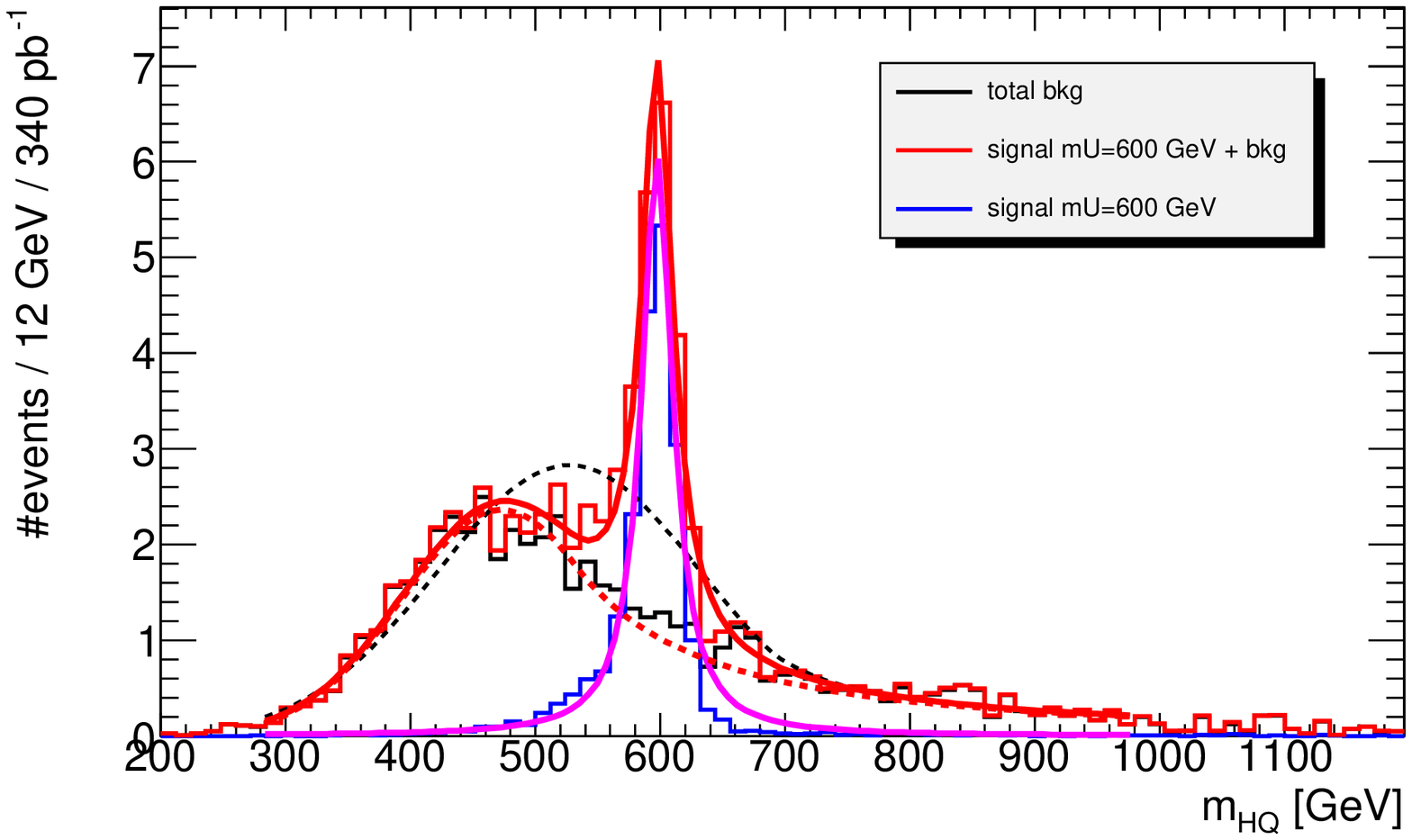}
\includegraphics[width=7.75cm,clip=true]{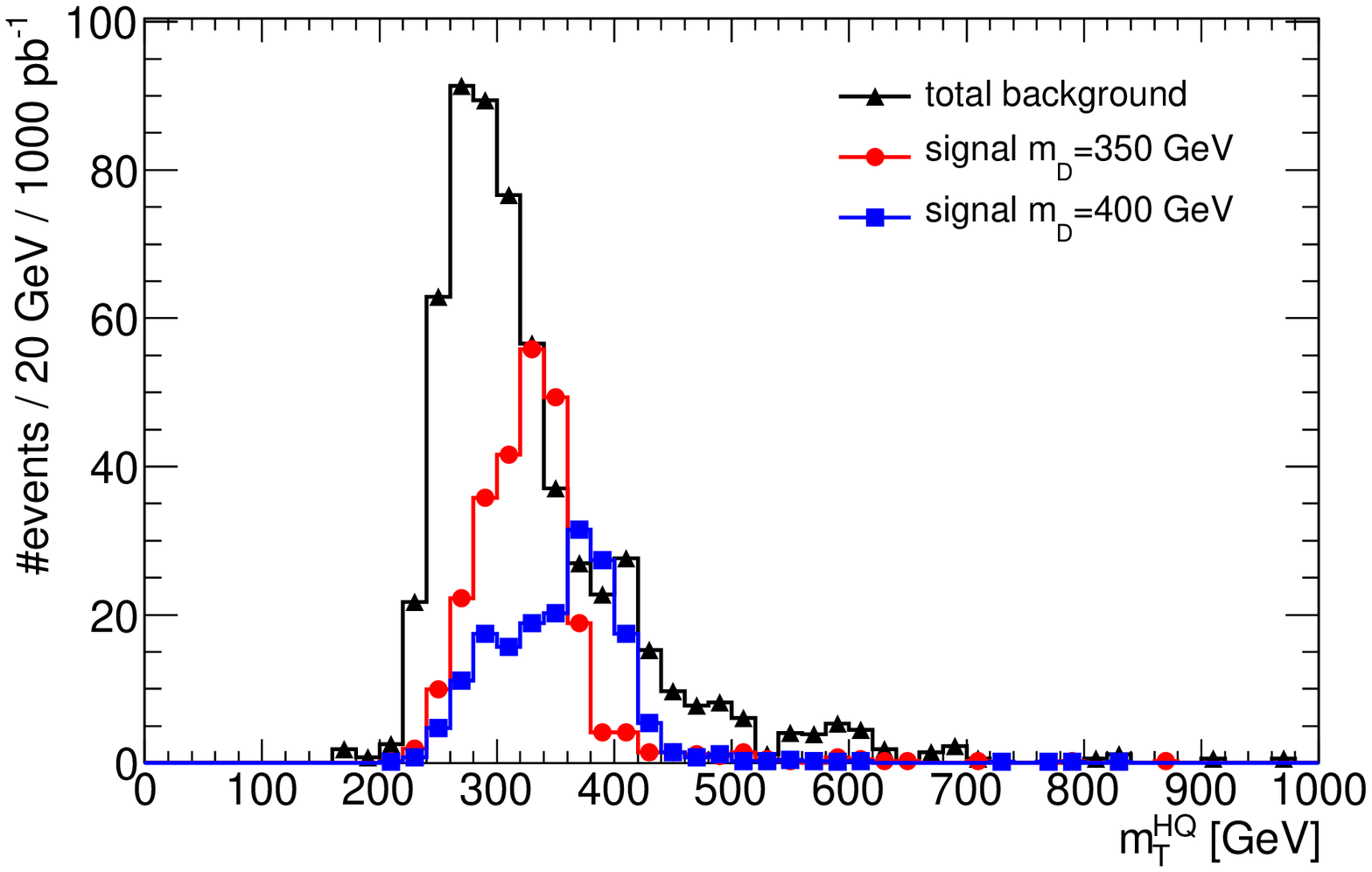}}
\caption {Reconstructed heavy quark distributions at the LHC with $\sqrt{s} = 7$ TeV. (a) left: invariant mass of the heavy quark of $U$ type and mass 600 GeV decaying leptonically via NC interactions. See text for a description of all the curves; (b) right: transverse mass of the heavy quark candidates of $D$ type with mass 350 and 400 GeV decaying invisibly into neutrinos via NC interactions (after all the selection criteria have been applied).} 
\label{fig:disthq-nc}
\end{figure}

The final heavy quark ($Q$) candidate's invariant mass is formed using
the reconstructed Z boson candidate and the jet with the highest
transverse momentum. The results of this are shown in
Fig.~\ref{fig:disthq-nc}(a) for a heavy $U$ type quark with mass of
600 GeV for $\sqrt{s} = 7$ TeV. The black, blue and red histograms
represent background, signal and signal plus background
respectively. In order to extract the signal significance, a series of
successive fits are performed to the signal+background sample. The
procedure starts with a Landau functional fit which should describe
the overall behaviour of the data set. The values of the parameters
obtained from the Landau fit are used as initial values in a second
fit which uses a Crystal Ball function~\cite{R-xtal-ball} . In the
absence of signal the Crystal Ball function is expected to describe
the sample behavior in an ideal way. The obtained curve is shown with
a dashed black line in Fig.~\ref{fig:disthq-nc}(a). If the signal
hypothesis is true, some mismatch between the fitted function and the
histograms is expected. Therefore a new fit is performed to
investigate the signal hypothesis. The fit function is the sum of two
components: a Crystal Ball term to represent the  background and a
Breit-Wigner or a Gauss term  term to represent the signal
resonance. The selection between Breit-Wigner and Gauss terms is made
based on the mass of the sought heavy quark: up to 1 TeV, the signal
is narrow, thus best represented by the former and after 1 TeV, the
signal distribution becomes fatter thus compatible with the
latter. The obtained curve is shown with a red solid line in the
invariant mass distributions of Fig.~\ref{fig:disthq-nc}(a). The
background and signal components of this function are also shown on
the same plot as dashed red and solid magenta lines,
respectively. Therefore a good match between the signal (solid blue
histogram) and its estimation (solid magenta line) shows the power of
the method described above.  

%%%
%%%

Finally, we mention some details about the signal simulation
sample. We have simulated signal samples for a $D$ type quark for the 7 TeV
analysis and a $U$ type quark for the 14 TeV analysis with
$\tilde\kappa = 1$ and $Br(Q \to qZ) = 100\%$. While the efficiencies
we list in Table~\ref{tab:selection_Z_ll} are from one quark sample
for each c.m. energy, we have checked that there is no significant
change in the efficiencies while using the other quark sample at the
same energy. Note that the choice of the quark sample and the specific
model parameters used does not affect the model-independent quantity
$S^Z$ inferred from an excess of signal events over the background at
a given luminosity. The choice of the quark sample does however affect the luminosity required for evidence or discovery of such a quark and the constraints on the various model parameters inferred from it.  

%%%%%%%%%%%%%%%%%%%%%%%%%%%%%%
\subsection{Neutral Current Channel with $Z$-Decays to Neutrinos}
\label{sec:hqcoll-ncnu}

In this section we study the case where the heavy quark decays into a jet and a $Z$ boson, with the  $Z$  boson decaying invisibly into neutrinos. An accurate reconstruction of the neutral channel in the $Z_{\nu\nu}+$jets final state is beyond the scope of this paper, largely due to difficulties in simulating the measurements of large missing transverse energy. Even with detailed GEANT4 simulations of the detectors, fake $E_{T}^{{\rm miss}}$ from event mismeasurements (due to non-Gaussian tails in the detector jet responses) is unlikely to be determined with confidence before studies are done with control samples in actual LHC data. However, this final state can still offer some supplementary evidence in the heavy quark searches as we show in this subsection. As an example, we consider the discovery potential for two relatively-low quark masses (350 and 400$\,$GeV) with 1$\,$fb$^{-1}$ of data at $7$ TeV c.m. energy and describe the analysis in detail below. 

 After all the selection criteria (as listed in Table$~$\ref{tab:selection_Z_nunu}), the dominant background is found to be the irreducible background from the Standard Model $Z_{\nu\nu}+$jets processes. A subdominant contribution is from the $W_{\ell\nu}+$jet events, where the charged lepton falls outside of the detector acceptance or is lost in the reconstruction. We also consider semileptonic $t\bar{t}$ and jet-associated single top as other sources of background and find them essentially negligible.

%%%%
\begin{table}\footnotesize
\caption{Selection criteria and the efficiencies (in percent) for the signal and background for the case of the neutral current channel with the $Z$ boson decaying to neutrinos. Note that $j_2$ here is the same as $j_t$. }
\begin{centering}
\begin{tabular}{|r||l|c|c|c|c|}
\hline
\multicolumn{1}{|c||}{} &\multicolumn{5}{|c|}{LHC 7 TeV} \tabularnewline
\hline 
selection & limit & $\epsilon_{350}$ & $\epsilon_{400}$ & $\epsilon_{Z_{\nu\nu}+jets}$ & $\epsilon_{W_{\ell\nu}+jets}$ \tabularnewline
\hline
\hline 
\# of $e/\mu$ $< $ & 1 &  99.5 & 99.4 & 99.5 & 10 \tabularnewline
\hline 
$\etmiss>$ & 100 & 73 & 81 & 13 & 11 \tabularnewline
\hline 
min \# of jets & 2 & 92 & 92 & 62 & 65 \tabularnewline
\hline 
$p^{j_1}_{T}>$ & 100 & 91 & 94 & 63 & 59 \tabularnewline
\hline 
$p^{j_2}_{T}>$ & 40 & 77 & 80 & 58 & 52 \tabularnewline
\hline 
$|\eta_{j_2}|>$ & 1.0 & 74 & 73 & 51 & 57 \tabularnewline
\hline 
$|\Delta\eta_{j_1,j_2}|>$ & 3.0 & 45 & 46 & 15 & 19 \tabularnewline
\hline 
$E_{j_2}>$ & 100 & 99 & 99 & 97 & 99 \tabularnewline
\hline 
$|\Delta\phi_{j_1,j_2}|<$ & 2.7 & 86 & 84 & 82 & 87 \tabularnewline
\hline 
$|\Delta\phi_{\ptmiss,j_{k=1,2,3}}| >$ & 0.5 & 79 & 78 & 73 & 59 \tabularnewline
\hline 
overall &  & 10.5 & 12.1 & 0.13 & 0.01 \tabularnewline
\hline 
\end{tabular}
\par\end{centering}
\label{tab:selection_Z_nunu}
\end{table}
%%%%
As can be seen in the table, the requirements on $|\Delta\phi|$ between the two leading jets and between any of the three leading jets and the missing transverse momentum appear to play no significant role in separating the signal from the main sources of background. However, we prefer to apply these cuts, since in Ref.~\cite{atlas-csc} they have been shown to suppress backgrounds from QCD processes. All the cuts are tested on Pythia di-jet samples generated at various hard scattering momenta, and while we fail to reproduce the results
from this reference due to limitations of the fast simulation, we observe that the cuts on the tag jet also provide an extra handle in reducing the QCD backgrounds to negligible levels. Note that for this channel, the tag jet also turns out to be the jet with the second highest transverse momentum {\it i.e.} $j_t \equiv j_2$ and the tag jet is denoted as $j_2$ in this section. 

Attributing the total missing momentum in the event to the $Z$ boson,
the transverse mass of the heavy quark candidate is reconstructed as
shown in Fig.~\ref{fig:disthq-nc}(b). While it is clear that the
signal and background distributions have distinct shapes and therefore
the signal could be extracted using a fitting technique, we
conservatively choose to do a simple counting measurement for this
final state. Since most of the background is from the
$Z_{\nu\nu}+$jets and to a lesser extent from the $W_{\ell\nu}+$jet
events, there are natural control samples obtained by reverting the
charged lepton veto done at the beginning of the event selection. For
example $Z_{\ell\ell}+$jets events can be used to determine the amount
of $Z_{\nu\nu}+$jets background. Therefore for an estimate of the
expected significance, we calculate $s/\sqrt{b^\prime}$, where $s$ is the
expected number of signal events and $b^\prime$ includes the expected
uncertainty in the background estimation. For simplicity, if we assume  
$b^\prime=b\times BR(Z\to\nu\nu)/BR(Z\to\ell\ell)$, where 
$\ell=e,\mu$ and $b$ is the expected total number of background
events, we find that the significance for the $350\ (400)$ GeV quark
signal is about $6.0\  (4.2) \sigma$ with $1$ fb$^{-1}$ of data. 

Next we give some details about the signal simulation sample. We
have simulated signal samples for $D$ type quarks with $\tilde\kappa = 1$
and $Br(D \to dZ) = 100\%$. From the studies of the neutral current
channel decaying to charged leptons and the charged current channel
decaying leptonically, we have found that  there is no significant
change in the efficiencies while using a $U$ quark sample. Hence,
using the same efficiencies as listed in
Table~\ref{tab:selection_Z_nunu} for a $U$ type quark, which has a
larger production cross section compared to the $D$ type quark, would
have resulted in a smaller luminosity required for evidence or
discovery. For simplicity, we only give the results from the $D$ type
quark as an illustration that  this channel is a viable mode for
evidence or discovery, without loss of generality, as discussed in
the previous section. 

%%%%%%%%%%%%%%%%%%%%%%%%%%%%%%
\subsection{Charged Current Channel}
\label{sec:hqcoll-cc}

In this section we study the case where the heavy quark decays into a jet and a $W$ boson, with the  $W$  boson decaying leptonically $W\to \ell \nu\ (\ell=e,\mu)$, and  investigate the prospects of the LHC in this channel. Once again, we adopt the mass parameters as in Eq.~(\ref{eqn:masses}), and describe the analysis in detail below. 

The simplest way of reconstructing the W boson candidate in the
charged channel is via its leptonic decays and hence the
reconstruction algorithm starts with the selection of events
containing at least one electron or muon ($\ell$) with the selection
criteria described in Eq.~(\ref{eq:basiccuts}). In the case where both
flavours of leptons are available, the lepton with the higher
transverse momentum is selected as the W decay product
candidate. After selecting events with a minimum missing transverse
energy ($\etmiss$), the four momentum of the $W$ boson is obtained
using the information from $\etmiss$ and $p^\ell_T$ and including the
W rest mass approximation. The two-fold ambiguity in the solution of
the W boson longitudinal momentum is handled by choosing the most
central (minimum $|\eta_W|$) solution. We find that, while this method
is only slightly better than a random choice at very high quark
masses, it reconstructs the true momentum about 65\% of the time for
low heavy quark masses, and improves the resolution of the
reconstructed heavy quark mass. The azimuthal angle between the
charged and neutral lepton is required to be smaller than a threshold
(again optimized for each mass point for the best signal significance)
to select the boosted $W$ bosons.  

The event is required to contain at least two jets isolated from leptons and other jets, the criteria for which are listed in Eq.~(\ref{eq:basiccuts}). The jet with the highest transverse momentum is assumed to be the decay product of the heavy quark. We identify the tag jet as the jet with the highest energy deposition in the event that is at least a minimum $|\Delta \eta|$ apart from both objects that we use to reconstruct the heavy quark candidate {\it i.e.} the tag jet has to be well separated from both the reconstructed $W$ boson and the highest $p_T$ jet. 

We introduce a new variable called "sum $p_T$" which is defined as the absolute value of the vector sum of the tagged jet and the reconstructed heavy quark transverse momenta to enhance the signal over backgrounds,
\beq
 p_T^{sum} \equiv \Bigl | \vec {p}^{\ j_t}_{\scriptstyle {T}} + \vec
 {p}^{\ Q}_{\scriptstyle {T}} \Bigr |.  \label{pT:sum:def}
\eeq
At the leading order, this variable is zero for the signal but detector effects and initial state radiation smear this distribution. However the distribution peaks at very small values for the signal while it is typically large for busy background events, in particular top pair production. We require the $p^{sum}_T$ to be smaller than a threshold to enhance the signal over the backgrounds. This cut reduces the top pair background significantly for the 14 TeV scenario while also having a reasonable impact for the 7 TeV analysis. The efficiencies of these selection criteria in the order they are applied for various signal and background samples are shown in Table~\ref{Tab:CC-effs} for the 7 TeV and 14 TeV c.m. energy scenarios and different mass values of the heavy quark. The superscript in the efficiencies of the backgrounds (for e.g. $i$ in $\epsilon^i_{BG}$) correspond to $W+$jets, diboson, top pair and single top for $i=0,1,2,3$ respectively. 

\begin{figure}[tb]
{\includegraphics[width=8.15cm,clip=true]{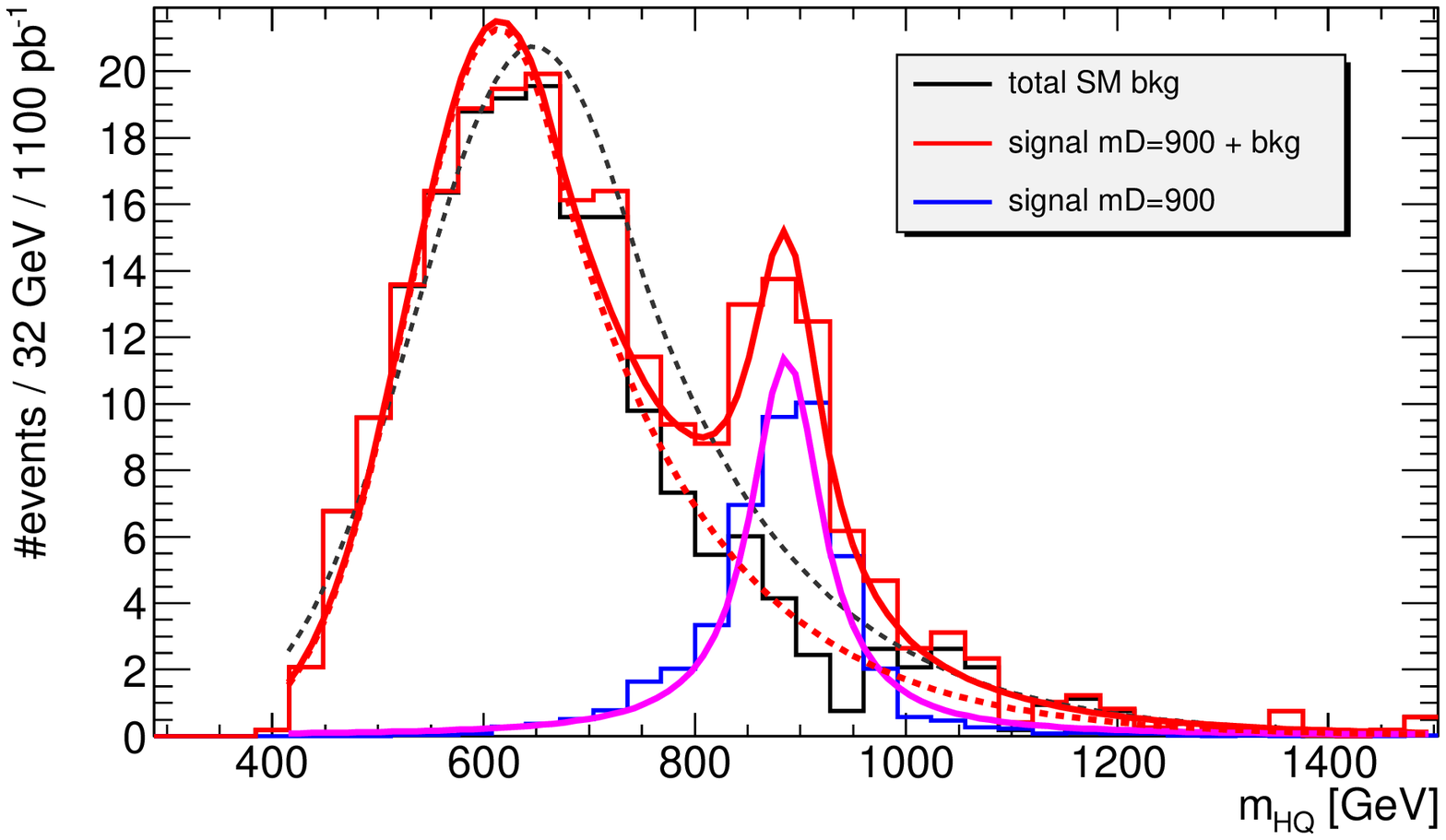}
\includegraphics[width=8.15cm,clip=true]{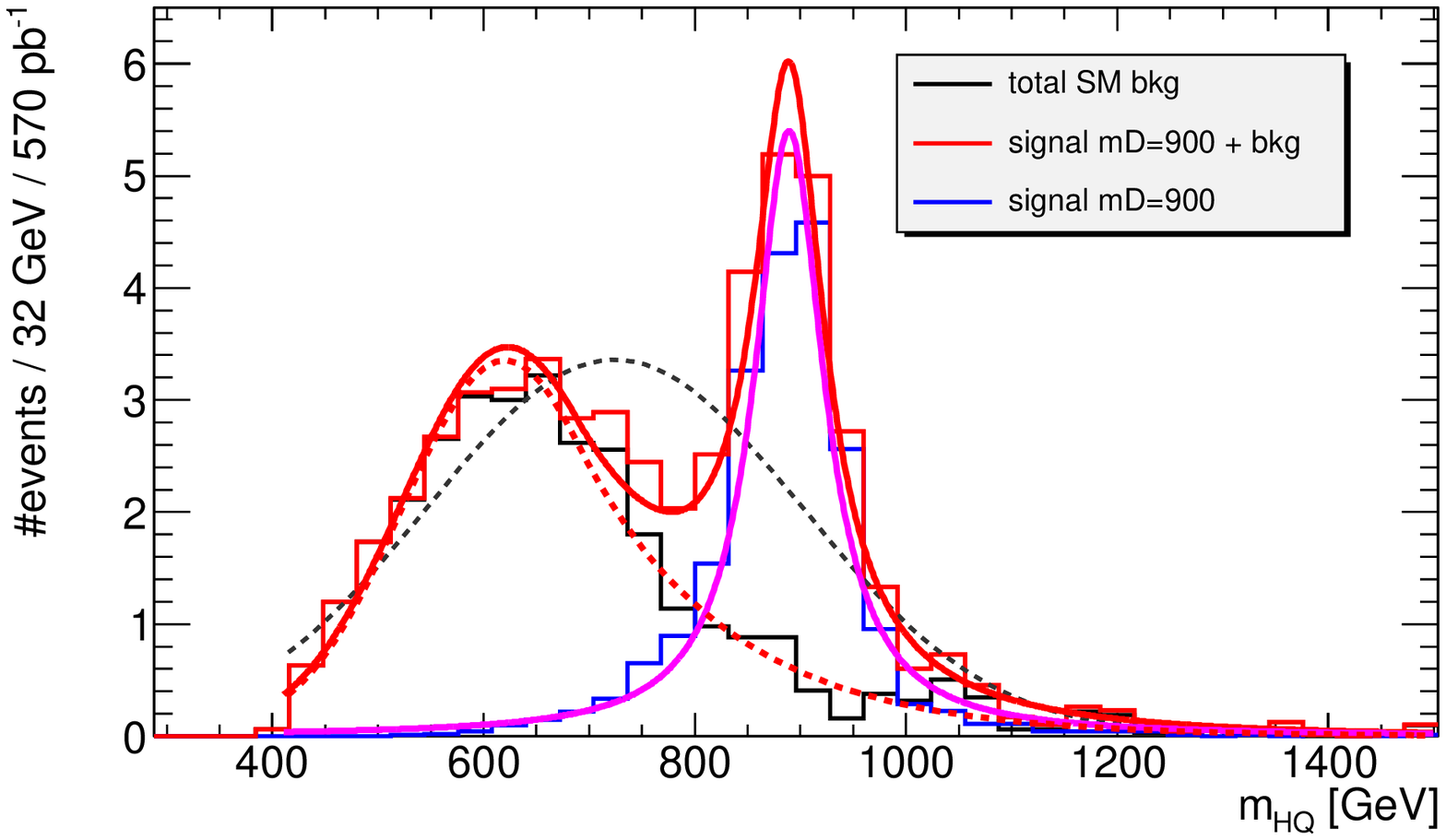}}
\caption {Reconstructed heavy quark distributions for a quark of type $D$ and mass 900 GeV decaying via CC interactions at the LHC with $\sqrt{s} = 7$ TeV. (a) left: no charge identification of the charged lepton has been made; (b) right: with charge identification of the negatively charged lepton. See Sec.~\ref{sec:hqcoll-nclep} for a description of all the curves. } 
\label{fig:disthq-cc}
\end{figure}

The leading jet and the $W$ boson are combined to obtain the reconstructed invariant mass of the heavy quark candidate. The procedure to extract the signal significance is the same as described earlier in Sec.~\ref{sec:hqcoll-nclep}. The invariant mass distribution for a heavy quark of $D$ type and mass 900 GeV  decaying leptonically via CC channel is shown in Fig.~\ref{fig:disthq-cc}. In Fig.~\ref{fig:disthq-cc}(a) we show the reconstructed heavy quark where the charge of the lepton is not identified and in Fig.~\ref{fig:disthq-cc}(b) when the charge of the negatively charged lepton coming from the decay of the heavy quark is identified. When we do not identify the charge of the lepton, we have to include signals from both quark and anti-quark and the background would include the contribution from both $W^\pm +$jets. This method adds a small component to the signal cross section as the anti-quark cross sections are much smaller than quark cross sections while adding significantly to the backgrounds as positively charged $W$ bosons are produced preferentially at a $pp$ collider such as the LHC. This reduces the signal significance. However identifying the charge of the negative lepton enhances the signal over the backgrounds and reduces the luminosity required for discovery by almost half. For the heavy quarks decaying into positively charged leptons (say $U$ and $X$), again the anti-quark contributions add very little to the signal but the reduction in the background is much smaller as $W^+$ are produced preferentially at the LHC.  However, there is still some improvement in the signal significance by identifying the charge of the positive lepton though not as much as in the case of the negative leptons. The effect of identifying the charge of the lepton (for both positive and negatively charged leptons) is shown in Table~\ref{Tab:CC-effs}. 

%%%
%%%

%%%%
\begin{table} \footnotesize
\caption{Selection efficiencies (in percent) for signal ($\epsilon_{m_Q}$) and background ($\epsilon^i_{BG}$) samples for the case of the charged current channel with leptonic decays into electrons and muons. The superscript in the efficiencies of the backgrounds (for e.g. $i$ in $\epsilon^i_{BG}$) correspond to $W+$jets, diboson, top pair and single top for $i=0,1,2,3$ respectively. }
\label{Tab:CC-effs}
{
%\scriptsize
%tiny
\centering{}\begin{tabular}{|r||c|c|c|c|c|c||c|c|c|c|c|c||c|c|c|c|c|c|}
\hline 
\multicolumn{1}{|c||}{} &\multicolumn{12}{|c||}{LHC 7 TeV} & \multicolumn{6}{|c|}{LHC 14 TeV}\tabularnewline
\hline
{\footnotesize  selection} &  {\footnotesize  limit} & { \footnotesize $\epsilon_{600}$} & {\footnotesize  $\epsilon_{BG}^{0}$} & {\footnotesize  $\epsilon_{BG}^{1}$} & {\footnotesize  $\epsilon_{BG}^{2}$} & { \footnotesize $\epsilon_{BG}^{3}$} & {\footnotesize  limit} & {\footnotesize  $\epsilon_{900}$} & {\footnotesize  $\epsilon_{BG}^{0}$} & {\footnotesize  $\epsilon_{BG}^{1}$} & {\footnotesize  $\epsilon_{BG}^{2}$} & {\footnotesize  $\epsilon_{BG}^{3}$} & {\footnotesize  limit} & {\footnotesize  $\epsilon_{3000}$} & {\footnotesize  $\epsilon_{BG}^{0}$} & {\footnotesize  $\epsilon_{BG}^{1}$} & {\footnotesize  $\epsilon_{BG}^{2}$} & {\footnotesize  $\epsilon_{BG}^{3}$}\tabularnewline
\hline
\hline 
{\footnotesize  min \# } &   &  &  & & & & & & & & & \tabularnewline
{\footnotesize  of $e/\mu$ } &  {\footnotesize  1} & {\footnotesize  96} & {\footnotesize  91} & {\footnotesize  91} & {\footnotesize  94} & {\footnotesize  92} & {\footnotesize  1} & {\footnotesize  97} & {\footnotesize  91} & {\footnotesize  91} & {\footnotesize  94} & {\footnotesize  92} &  { \footnotesize 1} &  {\footnotesize  99} &  {\footnotesize  98} &  {\footnotesize  90} &  { \footnotesize 96} &  {\footnotesize  96} \tabularnewline
\hline 
{\footnotesize  $\etmiss >$} &  {\footnotesize  30} & {\footnotesize  94} & {\footnotesize  76} & {\footnotesize  70} & {\footnotesize  80} & {\footnotesize  80} & {\footnotesize  50} & {\footnotesize  92} & {\footnotesize  57} & {\footnotesize  35} & {\footnotesize  54} & {\footnotesize  52} &  {\footnotesize  110} &  {\footnotesize  96} &  {\footnotesize  62} &  {\footnotesize  8} &  {\footnotesize  39} &  { \footnotesize 65} \tabularnewline
\hline 
{\footnotesize  $\nu$ solution} &  & {\footnotesize  82} & {\footnotesize  80} & {\footnotesize  74} & {\footnotesize  73} & {\footnotesize  74} &  & {\footnotesize  81} & {\footnotesize  80} & {\footnotesize  70} & {\footnotesize  71} & { \footnotesize 71} & &  {\footnotesize  67} &  {\footnotesize  78} &  {\footnotesize  76} &  {\footnotesize  77} &  {\footnotesize  77} \tabularnewline
\hline 
{\footnotesize  $|\Delta\phi_{\ell,\nu}|<$} & {\footnotesize  0.8} & {\footnotesize  86} & {\footnotesize  41} & {\footnotesize  13} & {\footnotesize  31} & {\footnotesize  19} & {\footnotesize  0.6} & {\footnotesize  91} & {\footnotesize  29} & {\footnotesize  13} & {\footnotesize  25} & {\footnotesize  14} &  {\footnotesize  0.5} &  {\footnotesize  99} &  {\footnotesize  84} &  {\footnotesize  50}  &  {\footnotesize  63} &  {\footnotesize  71} \tabularnewline
\hline 
{\footnotesize  min \# } & & & & & & & & & & & & & & & & & &\tabularnewline
{\footnotesize  of jets}  & { \footnotesize 2} & {\footnotesize  91} & {\footnotesize  63} & {\footnotesize  89} & {\footnotesize  100} & {\footnotesize  95} & {\footnotesize  2} & {\footnotesize  90} & {\footnotesize  63} & {\footnotesize  90} & {\footnotesize  100} & {\footnotesize  95} &  {\footnotesize  2} &  {\footnotesize  90} &  {\footnotesize  96} &  {\footnotesize  95}  &  {\footnotesize  99} &  {\footnotesize  95} \tabularnewline
\hline 
{\footnotesize  $|\Delta\eta_{j_t,W}|$,} & & & & & & & & & & & & & & & & & &\tabularnewline
{\footnotesize $|\Delta\eta_{j_t,j_1}|>$} &  {\footnotesize  1.8} & {\footnotesize  68} & {\footnotesize  28} & {\footnotesize  20} & {\footnotesize  30} & { \footnotesize 28} & {\footnotesize  2} & {\footnotesize  65} & { \footnotesize 24} & {\footnotesize  17} & {\footnotesize  25} & { \footnotesize 23} &  {\footnotesize  2.3} &  {\footnotesize  72} &  {\footnotesize  31} &  {\footnotesize  25} &  {\footnotesize  32} &  {\footnotesize  30} \tabularnewline
\hline 
{\footnotesize  $p^{j_1}_{T}>$} &  {\footnotesize  200} & {\footnotesize  82} & {\footnotesize  25} & {\footnotesize  26} & {\footnotesize  18} & {\footnotesize  15} & {\footnotesize  250} & {\footnotesize  90} & {\footnotesize  17} & {\footnotesize  24} & {\footnotesize  14} & {\footnotesize  11} &  {\footnotesize  450} &  {\footnotesize  98} &  {\footnotesize  67} &  {\footnotesize  53} &  {\footnotesize  24} &  {\footnotesize  70} \tabularnewline
\hline 
{\footnotesize  $E_{j_t}>$} &  {\footnotesize  200} & {\footnotesize  78} & {\footnotesize  61} & {\footnotesize  60} & {\footnotesize  71} & {\footnotesize  62} & {\footnotesize  300} & {\footnotesize  74} & {\footnotesize  52} & {\footnotesize  47} & {\footnotesize  60} & {\footnotesize  52} &  {\footnotesize  270} &  {\footnotesize  95} &  {\footnotesize  72} &  {\footnotesize  73} &  {\footnotesize  79} &  {\footnotesize  70} \tabularnewline
\hline 
{\footnotesize  $|\eta_{j_t}|>$} &   {\footnotesize  0.6} &  {\footnotesize  100} & {\footnotesize  100} & {\footnotesize  100} & {\footnotesize  100} & {\footnotesize  100}  &  {\footnotesize  0.6} & {\footnotesize  100} & {\footnotesize  100} & {\footnotesize  100} & {\footnotesize  100} & {\footnotesize\footnotesize  100} &  {  \footnotesize 0.6} &  {\footnotesize  100} &  {\footnotesize  100} &  {\footnotesize  100} &  {\footnotesize  100} &  {\footnotesize  100}  
\tabularnewline
\hline
{\footnotesize  $p_T^{sum}$} & {\footnotesize  100}  &  {\footnotesize  94} & {\footnotesize  76} & {\footnotesize  56} & {\footnotesize  47} & {\footnotesize  47} & {\footnotesize  100} & {\footnotesize  93} &{\footnotesize  74} & {\footnotesize  56} & {\footnotesize  41} & {\footnotesize  40}  &  {\footnotesize  100} &  {\footnotesize  74} &  {\footnotesize  48} &  {\footnotesize  38} &  {\footnotesize  18} &  {\footnotesize  18} \tabularnewline
%\hline
%{\footnotesize  $|\vec {p}^{\ j_t}_{\scriptstyle {T}} + \vec {p}^{\ Q}_{\scriptstyle {T}}|$} & {\footnotesize  100}  &  {\footnotesize  94} & {\footnotesize  76} & {\footnotesize  56} & {\footnotesize  47} & {\footnotesize  47} & {\footnotesize  100} & {\footnotesize  93} &{\footnotesize  74} & {\footnotesize  56} & {\footnotesize  41} & {\footnotesize  40}  &  {\footnotesize  100} &  {\footnotesize  74} &  {\footnotesize  48} &  {\footnotesize  38} &  {\footnotesize  18} &  {\footnotesize  18} \tabularnewline
\hline
 %{\footnotesize  overall } &  & {\footnotesize  24} & {\footnotesize  0.47} & {\footnotesize  0.096} & {\footnotesize  0.31} & {\footnotesize  0.11} &  & {\footnotesize  24} & {\footnotesize  0.12} & {\footnotesize  0.028} & {\footnotesize  0.074} & { \footnotesize 0.025} & &  {\footnotesize  28} &  {\footnotesize  2.8} &  {\footnotesize  0.090} &  {\footnotesize  0.20} &  { \footnotesize 0.85}  \tabularnewline
%\hline
\hline 
  {\footnotesize  $Q_\ell=-1$ } &   & {\footnotesize  85} & {\footnotesize  31} & {\footnotesize  39} & {\footnotesize  49} & {\footnotesize  28} &  & {\footnotesize  88} & {\footnotesize  31} & {\footnotesize  50} & {\footnotesize  47} & {\footnotesize  36} & &  {\footnotesize  93} &  {\footnotesize  29} &  {\footnotesize  45} &  {\footnotesize  45} &  {\footnotesize  29} \tabularnewline
\hline 
 { \footnotesize overall } &  & {\footnotesize  20} & {\footnotesize  0.15} & {\footnotesize  0.037} & {\footnotesize  0.15} & {\footnotesize  0.031} &  & { \footnotesize 21} & { \footnotesize 0.036} & {\footnotesize  0.014} & {\footnotesize  0.035} & {\footnotesize 0.0089} & &  {\footnotesize  26} &  {\footnotesize  0.82} &  {\footnotesize  0.041} &  {\footnotesize  0.088} &  {\footnotesize  0.25} \tabularnewline
\hline
{  $Q_\ell=+1$ } &   & 90 & 69 & 61 & 51 & 72 &  & 94 & 69 & 50 & 53 & 64 & & 97 & 71 & 55 & 55 & 71 \tabularnewline
\hline 
 {  overall } &   & 24 & 0.32 & 0.059 & 0.16 & 0.079 &  & 25 & 0.083 & 0.014 & 0.039 & 0.016 & & 29 & 1.99 & 0.050 & 0.11 & 0.60 \tabularnewline
\hline
\end{tabular}
}
\end{table}
%%%%

\begin{figure}[tb]
{\includegraphics[width=0.495\textwidth,clip=true]{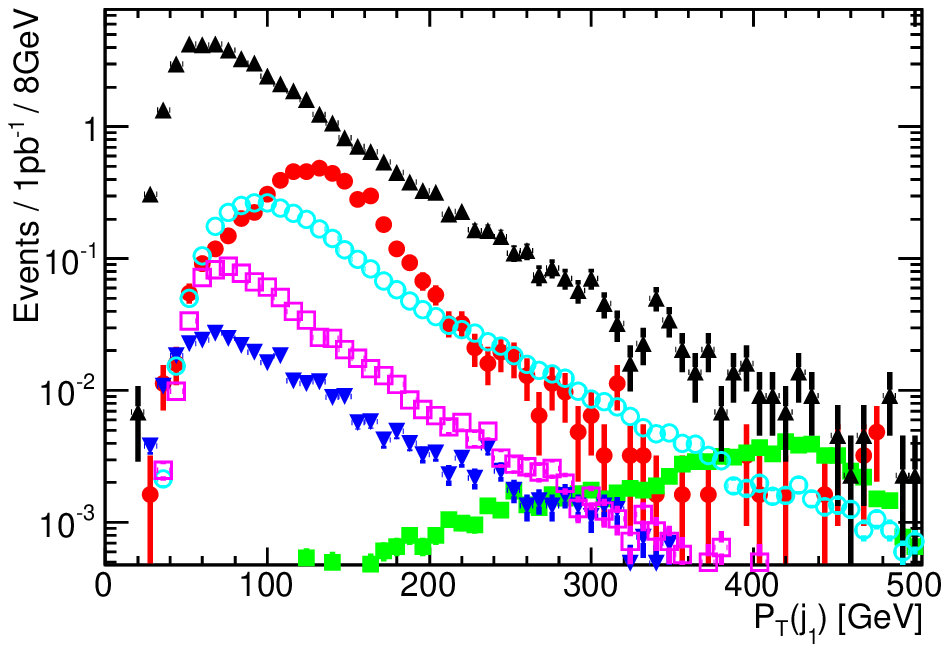}
\includegraphics[width=0.495\textwidth,clip=true]{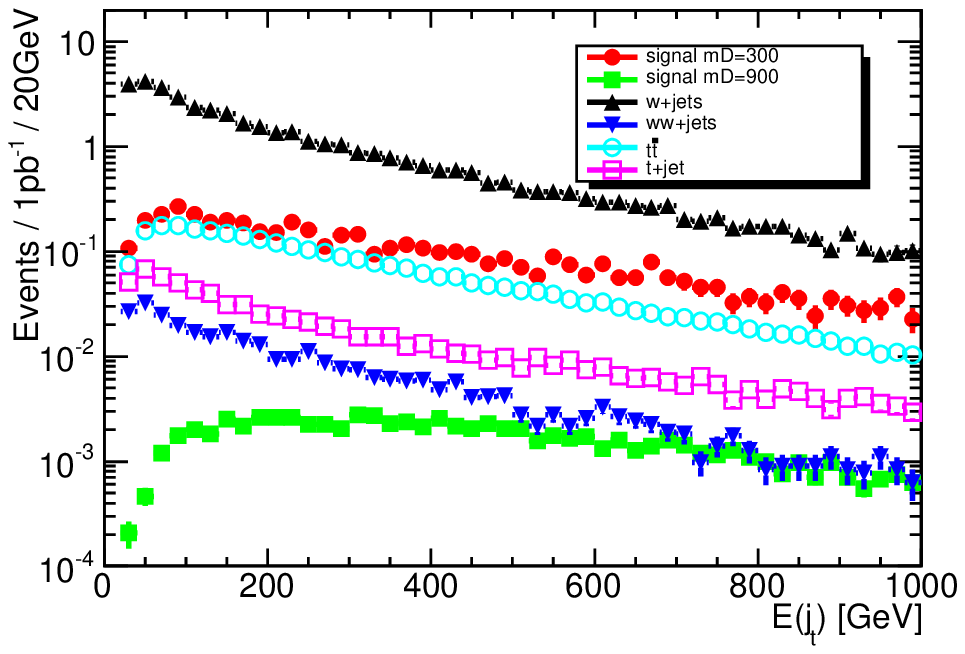}
\includegraphics[width=0.495\textwidth,clip=true]{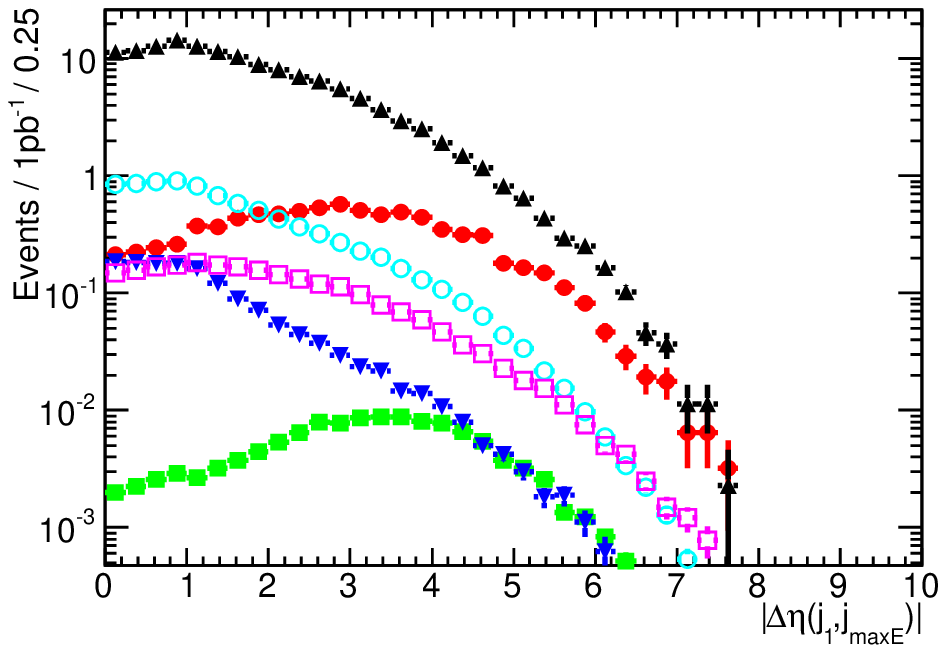}
\includegraphics[width=0.495\textwidth,clip=true]{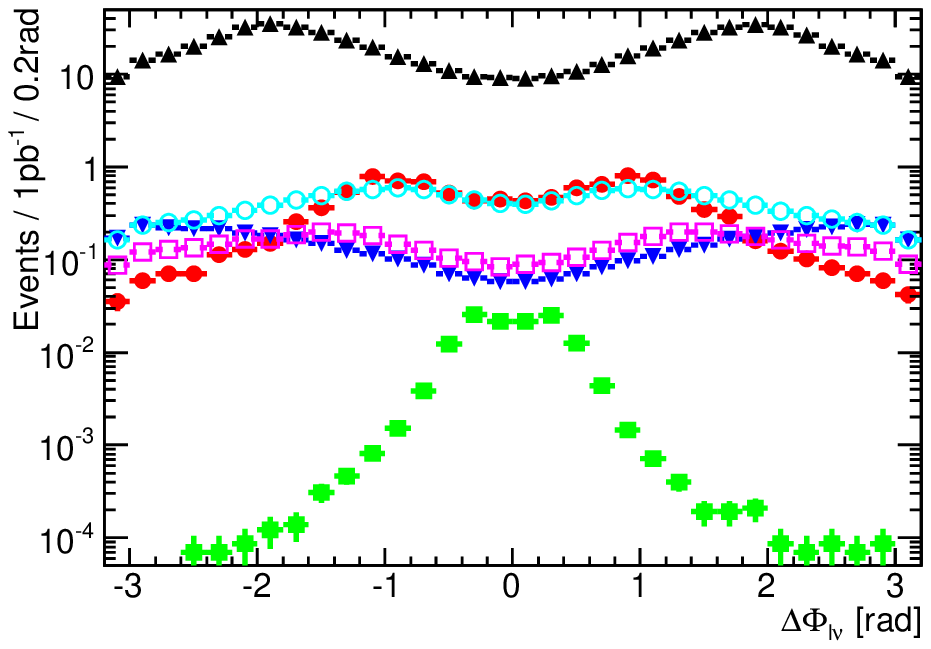}}
\caption{ (a) Top left: $p_T$ of the hardest jet. This jet is used in reconstructing the heavy quark mass; (b) top right: total energy of the forward, tagged jet (c) bottom left: difference in pseudo-rapidity between the hardest jet and the tagged jet (d) bottom right: azimuthal angle between $p_T(\ell)$ and $\ptmiss$.}  
\label{fig:dist}
\end{figure}

We show some sample distributions for a few of the relevant kinematic
quantities which are unique to single production of heavy quarks and
enable us to separate the signal from the background in
Fig.~\ref{fig:dist}. The $p_T$ distribution of the hard jet and the
energy distribution of the forward tagged jet are shown in
Fig.~\ref{fig:dist}(a) and Fig.~\ref{fig:dist}(b) respectively. The
difference in pseudo-rapidity between the hard jet and the forward
tagged jet is shown in Fig.~\ref{fig:dist}(c) and the azimuthal angle
separation between the leptons ($p^\ell_T$ and $\ptmiss$) is shown in
Fig.~\ref{fig:dist}(d). All distributions in Fig.~\ref{fig:dist} are
normalized to 1 pb$^{-1}$ integrated luminosity. 

%%%
%%%

We mention some details about the signal simulation sample. We simulate signal samples for both $D$ and $U$ type quarks with $\tilde\kappa = 1$ and $Br(Q \to qW) = 100\%$. The efficiencies we list for the various cuts (except the charge identification cut) in Table~\ref{Tab:CC-effs} are from the $D$ quark sample. However, we have checked that there is no significant change in the efficiencies while using a $U$ quark sample, except for the cut on $E_{j_t}$ where a $U$ type quark has a higher signal efficiency compared to the $D$ type quark. The reason is as follows: from Eq.~(\ref{eq:Qprod}) we see that the dominant contribution comes from the case where the tag jet is a SM $d$ quark for $D$ type quark and a SM $u$ quark for the $U$ type quark. Due to the PDFs, the $u$ quark has a slightly harder energy distribution compared to the $d$ quark, which results in a higher efficiency for the tag jet in the case of the signal with $U$ type quark compared to the $D$ type quark. While the numbers we list in Table~\ref{Tab:CC-effs} are for the $D$ type quark we use the actual values for the $U$ type quark when extracting the signal and background events for evidence or discovery reach. 

In the last few rows we show the efficiency when the charge of the lepton coming from the decay of the heavy quark is identified. For $Q_\ell = -1$ we use the $D$ type quark sample and for $Q_\ell = +1$ we use the $U$ type quark sample. Note that the efficiency for identifying a lepton correctly as positively or negatively charged is the same irrespective of origin of the lepton. In fact in our analysis the efficiency of identifying the charge of a lepton itself is essentially 100\% (see Sec.~\ref{sec:results} for details). However from Table~\ref{Tab:CC-effs} it is evident that a $U$ type quark has a slightly higher efficiency in identifying the charge of the lepton. The reason is that we start with a $Q+\overline{Q}$ sample and then identify the charge of the positive or negatively charged leptons in this sample. Given that $U$ type quark has a smaller anti-quark component (which gives the wrong sign leptons) of the total cross section compared to the $D$ type quark, we get a slightly higher efficiency for charge identification for the $U$ type quark. 

Finally, note that the choice of the quark sample and the specific model parameters used does not affect the model-independent quantity $S^{W^\pm}$ inferred from an excess of signal events over the background at a given luminosity. The model-independent parameterization discussed in Sec.~\ref{sec:hqdecay} was designed such that the effects of $\tilde\kappa$ and the branching ratio of the heavy quark decay mode could be factored out. The choice of the quark sample does however affect the luminosity required for evidence or discovery of such a quark and the constraints on the various model parameters inferred from it. 

%%%%%%%%%%%%%%%%%%%%%%%%%%%%%%
\subsection{Identification of the Heavy Quark}

In this section we discuss the interesting possibility of determining the identity 
of the heavy quark produced. In $p\bar{p}$ collisions  such as at the Tevatron, 
this is rather straight forward as demonstrated in Ref.~\cite{Atre:2008iu}: Defining the proton beam as the $+\hat z$ direction, the forward (backward) tagged jet should be correlated with the heavy anti-quark (quark) production. Moreover, this can be used as an indication for down-type or up-type heavy quark production by identifying the electromagnetic charge of the lepton from the charged current decay. For e.g., an event with a backward tagged jet and a positive (negative) lepton would indicate production of $U$ ($D$) heavy quark. Similarly, an event with a forward tagged jet and a positive (negative) lepton would indicate production of $\overline D$ ($\overline U$) heavy quark. However, note that one cannot distinguish between a charge $2/3$ and charge $5/3$ quark (or a charge $-1/3$ and charge $-4/3$ quark) by this procedure. 

At a $pp$ machine such as the LHC, the situation is somewhat different. First of all, as seen from the total cross section curves in Figs.~(\ref{fig:xsecd}) - (\ref{fig:xsecxy}), a heavy quark production is substantially larger than that for an anti-quark, due to the $u,d$ valence quark dominance. There is a further subtle difference between the production of a heavy quark and an anti-quark. A valence quark parton has a harder distribution in the energy fraction $x$, while an anti-quark parton distribution is a lot softer. Consequently, a heavy quark produced will be more strongly boosted. Finally, the leptonic decay of a heavy quark via charged-current may still be utilized to specify the electric charge of $U$ or $D$. 

Another way to exploit the difference in the cross section of a heavy quark and anti-quark to gain sensitivity to the charge of the heavy quark produced is described below. We define a charge asymmetry which will be independent of the couplings $\tilde \kappa$ or of the branching ratios: 
\beq
A = \frac{N_+ - N_-}{N_+ + N_-},
\eeq
where $N_{\pm}$ is the cross section measured for the $W^\pm j$ channel. For example, at $\sqrt{s}=7$ TeV, based on the values given in Table~\ref{tab:singl:alt}, one finds $A=(-0.77,0.84,0.95,-0.41)$ for quarks of mass 900 GeV and of type (D,U,X,Y). 

Note however, that both of the methods described above work best when the quarks are non-degenerate and the resonances can be distinguished experimentally. In that case, the charge asymmetry  and/or the rapidity of the reconstructed quark along with the charge of the observed lepton could serve to distinguish between the quarks with different electromagnetic charges. When the states are exactly degenerate (equal mass and equal couplings) these methods have a contamination from the anti-quarks from a few percent up to few tens of percent depending on the model and the heavy quark under consideration. For example, in the case of doublets with hypercharges 1/6 and 7/6, upon observing a $5\sigma$ excess in the $S^{W-}$ channel for a mass of 900 GeV, one would have a contamination from the $\overline {X}$ quark $\sim5\%$ in the case of degenerate bidoublets. If the quarks were non-degenerate with couplings also varying by 10\%, then the contamination is $\cal{O}$$(10\%)$. If the observation were to be in the $S^{W+}$ channel for the hypercharge 1/6 and -5/6 doublets, the contribution from the anti-quarks increases to $\sim20\%$ for the degenerate case and further goes up  to  $\cal {O}$$(30\%)$ for close by resonances with slightly unequal couplings. In summary, the methods described for charge identification are best applied for the case of non-degenerate quarks and caution must be exercised when using the same methods for degenerate (and very close in mass and coupling) quarks. 

%%%%%%%%%%%%%%%%%%%%%%%%%%%%%%%%%%%%%%%%
\section{Results}
\label{sec:results}

In this section we will consider the current constraints from direct searches, the sensitivity of the search for new heavy vector-like quarks at the LHC, the constraints on some example models at the LHC and finally discuss some systematic effects. 

%%%%%%%%%%%%%%%%%%%%%%%%%%%%%%
\subsection{Current Constraints from Direct Searches}
\label{sec:results_constraints}

The current bound from direct searches at the Tevatron experiments is 
\bea
m_Q > 335\ (268)\ {\rm GeV\ at\ 95\%\ C.L. }
\eea
for heavy up (down) type quarks produced in pairs and decaying via CC (NC) interactions \cite{mqbound, Aaltonen:2007je}. Recently both CDF and D0 experiments at the Tevatron did a search for
heavy quarks with sizable couplings to first generation SM quarks in
the single production channel \cite{cdf_hq, Abazov:2010ku}. They presented their results in a model-independent fashion in the plane of mass of the heavy quark vs cross section times branching ratio (a parameter similar to $S^V$). From these analyses the heavy quarks have limits as given below. 
\bea
\nonumber
{\rm CDF:}\ \ \ \sigma(p\bar p \to q Q) &\times& Br(Q \to qW) \lsim 240\  {\rm fb\ for\ } m_Q = 530\ {\rm GeV} \\
\nonumber
{\rm D0:}\ \ \ \sigma(p\bar p \to q Q) &\times& Br(Q \to qW) \lsim 40\  {\rm fb\ for\ } m_Q = 690\ {\rm GeV} \\
{\rm D0:}\ \ \ \sigma(p\bar p \to q Q) &\times& Br(Q \to qZ) \lsim 120\  {\rm fb\ for\ } m_Q = 550\ {\rm GeV},  
\eea
where $Q$ is a generic heavy quark. The charge of the lepton was not identified in these analyses and hence the correlation with our $S^V$ parameters is slightly different and is given by
\bea
\nonumber
\sigma(p\bar p \to q Q) \times Br(Q \to qW) &=& S^{W^-}+S^{W^+}\ \  {\rm for\  CC,}\\
\nonumber
\sigma(p\bar p \to q Q) \times Br(Q \to qZ) &=& S^{Z}\ \ \ \ \ \  \ \ \ \ \ \ \ \  {\rm for\  NC.}
\eea
The Tevatron being a $p\bar p$ machine, heavy quarks and anti-quarks are produced in equal numbers with the result that $S^{W^-} = S^{W^+}$. The constraints from the Tevatron (on $S^{W^+} + S^{W^-}$) can hence be translated into constraints on $S^{W^+}$ or $S^{W^-}$.

While it is very useful to have model-independent limits due to the wide applicability of results, it is also illustrative to interpret the results in the case of a particular model. For example, interpreting these results in the context of degenerate doublets (with hypercharges $1/6$ and $7/6$) with $\tilde\kappa_{uD} = 1$ and decaying 100\% via CC currents, the results from CDF exclude heavy $D$ type quarks up to mass 530 GeV while the D0 constraints are more stringent and exclude heavy $D$ type quarks up to 690 GeV. In the same model for a $U$ type quark with $\tilde\kappa_{uU} = \sqrt{2}$ and decaying 100\% via NC currents, the sensitivity is to a 550 GeV quark. Note that in the context of this model, since all the quarks are degenerate, a $U$ type quark of mass 690 GeV decaying via NC is also excluded. However in other models the interpretation can be quite different. 

%%%%%%%%%%%%%%%%%%%%%%%%%%%%%%
\subsection{Sensitivity at the LHC}
\label{results_sensitivity}

We have simulated signal and background events for two different scenarios: an early run with $\sqrt{s}=7$ TeV and an integrated luminosity of 1 fb$^{-1}$, and a longer run with $\sqrt{s}=14$ TeV and 100 fb$^{-1}$ integrated luminosity. The estimated number of signal and background events are extracted using the fitted function to the total number of events as described in Sec.~\ref{sec:hqcoll-nclep}. The statistical significance is calculated as ~\cite{CMS_significance}
\beq
{\cal S}=\sqrt{2\times[(s+b)\ln(1+\frac{s}{b})-s]}, 
\label{eq:statsig}
\eeq
where $s(b)$ is the number of signal (background) events determined from the Breit-Wigner (Crystal Ball) term of the fitted function by integrating over $2\sigma$ of the most probable value of the Breit-Wigner component. The estimated number of signal and background events extracted from
the fits to the total number of events for the LHC with $\sqrt{s} = 7$
TeV and $14$ TeV are presented in Table.~\ref{tab:results7} and
Table.~\ref{tab:results14} respectively. For each case, the number of
signal and background events giving rise to 3$\sigma$ evidence and
5$\sigma$ discovery (in parentheses) are also given. In the same
tables are also given the luminosities required for 3$\sigma$ evidence
and 5$\sigma$ discovery of heavy quarks corresponding to three
different channels, namely, $\ell^+\ell^-\ 2j$, $\ell^- \etmiss\ 2j$
and $\ell^+ \etmiss\ 2j$. While the number of signal and background
events are model-independent quantities, the luminosity required is
model-dependent. The luminosities shown are for a $U, D$ and $U$ type
quark for the $\ell^+\ell^-\ 2j$, $\ell^- \etmiss\ 2j$ and $\ell^+
\etmiss\ 2j$ channels respectively. For each of these quarks, the
model-dependent parameter, $\tilde\kappa = 1$ while the decay of the
heavy quark in the respective gauge boson mode is 100\%. The choice of
these parameters is for illustrative purposes only and corresponds to
the degenerate bidoublet scenario.  

%%%%
\begin{table}\footnotesize
\caption{The estimated number of signal and background events extracted from
the fits to the total number of events for the LHC with $\sqrt{s} = 7$ TeV. For each case, the values for 3$\sigma$ (and 5$\sigma$ in parentheses) statistical significance is also given.}
\centering{}
\begin{tabular}{|c|c|c|c|}
\hline 
\multicolumn{4}{|c|}{Channel : $\ell^+\ell^-\ 2j$}\tabularnewline
\hline
$m_Q$ (GeV) & 300 & 600 & 900\tabularnewline
\hline
Signal events & 10.7 (29.2) & 7.10 (20.3) & 6.80 (18.9) \tabularnewline
\hline 
Background events & 9.30 (25.4) & 3.70 (11.0) & 3.30 (9.20)\tabularnewline
\hline 
$\int$Luminosity (pb$^{-1}$) & 22.0 (60.0) & 116 (340) & 540 (1500) \tabularnewline
\hline
\multicolumn{4}{|c|}{Channel: $\ell^- \etmiss\ 2j$}\tabularnewline
\hline 
$m_Q$ (GeV) & 300 & 600 & 900\tabularnewline
\hline
Signal events &  12.7 (35.3) & 7.60(21.1) & 6.48(18.0) \tabularnewline
\hline 
Background events & 14.1(39.3) & 4.28(11.9) & 2.77(7.70) \tabularnewline
\hline 
$\int$Luminosity (pb$^{-1}$) & 6.80(18.9) & 37.1(103) & 205 (570) \tabularnewline
\hline
\multicolumn{4}{|c|}{Channel: $\ell^+ \etmiss\ 2j$}\tabularnewline
\hline 
$m_Q$ (GeV) & 300 & 600 & 900\tabularnewline
\hline
Signal events & 17.8(49.4) & 12.8(35.5) & 8.78(24.4)\tabularnewline
\hline 
Background events & 29.7(82.5) & 14.1(39.1) & 6.08(16.9) \tabularnewline
\hline 
$\int$Luminosity (pb$^{-1}$) & 11.2(31.1) & 75.6(210) & 283(785) \tabularnewline
\hline
\end{tabular}
\label{tab:results7}
\end{table}
%%%%

%%%%
\begin{table}\footnotesize
\caption{Same as Table~\ref{tab:results7} but for 14 TeV.}
\centering{}
\begin{tabular}{|c|c|c|c|}
\hline 
\multicolumn{4}{|c|}{Channel : $\ell^+\ell^-\ 2j$}\tabularnewline
\hline
$m_Q$ (GeV) & 900 & 1800 & 2400\tabularnewline
\hline
Signal events & 10.1 (28.3)  & 8.60 (24.1)  & 13.1 (37.0)  \tabularnewline
\hline 
Background events & 8.20 (23.0) & 5.80 (16.3) & 15.1 (42.8) \tabularnewline
\hline 
$\int$Luminosity (fb$^{-1}$) & 0.15 (0.42) & 2.64 (7.40) & 35.0 (99.0) \tabularnewline
\hline
\multicolumn{4}{|c|}{Channel: $\ell^- \etmiss\ 2j$}\tabularnewline
\hline 
$m_Q$ (GeV) & 900 & 1800 & 3000 \tabularnewline
\hline
Signal events &  9.29(25.8) & 4.72(13.1) & 4.18(11.6) \tabularnewline
\hline 
Background events &  6.70(18.6)& 1.22 (3.40) & 0.720(2.00) \tabularnewline
\hline 
$\int$Luminosity (fb$^{-1}$) &  0.0450(0.125) & 0.360(1.00) &  6.12(17.0) \tabularnewline
\hline
\multicolumn{4}{|c|}{Channel: $\ell^+ \etmiss\ 2j$}\tabularnewline
\hline 
$m_Q$ (GeV) & 900 & 1800 & 3000 \tabularnewline
\hline
Signal events & 15.7(43.7) & 13.1(36.3) & 6.48(18.0) \tabularnewline
\hline 
Background events & 23.6(65.6) & 14.4(40.0) & 2.41(6.70)\tabularnewline
\hline 
$\int$Luminosity (fb$^{-1}$) &0.0864(0.240) & 1.15(3.20) &11.5(32.0) \tabularnewline
\hline
\end{tabular}
\label{tab:results14}
\end{table}
%%%%

\begin{figure}[tb]
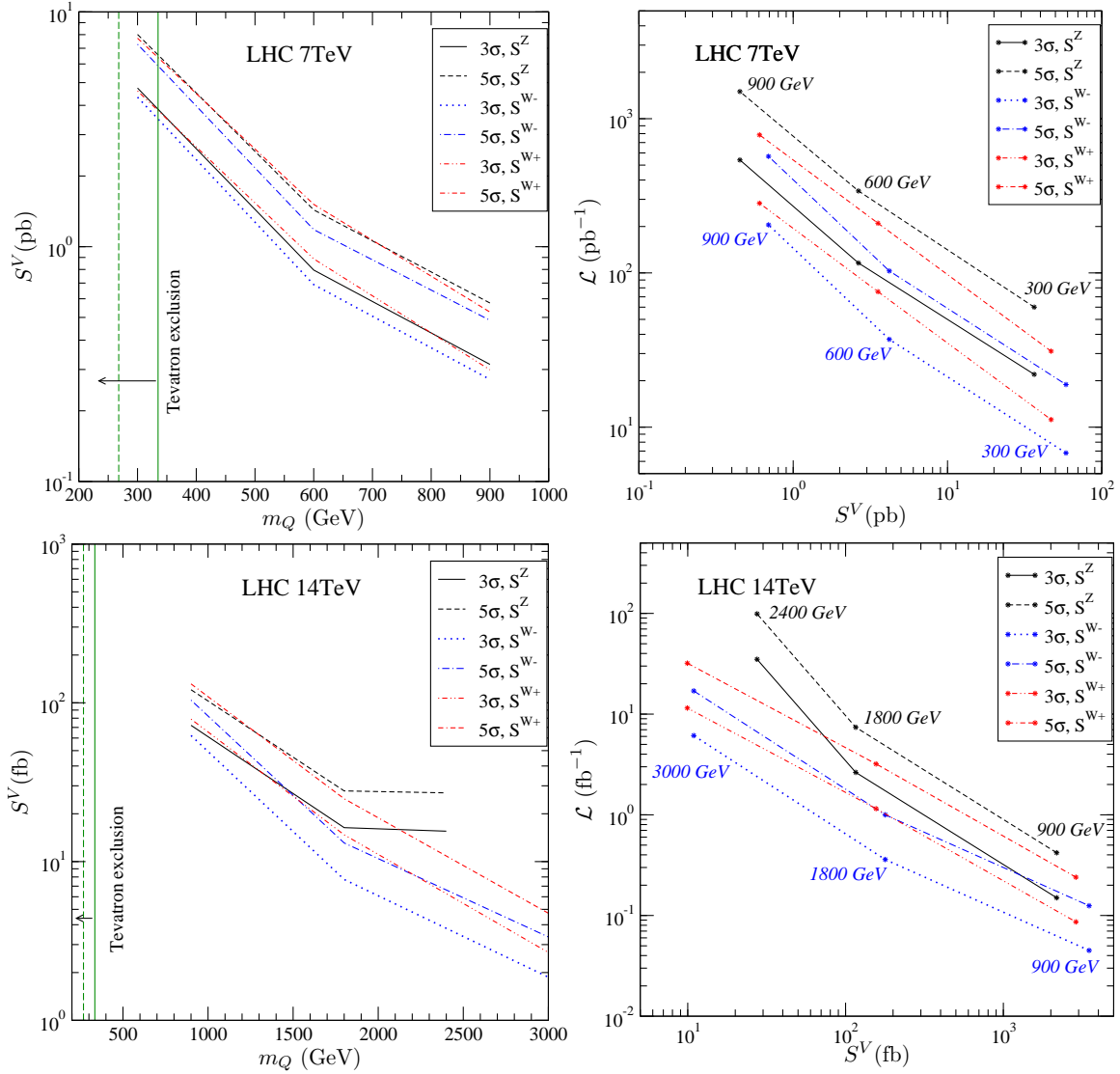

\psfrag{mQ}[t][t][0.8]{$m_Q\ \mathrm{(GeV)}$}
\psfrag{lumpb}[b][b][0.8]{$\cal{L}$ $\mathrm{(pb}^{-1}\mathrm{)}$}  
\psfrag{lumfb}[b][b][0.8]{$\cal{L}$ $\mathrm{(fb}^{-1}\mathrm{)}$} 
\psfrag{SVpb}[b][b][0.8]{$S^V \mathrm{(pb)}$} 
\psfrag{SVfb}[b][b][0.8]{$S^V \mathrm{(fb)}$}
\psfrag{nc3x}[t][t][0.8]{$3\sigma, U$}
\psfrag{nc5x}[t][t][0.8]{$5\sigma, U$}
\psfrag{wm3}[t][t][0.8]{$3\sigma, D$}
\psfrag{wm5}[t][t][0.8]{$5\sigma, D$}
\psfrag{wp3}[t][t][0.8]{$3\sigma, U$}
\psfrag{wp5}[t][t][0.8]{$5\sigma, U$}
\psfrag{sz3xxx}[t][t][0.8]{$3\sigma, S^Z$}
\psfrag{sz5xxx}[t][t][0.8]{$5\sigma, S^Z$}
\psfrag{swm3x}[t][t][0.8]{$3\sigma, S^{W^-}$}
\psfrag{swm5x}[t][t][0.8]{$5\sigma, S^{W^-}$}
\psfrag{swp3x}[t][t][0.8]{$3\sigma, S^{W^+}$}
\psfrag{swp5x}[t][t][0.8]{$5\sigma, S^{W^+}$}
{\includegraphics[width=0.465\textwidth,clip=true]{mQvsSV7.eps}
\includegraphics[width=0.45\textwidth,clip=true]{SVvsLum7.eps}
\includegraphics[width=0.465\textwidth,clip=true]{mQvsSV14.eps}
\includegraphics[width=0.45\textwidth,clip=true]{SVvsLum14.eps}}
\caption{ 
(a) Top left: Sensitivity plot for 3$\sigma$ evidence and 5$\sigma$ discovery in the plane of the parameters $S^{W-}, S^{W+}$ and $S^{Z}$ and heavy quark mass $m_Q$ with 1 fb$^{-1}$ of data at $\sqrt{s} = 7$ TeV; (b) top right: luminosity needed for $3 \sigma$ evidence and $5 \sigma$ discovery vs the parameters $S^{W-}, S^{W+}$ and $S^{Z}$ at the LHC with $\sqrt{s} = 7$ TeV. The three points on each curve represented by black (blue or red) stars correspond to the mass for each $S^Z (S^{W^-}$ or $S^{W^+})$; (c) bottom left: same as (a) but for $\sqrt{s} = 14$ TeV and 100 fb$^{-1}$ of data; (d) bottom right: same as (b) but for $\sqrt{s} = 14$ TeV. The vertical solid green and vertical solid dashed green bounds are the 95\% C.L exclusion limits from Tevatron for the heavy quarks produced in pairs via QCD interactions and decaying via CC and NC decay modes respectively.}
\label{fig:sensjq}
\end{figure}

We present the model-independent results of our analysis for the LHC
in Fig.~\ref{fig:sensjq} for the different channels. There are three
parameters - $S^V, m_Q$ and the luminosity -  in our analysis and we
present our results in the planes defined by two of these parameters.
In Fig.~\ref{fig:sensjq}(a)  and Fig.~\ref{fig:sensjq}(c) we present
the sensitivity plot for 3$\sigma$ evidence and 5$\sigma$ discovery in
the plane of the parameters $S^{W-}, S^{W+}$ and $S^{Z}$ and heavy
quark mass $m_Q$ for $\sqrt{s} = 7$ TeV with 1 fb$^{-1}$ of data and
for $\sqrt{s} = 14$ TeV with 100 fb$^{-1}$ of data respectively. In
Fig.~\ref{fig:sensjq}(b)  and Fig.~\ref{fig:sensjq}(d) we present the
luminosity needed for $3 \sigma$ evidence and $5 \sigma$ discovery vs
$S^V$ at the LHC with $\sqrt{s} = 7$ TeV and
$\sqrt{s} = 14$ TeV respectively. In Figs.~\ref{fig:sensjq} (b) and
(d) the quark masses change along the different curves. For reference
we show with stars three different values of quark masses on each
curve.  

%%%
%%%

The optimal channel for discovery in an early LHC run is $m_Q\approx 1$ TeV for
$S^{W^-}\approx 0.41$pb for $\sqrt{s}=7$ TeV and 1 fb$^{-1}$ integrated
luminosity. This is comparable to the final Tevatron reach for single
production of these quarks. Of course, the
reach increases dramatically for a longer LHC run. We have estimated 
that with 100 fb$^{-1}$ at 14 TeV and $S^{W^-}=2.6$ fb, the LHC can
discover a new $Q$ quark with a mass up to 
$m_Q \approx 3.7$ TeV.

%%%%%%%%%%%%%%%%%%%%%%%%%%%%%%
\subsection{Constraints on Specific Models}
\label{sec:results_modelconstraints}

\begin{figure}[tb]
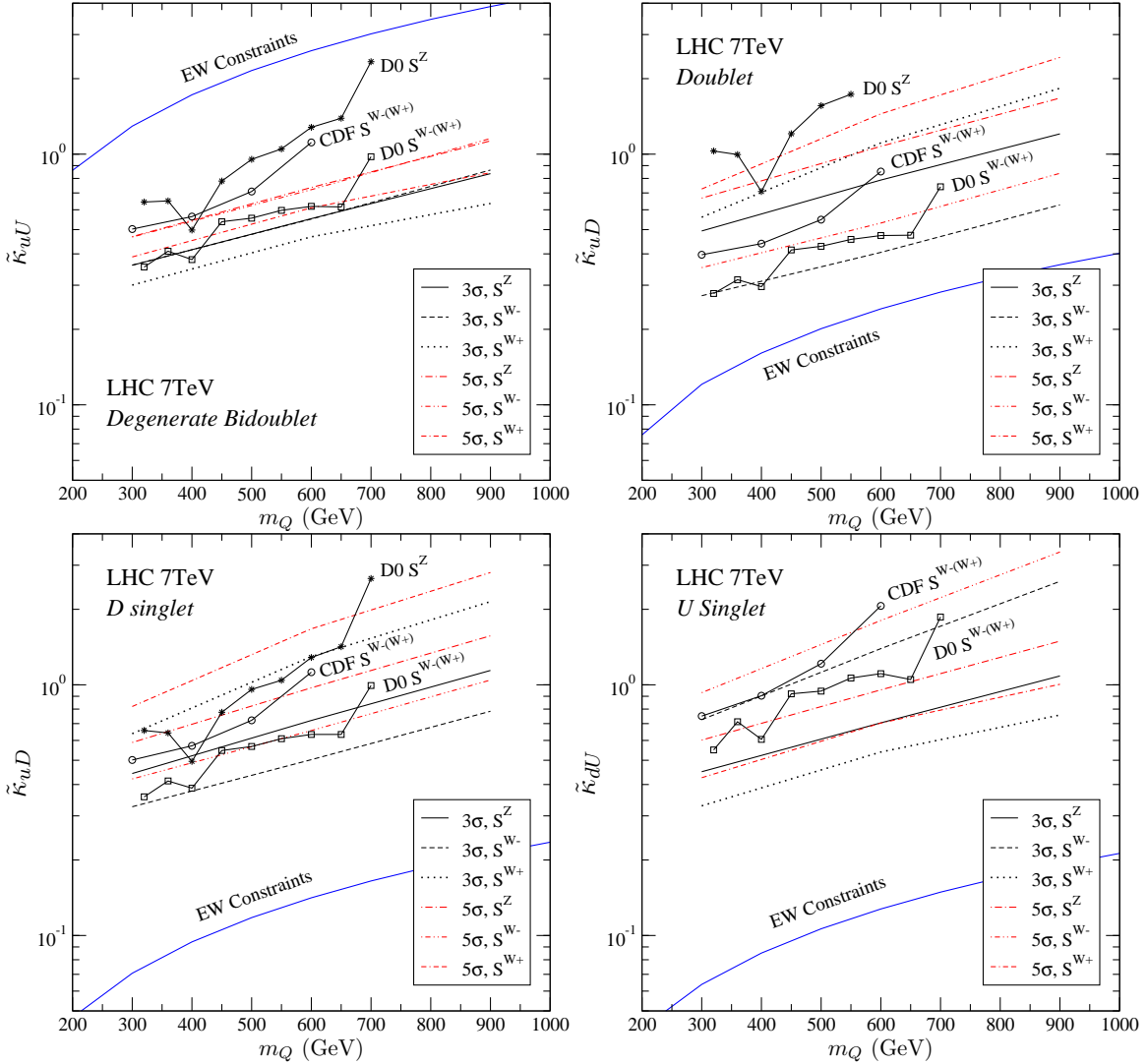

\psfrag{mQ}[t][t][0.8]{$m_Q\ \mathrm{(GeV)}$}
\psfrag{ktildeuU}[b][b][1.0]{$\tilde\kappa_{uU}$}
\psfrag{ktildeuD}[b][b][1.0]{$\tilde\kappa_{uD}$}
\psfrag{ktildedU}[b][b][1.0]{$\tilde\kappa_{dU}$}
\includegraphics[width=0.46\textwidth,clip=true]{mQvsktilde-bidoublet7.eps}
\includegraphics[width=0.46\textwidth,clip=true]{mQvsktilde-doublet7.eps}
\includegraphics[width=0.46\textwidth,clip=true]{mQvsktilde-Dsinglet7.eps}
\includegraphics[width=0.46\textwidth,clip=true]{mQvsktilde-Usinglet7.eps}
\caption{ 
(a) Top left: sensitivity plot for 3$\sigma$ evidence and 5$\sigma$
  discovery in the plane of the model parameters $\tilde\kappa_{uU}$
  and heavy quark mass $m_Q$ at the LHC with 1 fb$^{-1}$ of data and $\sqrt{s} =
  7$ TeV for the model with degenerate bidoublets; (b) top right: same
  as (a) but for $\tilde\kappa_{uD}$ for the model with one doublet;
  (c) bottom left: same as (a) but for $\tilde\kappa_{uD}$ for the
  model with $D$ type singlet (d) bottom right: same as (a) but for
  $\tilde\kappa_{dU}$ for the model with $U$ type singlet. The curves
  with data points show the $95 \%$ C.L. constraints on $\tilde\kappa_{qQ}$ for
  each model coming from the direct searches by CDF and D0 experiments
  in the channel where heavy quarks are produced singly via EW
  interactions. The region above the blue
  solid line denoted "EW Constraints" is excluded by the EW
  precision tests described in section \ref{model:considerations}. The bounds on the other couplings in each model can be
  obtained from the relations between couplings listed in
  Sec.~\ref{sec:results_modelconstraints}.} 
\label{fig:sensmod7}
\end{figure}

\begin{figure}[tb]
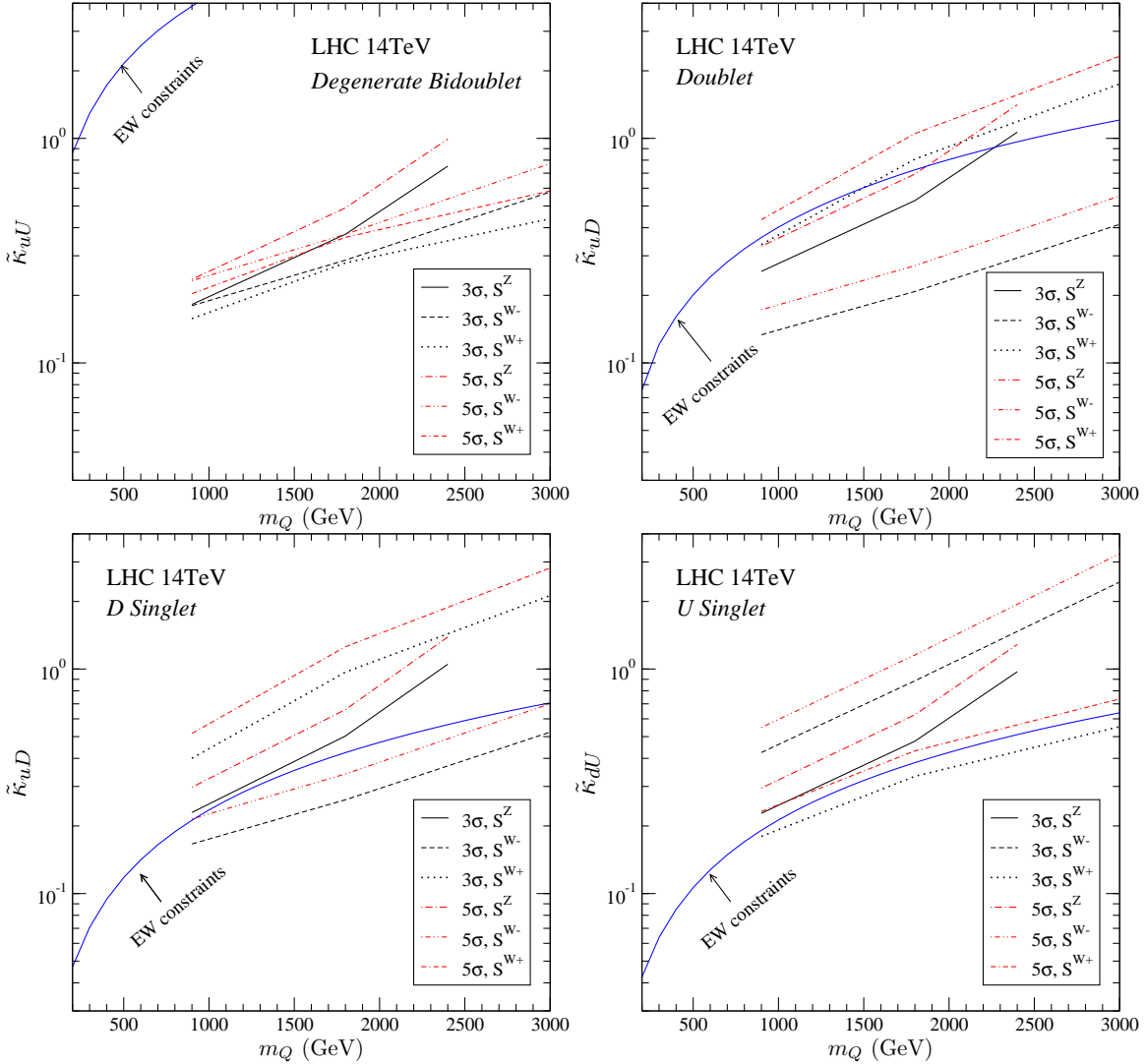

\psfrag{mQ}[t][t][0.8]{$m_Q\ \mathrm{(GeV)}$}
\psfrag{ktildeuU}[b][b][1.0]{$\tilde\kappa_{uU}$}
\psfrag{ktildeuD}[b][b][1.0]{$\tilde\kappa_{uD}$}
\psfrag{ktildedU}[b][b][1.0]{$\tilde\kappa_{dU}$}
\includegraphics[width=0.46\textwidth,clip=true]{mQvsktilde-bidoublet14.eps}
\includegraphics[width=0.46\textwidth,clip=true]{mQvsktilde-doublet14.eps}
\includegraphics[width=0.46\textwidth,clip=true]{mQvsktilde-Dsinglet14.eps}
\includegraphics[width=0.46\textwidth,clip=true]{mQvsktilde-Usinglet14.eps}
\caption{ Same as Fig.~\ref{fig:sensmod7} but for $\sqrt{s} = 14$ TeV and 100 fb$^{-1}$ of data.}
\label{fig:sensmod14}
\end{figure}

We interpret the model-independent results presented in
Fig.~\ref{fig:sensjq} in terms of some specific models next. We present the
sensitivity of the model dependent parameter $\tilde\kappa_{qQ}$ as a
function of the mass of the heavy quark ($m_Q$) for some popular
scenarios. We present the 3$\sigma$ and 5$\sigma$
constraints on $\tilde\kappa_{qQ}$ for $\sqrt{s} = 7$ TeV with 1
fb$^{-1}$  of data and $\sqrt{s} = 14$ TeV with 100 fb$^{-1}$ of
data in Fig.~\ref{fig:sensmod7} and Fig.~\ref{fig:sensmod14}
respectively.  
We also show the model dependent
 constraints from electroweak precision observables, which correspond
 to 
 \beq
 {\kappa}_{qQ} \geq 0.75,\ 0.066, \ 0.041, \ 0.037,
 \eeq
  for the
 degenerate bidoublet, doublet, D singlet and U singlet models,
 respectively. We also show in Fig.~\ref{fig:sensmod7} the current
 constraints from the recent Tevatron searches. 
Several conclusions can be obtained from our
 results. First, as expected, only for the degenerate bidoublet model
 can the Tevatron and the early LHC run probe regions of parameter
 space which are not excluded by electroweak precision tests. Second,
 the early LHC run reach is comparable and in general better than the
 Tevatron reach. In some cases, like the neutral channel in the
 $U$ singlet model, the Tevatron is unable to constrain the model
 (the probed cross sections cannot be generated due to the bound from
 the unitarity of the mixing) whereas they are all probed by the LHC.
 Finally, the reach of the LHC run at $\sqrt{s}=14$ TeV and 100 fb$^{-1}$ integrated luminosity is
 sensibly better, probing regions of parameter space not excluded by
 electroweak precision tests for all four models (for large enough
 masses). In particular, masses in the $3-4$ TeV region can be probed
 in all four models provided we have
 $\tilde{\kappa}_{qQ}\gtrsim 0.4$. Recall that for these large masses,
 $\tilde{\kappa}_{qQ}$ can be directly related to a Yukawa coupling in
 particular models. Next we consider a few example models in detail.

\subsubsection{$D$ type singlet}

The first model contains an extra vector-like electroweak quark
singlet with 
 hypercharge $-1/3$.
We assume the singlet mixes only with the first generation (doublet)
in the basis in which all SM flavour occurs in the up quark
sector. The model can be fully parameterized in terms of the physical down
quark mass, $m_d$, the physical heavy down quark mass, $m_D$, and the
coupling between $D$ and $u$, $\kappa_{uD}^L \equiv V_{ud} s_L$, with
$V$ the CKM matrix and $s_L$ is used to denote the sine of an angle
(and is therefore smaller than one). The decoupling limit occurs for
$m_D \to \infty$, $s_L\to 0$ with $s_L m_D$ constant. In that limit
all effects go to zero. The non-zero couplings apart from
$\kappa_{uD}^L$, are
\beq
\kappa_{dD}^L = s_L c_L,
\eeq
which is therefore smaller than $1/2$,
and Yukawa couplings given in the notation of
Appendix~\ref{partialwidths}
\beq
Y_{dD}=s_L c_L \frac{m_D}{v}, \quad
Y_{Dd}=s_L c_L \frac{m_d}{v}. 
\eeq
These Yukawa couplings are needed to compute the corresponding 
branching fractions.

\subsubsection{$U$ type singlet}

A very similar model consists of an extra singlet $U$ with hypercharge
$2/3$ and we assume that it mixes only with the first generation (doublet)
in the basis in which all SM flavour occurs in the down quark
sector. The model can be fully parameterized in terms of the physical up
quark mass, $m_u$, the physical heavy up quark mass, $m_U$, and the
coupling between $U$ and $d$, $\kappa_{dU}^L \equiv V_{ud} s_L$, with
$V$ the CKM matrix and $s_L$ is used to denote the sine of an angle
(and is therefore smaller than one). The decoupling limit occurs for
$m_U \to \infty$, $s_L\to 0$ with $s_L m_U$ constant. In that limit
all effects go to zero. The non-zero couplings apart from
$\kappa_{dU}^L$, are
\beq
\kappa_{uU}^L = s_L c_L,
\eeq
which is therefore smaller than $1/2$,
and Yukawa couplings given in the notation of
Appendix~\ref{partialwidths} 
\beq
Y_{uU}=s_L c_L \frac{m_U}{v}, \quad
Y_{Uu}=s_L c_L \frac{m_u}{v}. 
\eeq
These Yukawa couplings are needed to compute the corresponding 
branching fractions.

\subsubsection{Doublet with Hypercharge $1/6$}

We now consider one hypercharge $1/6$ doublet that only mixes with
$u_R$ in the basis in which all SM flavor occurs in the down
sector. Everything can be parameterized in terms of the up quark mass,
$m_u$, the $D$ quark mass, $m_D$, and $\kappa_{uD}=-s_R$. The
other non-vanishing couplings are
\beq
\kappa_{uU}=-s_R c_R,
\quad
\kappa_{dU}^L=-\kappa^L_{uD}=s_L,
\eeq
and a Yukawa coupling between $u$ and $U$ that we use to compute the
branching fractions
\beq
Y_{uU}=s_R c_R \frac{m_u}{v}, \quad
Y_{Uu}=s_R c_R \frac{m_U}{v},
\eeq
where the mass of the heavy $U$ quark is
\beq
m_U=\frac{\sqrt{m_D^2-s_R^2m_u^2}}{c_R}.
\eeq
The two LH couplings are irrelevant in practice, since we have the
relation
\beq
s_L = \frac{s_R m_u}{m_D}\approx 0,
\eeq
for $m_D$ above current experimental limits.

\subsubsection{Degenerate Bidoublet with Hypercharge $1/6$ and $7/6$}

This model discussed in detail in Ref.~\cite{Atre:2008iu} and in Sec.~\ref{sec:simpmod}, consists of two new vector-like quark electroweak doublets, with hypercharges $1/6$ and $7/6$ respectively, that mix only with the up quark, in the basis of diagonal Yukawa couplings for the charge $2/3$ SM quarks. The two doublets are degenerate, \textit{i.e.} they have the same mass and Yukawa couplings to $u$ before EWSB, and the particle content of these two doublets is two charge $2/3$, one charge $-1/3$ and one charge $5/3$ quark. The three quarks of types $U$, $D$ and $X$ in our notation, have the following values of $\kappa$ 
\beq
\kappa_{uU}=\sqrt{2}\kappa_{uD}=\sqrt{2} \kappa_{uX}= s_R,
\eeq
where $s_R$ is the sine of the corresponding mixing angle which depends on the particular values of the model parameters. In this model, the generic bound from the unitarity of the mixing is saturated for 
$\kappa_{uU}\leq 1$ and is a bit more stringent for  
$\kappa_{uD}=\kappa_{uX} \leq 1/\sqrt{2}$. Also the constraints from
the oblique parameters impose a mild bound of $\kappa_{uU} \lesssim
0.75$ (assuming $\Delta S\leq 0.2$) in the model under consideration. 

In this model each of the quark decays with $100\%$ branching ratio to the respective gauge boson, namely Br$(U \to uZ) =$ Br$(D \to uW) =$ Br$(X \to uW) = 100\%$.

%%%
%%%

%%%
%%%

%%%%%%%%%%%%%%%%%%%%%%%%%%%%%%
\subsection{Systematic Effects}
\label{sec:results_systeffects}

The analysis reported here gives only an estimate of the LHC reach. Although better results could be obtained if the signal selection was based on a Likelihood or Discriminant Variable technique, it must be noted that a number of systematic effects, which are difficult to estimate by simulation, have been
 neglected. In particular,
\begin{itemize}
\item 
No k-factors have been taken into account for the signals and backgrounds.
\item
Since the charge identification plays a significant role in the CC
analysis and Delphes does not simulate charge misidentification, the
likelihood of charge mismeasurement was studied at the generator
level. As an example, $p_T$ and $\eta$ dependent parameterization of
the ATLAS tracker response was taken from ~\cite{atlas-csc}. We find
the misidentification fraction to be very small, only going up to
1.2\% (0.9\%) for the electrons (muons) from the heaviest quark (3TeV)
studied in this paper. 
\item Multijet events, which are abundant at the LHC, can fake the CC decay
signal if jets are misreconstructed as leptons and if jet transverse
energies are poorly measured, leading to a presence of missing energy
in the events. These effects are very difficult to estimate by fast
Monte Carlo simulation since they depend strongly on the criteria used
for lepton identification and jet measurements. Nevertheless, we
expect that with optimized experimental methods this source of
background will be small. Indeed, for the D0 and CDF analyses
\cite{cdf_hq,Abazov:2010ku} it is almost negligible.

\item 
The selection cuts have been optimized for specific mass values. In
practice, a sliding window search would have these cuts re-optimized
at intermediate mass points. In some cases, because of experimental
uncertainties and to have more robust results, the cuts could be
loosened. For example, the $p_T^{sum}$ cut on the tag jet plus
heavy quark candidate system, as given in Eq. (\ref{pT:sum:def}), 
could be raised to 150 GeV for the 14 TeV
analysis, at the cost of $\sim 10\%$ loss in significance.  
\item 
At high luminosity, pileup effects can become very important. Pileup will affect the purity of the signal. The isolation efficiency and resolution of leptons will deteriorate and fake forward jets would be generated.
\end{itemize}

%%%%%%%%%%%%%%%%%%%%%%%%%%%%%%%%%%%%%%%%
\section{Summary}
\label{sec:concl}

The Fermilab Tevatron has been leading the way for searches for new
heavy particles, and the historical mission of the LHC at the
energy and luminosity frontier  will take the field of high energy
physics to a new era. Among many highly motivated new particles in the
TeV region, are new vector-like quarks that can have sizable couplings to first
generation quarks without conflicting with current experimental
constraints. The coupling with valence quarks and unique kinematics
make single production at hadron colliders the optimal discovery
process, potentially superior to the pair production mechanism via QCD
interactions.  
 
In this article, we first motivated the consideration for the new heavy
quarks by presenting a few theoretical models. We then discussed the
existing experimental bounds on the quark mass and their couplings to
the SM fields.  
In order to make our
study as widely applicable as possible, we have performed a model
independent-analysis of the discovery reach at the Large Hadron
Collider for new vector-like quarks considering single production and
subsequent decays via electroweak interactions. After an exploration
of the unique features of the signal kinematics, we performed an
exhaustive optimization for signal extraction over the SM
backgrounds. With this optimized analysis, we determined the reach for
the signal as a function of the heavy quark mass. The result is given
in terms of model independent quantities that can be identified with
production cross section times branching ratio of the heavy quark decay to the corresponding
electroweak gauge boson and a SM quark. We then translated these model
independent bounds into bounds on the parameters of four popular
models, with two degenerate doublets of hypercharges 1/6 and 7/6;
one single doublet with hypercharge 1/6; one down type
singlet and one up type singlet.

We considered two scenarios, one with a short LHC run with 1
fb$^{-1}$ integrated luminosity at $\sqrt{s}=7$ TeV and a longer run
with 100 fb$^{-1}$ at $\sqrt{s}=14$ TeV. The reach of the former has
been found to be comparable, although typically slightly better, to
the final Tevatron reach with 10 fb$^{-1}$ integrated luminosity. It is only in the case of the
degenerate doublets the probed region by the Tevatron and the short
LHC run are unconstrained by precision data. However, the reach dramatically improves both in
terms of mass and couplings for the longer run, probing in all four
models considerable regions of parameter space not constrained by
electroweak precision observables . In particular, masses up
to $\sim 3.7$ TeV can be probed with couplings of order one or
equivalently, production cross sections times branching ratios as
small as 2.6 fb. It is evident from these results that the discovery
potential for new heavy quarks at the LHC is very encouraging.

%%%%%%%%%%%%%%%%%%%%%%%%%%%%%%%%%%%%%%%%
\begin{acknowledgments}
We acknowledge interesting discussions with R. Sekhar Chivukula,
Joseph Lykken and
Elizabeth Simmons. Fermilab is operated by Fermi Research Alliance,
LLC under Contract  No. DE-AC02-07CH11359 with the United States
Department of  Energy. AA was supported in part by the US National Science Foundation under grant PHY-0854889. The work of TH is supported in part by the
United States Department of Energy under grant DE-FG02-95ER40896. JS
is partially supported by projects FPA2006-05294, FQM 101, FQM 03048
and by MICINN through a Ram\'on y Cajal contract and would like to
thank CERN TH division for hospitality during completion of this work.  
\end{acknowledgments}

%%%%%%%%%%%%%%%%%%%%%%%%%%%%%%
\appendix
%%%%%%%%%%%%%%%%%%%%%%%%%%%%%%
\section{Explicit calculation of S}
\label{app:calcS}

The one loop contribution to the S parameter from an arbitrary number
of fermions is given by~\cite{Carena:2007ua}
\beq
S=\frac{3}{4\pi} \sum_{i,j} \Big[ 
\Big( U^L_{ij} Y^L_{ji}
+ U^R_{ij} Y^R_{ji} \Big) \bar{\chi}_+(
\mathcal{M}_{ii},\mathcal{M}_{jj})
+
\Big( U^L_{ij} Y^R_{ji}
+ U^R_{ij} Y^L_{ji} \Big) \bar{\chi}_-(
\mathcal{M}_{ii},\mathcal{M}_{jj})
\Big],\label{S}
\eeq
where the sum runs over all fermions, $\mathcal{M}_{ij}$ is the physical (diagonal) mass matrix, $U^{L,R}_{ij}$ are the couplings of left and right handed fermions to $W^3_\mu$, respectively and $Y^{L,R}_{ij}$ the corresponding couplings to $B_\mu$, all in the physical basis. We have also defined the following functions
\bea
\nonumber
\bar{\chi}_+(y_1,y_2) &=&
\frac{5(y_1^4+y_2^4)-22 y_1^2 y_2^2}{9(y_1^2-y_2^2)^2}
+\frac{2 y_1^2 y_2^2(y_1^2+y_2^2)-y_1^6 -y_2^6}{3(y_1^2-y_2^2)^3} 
\log \left(\frac{y_1^2}{y_2^2}\right) \\
&-& \frac{1}{3} \left[
 \log \left( \frac{y_1^2}{\mu^2} \right)
+ \log \left( \frac{y_2^2}{\mu^2} \right)
\right], \\
\bar{\chi}_-(y_1,y_2) &=&
\frac{y_1 y_2}{(y_1^2-y_2^2)^3} \left[ y_1^4 -y_2^4 -2 y_1^2 y_2^2
  \log\left( \frac{y_1^2}{y_2^2}\right)\right].
\eea

The model with two exactly degenerate doublets that only mix with the up quark can be parameterized by three parameters which are usually taken to be the up Yukawa coupling, the Yukawa coupling between the up and the two doublets and the (common) mass of the two doublets. All these parameters are defined in the current eigenstate bases, i.e. before diagonalizing the mass matrix. It is however much more advantageous to define physical parameters which, in our case can be taken to be the (physical) mass of the up quark $m_u$, the physical mass of the three degenerate heavy states (which coincides with $M$) and the coupling $\kappa_{uU}=s_R$, which coincides with the sine of the RH rotation needed to diagonalize the mass matrix. Full details of this parameterization can be found in Ref.~\cite{delAguila:2010es}. The physical masses, written in terms of these parameters and in a basis with first the (three) charge $2/3$ quarks, then the two charge $-1/3$ quarks and finally the charge $-5/3$ one are, 
\beq
\mathcal{M}=\mathrm{diag}(m_u,M,\frac{\sqrt{M^2-m_u^2 s_R^2}}{c_R},
m_d,M,M),
\eeq
whereas the couplings relevant for the calculation of the $S$ parameter are, in the same basis
\beq
U^L=
\begin{pmatrix} 
c_L^2 & -s_L & s_L c_L & 0 & 0 & 0 \\
-s_L  & 0    & c_L     & 0 & 0 & 0 \\
s_L c_L & c_L & s_L^2  & 0 & 0 & 0 \\
0 & 0 & 0 & -1 & 0 & 0 \\
0& 0 & 0 & 0 & -1  & 0 \\
0& 0& 0 & 0 & 0 & 1
\end{pmatrix},
\quad
Y^L=
\begin{pmatrix} 
\frac{c_L^2}{6}+\frac{2s_L^2}{3} & \frac{s_L}{2} & 
-\frac{s_L c_L}{2} & 0 & 0 & 0 \\
\frac{s_L}{2}  & \frac{2}{3}    & -\frac{c_L}{2}    & 0 & 0 & 0 \\
-\frac{s_L c_L}{2} & -\frac{c_L}{2} & 
\frac{s_L^2}{6} + \frac{2 c_L^2}{3}  & 0 & 0 & 0 \\
0 & 0 & 0 & \frac{1}{6} & 0 & 0 \\
0& 0 & 0 & 0 & \frac{1}{6}  & 0 \\
0& 0& 0 & 0 & 0 & \frac{7}{6}
\end{pmatrix},
\eeq
and
\beq
U^R=
\begin{pmatrix} 
0 & -s_R & 0 & 0 & 0 & 0 \\
-s_R  & 0    & c_R     & 0 & 0 & 0 \\
0 & c_R & 0  & 0 & 0 & 0 \\
0 & 0 & 0 & 0 & 0 & 0 \\
0& 0 & 0 & 0 & -1  & 0 \\
0& 0& 0 & 0 & 0 & 1
\end{pmatrix},
\quad
Y^R=
\begin{pmatrix} 
\frac{2}{3} & \frac{s_R}{2} & 
0 & 0 & 0 & 0 \\
\frac{s_R}{2}  & \frac{2}{3}    & -\frac{c_R}{2}    & 0 & 0 & 0 \\
0 & -\frac{c_R}{2} & 
\frac{2}{3}  & 0 & 0 & 0 \\
0 & 0 & 0 & -\frac{1}{3} & 0 & 0 \\
0& 0 & 0 & 0 & \frac{1}{6}  & 0 \\
0& 0& 0 & 0 & 0 & \frac{7}{6}
\end{pmatrix},\label{Right:couplings}
\eeq
where the LH rotation is given in terms of our input parameters by 
$s_L=s_R m_u/M$ (see~\cite{delAguila:2010es}). Eq.~(\ref{Right:couplings}) shows that we indeed have $\kappa_{uU}=s_R$. We can now insert the explicit form of the couplings and masses in Eq.~(\ref{S}) and expand in the small ratio $m_u/M \lesssim 10^{-6}$. The resut is
\beq
S =S_{\mathrm{SM}}- \frac{
6 s_R^2 -9 s_R^4+5 s_R^6+ 3(2-4s_R^2+3 s_R^4) \log[1-s_R^2]
}{3\pi s_R^6}+ \mathcal{O}\left(\frac{m_u^2}{M^2}\right),
\eeq
where 
\beq
S_{\mathrm{SM}}=
\frac{3-\log(m_u^2/m_d^2)}{6\pi},
\eeq
is the SM contribution from the up and down quarks that we have to
subtract to compute the new physics contribution to the S
parameter. Thus, we see that, up to tiny corrections
$\mathcal{O}(m_u^2/M^2) \lesssim 10^{-12}$, the new physics
contribution to S in this model is independent of $M$ for fixed values
of $\kappa_{uU}$. 

\section{Heavy fermion decay widths}
\label{partialwidths}

In this appendix we collect the general expressions for the decay
widths needed in the calculation of the branching ratios for arbitrary
models. 
The decay width for a two body decay of a heavy
fermion, $Q$ with mass $m_Q$,
into a light fermion, $q$ with mass $m_q$ and a massive 
gauge boson, $V$ with mass $m_V$, and couplings
\beq
V_\mu 
\Big[
\bar{q} \gamma^\mu(g_L \mathcal{P}_L + g_R \mathcal{P}_R )
Q + \mathrm{h.c.}\Big],
\eeq
reads
\bea
\Gamma(Q\to V q) &=& \frac{
\sqrt{m_q^4-2m_q^2(m_Q^2+m_V^2)+(m_Q^2-m_V^2)^2}}{32 \pi m_Q^3 m_V^2}
\nonumber \\
& \times& \Big\{
(|g_L|^2 +|g_R|^2) [(m_Q^2-m_q^2)^2+m_V^2(m_Q^2+m_q^2) -2 m_V^4]
\nonumber \\
&&-6 (g_L g_R^\ast +g_L^\ast g_R) m_Q m_q m_V^2 \Big\}
\nonumber  \\
& \approx & 
\frac{(g_L^2+g_R^2)}{32\pi}\frac{m_Q^3}{m_V^2}
\left[1-3\frac{m_V^2}{m_Q^2}+2\frac{m_V^6}{m_Q^6}\right], 
\mbox{ for } (m_Q , m_V \gg m_q),
\eea
Similarly the decay into a quark and a scalar, with couplings,
\beq
\frac{1}{\sqrt{2}}H \Big[ Y_{Qq}\bar{Q}_L q_R + Y_{qQ} \bar{q}_L Q_R
  +\mathrm{h.c.} 
\Big],
\eeq
is given by
\bea
\Gamma(Q\to H q) &=& \frac{
\sqrt{m_H^4+(m_Q^2-m_q^2)^2-2m_H^2(m_Q^2+m_q^2)}}{64 \pi m_Q^3}
\\
& \times& \Big[
(Y_{Qq}^2 +Y_{qQ}^2) [m_Q^2+m_q^2-m_H^2]
+2 (Y_{Qq} Y_{qQ}^\ast+Y_{Qq}^\ast Y_{qQ}) m_Q m_q \Big]
\nonumber \\
&\approx& \frac{Y^2_{Qq}+Y^2_{qQ}}{64 \pi} m_Q
\left[1-2\frac{m_H^2}{m_Q^2}+\frac{m_H^4}{m_Q^4}\right], 
\mbox{ for } (m_Q , m_H \gg m_q). \nonumber
\eea
%

%%%%%%%%%%%%%%%%%%%%%%%%%%%%%%
%Bibliography
%\begin{center}
%\rule{10cm}{0.25pt}
%\end{center}
%\vspace{-1.5cm}
%\bibliographystyle{apsrev}
%\bibliography{hq-lhc.bib}
%\begin{thebibliography}{49}
%\end{thebibliography}

\bibliography{}

\begin{thebibliography}{50}
\expandafter\ifx\csname natexlab\endcsname\relax\def\natexlab#1{#1}\fi
\expandafter\ifx\csname bibnamefont\endcsname\relax
  \def\bibnamefont#1{#1}\fi
\expandafter\ifx\csname bibfnamefont\endcsname\relax
  \def\bibfnamefont#1{#1}\fi
\expandafter\ifx\csname citenamefont\endcsname\relax
  \def\citenamefont#1{#1}\fi
\expandafter\ifx\csname url\endcsname\relax
  \def\url#1{\texttt{#1}}\fi
\expandafter\ifx\csname urlprefix\endcsname\relax\def\urlprefix{URL }\fi
\providecommand{\bibinfo}[2]{#2}
\providecommand{\eprint}[2][]{\url{#2}}

\bibitem[{\citenamefont{del Aguila et~al.}(2000{\natexlab{a}})\citenamefont{del
  Aguila, Perez-Victoria, and Santiago}}]{delAguila:2000aa}
\bibinfo{author}{\bibfnamefont{F.}~\bibnamefont{del Aguila}},
  \bibinfo{author}{\bibfnamefont{M.}~\bibnamefont{Perez-Victoria}},
  \bibnamefont{and} \bibinfo{author}{\bibfnamefont{J.}~\bibnamefont{Santiago}},
  \bibinfo{journal}{Phys. Lett.} \textbf{\bibinfo{volume}{B492}},
  \bibinfo{pages}{98} (\bibinfo{year}{2000}{\natexlab{a}}),
  \eprint{hep-ph/0007160}.

\bibitem[{\citenamefont{del Aguila et~al.}(2000{\natexlab{b}})\citenamefont{del
  Aguila, Perez-Victoria, and Santiago}}]{delAguila:2000rc}
\bibinfo{author}{\bibfnamefont{F.}~\bibnamefont{del Aguila}},
  \bibinfo{author}{\bibfnamefont{M.}~\bibnamefont{Perez-Victoria}},
  \bibnamefont{and} \bibinfo{author}{\bibfnamefont{J.}~\bibnamefont{Santiago}},
  \bibinfo{journal}{JHEP} \textbf{\bibinfo{volume}{09}}, \bibinfo{pages}{011}
  (\bibinfo{year}{2000}{\natexlab{b}}), \eprint{hep-ph/0007316}.

\bibitem[{\citenamefont{del Aguila et~al.}(2008{\natexlab{a}})\citenamefont{del
  Aguila, de~Blas, and Perez-Victoria}}]{delAguila:2008pw}
\bibinfo{author}{\bibfnamefont{F.}~\bibnamefont{del Aguila}},
  \bibinfo{author}{\bibfnamefont{J.}~\bibnamefont{de~Blas}}, \bibnamefont{and}
  \bibinfo{author}{\bibfnamefont{M.}~\bibnamefont{Perez-Victoria}},
  \bibinfo{journal}{Phys. Rev.} \textbf{\bibinfo{volume}{D78}},
  \bibinfo{pages}{013010} (\bibinfo{year}{2008}{\natexlab{a}}),
  \eprint{0803.4008}.

\bibitem[{\citenamefont{Atre et~al.}(2009)\citenamefont{Atre, Carena, Han, and
  Santiago}}]{Atre:2008iu}
\bibinfo{author}{\bibfnamefont{A.}~\bibnamefont{Atre}},
  \bibinfo{author}{\bibfnamefont{M.}~\bibnamefont{Carena}},
  \bibinfo{author}{\bibfnamefont{T.}~\bibnamefont{Han}}, \bibnamefont{and}
  \bibinfo{author}{\bibfnamefont{J.}~\bibnamefont{Santiago}},
  \bibinfo{journal}{Phys. Rev.} \textbf{\bibinfo{volume}{D79}},
  \bibinfo{pages}{054018} (\bibinfo{year}{2009}), \eprint{0806.3966}.

\bibitem[{\citenamefont{Agashe et~al.}(2006)\citenamefont{Agashe, Contino,
  Da~Rold, and Pomarol}}]{Agashe:2006at}
\bibinfo{author}{\bibfnamefont{K.}~\bibnamefont{Agashe}},
  \bibinfo{author}{\bibfnamefont{R.}~\bibnamefont{Contino}},
  \bibinfo{author}{\bibfnamefont{L.}~\bibnamefont{Da~Rold}}, \bibnamefont{and}
  \bibinfo{author}{\bibfnamefont{A.}~\bibnamefont{Pomarol}},
  \bibinfo{journal}{Phys. Lett.} \textbf{\bibinfo{volume}{B641}},
  \bibinfo{pages}{62} (\bibinfo{year}{2006}), \eprint{hep-ph/0605341}.

\bibitem[{\citenamefont{Carena et~al.}(2006)\citenamefont{Carena, Ponton,
  Santiago, and Wagner}}]{Carena:2006bn}
\bibinfo{author}{\bibfnamefont{M.~S.} \bibnamefont{Carena}},
  \bibinfo{author}{\bibfnamefont{E.}~\bibnamefont{Ponton}},
  \bibinfo{author}{\bibfnamefont{J.}~\bibnamefont{Santiago}}, \bibnamefont{and}
  \bibinfo{author}{\bibfnamefont{C.~E.~M.} \bibnamefont{Wagner}},
  \bibinfo{journal}{Nucl. Phys.} \textbf{\bibinfo{volume}{B759}},
  \bibinfo{pages}{202} (\bibinfo{year}{2006}), \eprint{hep-ph/0607106}.

\bibitem[{\citenamefont{Cacciapaglia et~al.}(2007)\citenamefont{Cacciapaglia,
  Csaki, Marandella, and Terning}}]{Cacciapaglia:2006gp}
\bibinfo{author}{\bibfnamefont{G.}~\bibnamefont{Cacciapaglia}},
  \bibinfo{author}{\bibfnamefont{C.}~\bibnamefont{Csaki}},
  \bibinfo{author}{\bibfnamefont{G.}~\bibnamefont{Marandella}},
  \bibnamefont{and} \bibinfo{author}{\bibfnamefont{J.}~\bibnamefont{Terning}},
  \bibinfo{journal}{Phys.Rev.} \textbf{\bibinfo{volume}{D75}},
  \bibinfo{pages}{015003} (\bibinfo{year}{2007}), \eprint{hep-ph/0607146}.

\bibitem[{\citenamefont{Contino et~al.}(2007)\citenamefont{Contino, Da~Rold,
  and Pomarol}}]{Contino:2006qr}
\bibinfo{author}{\bibfnamefont{R.}~\bibnamefont{Contino}},
  \bibinfo{author}{\bibfnamefont{L.}~\bibnamefont{Da~Rold}}, \bibnamefont{and}
  \bibinfo{author}{\bibfnamefont{A.}~\bibnamefont{Pomarol}},
  \bibinfo{journal}{Phys.Rev.} \textbf{\bibinfo{volume}{D75}},
  \bibinfo{pages}{055014} (\bibinfo{year}{2007}), \eprint{hep-ph/0612048}.

\bibitem[{\citenamefont{Carena et~al.}(2007)\citenamefont{Carena, Ponton,
  Santiago, and Wagner}}]{Carena:2007ua}
\bibinfo{author}{\bibfnamefont{M.~S.} \bibnamefont{Carena}},
  \bibinfo{author}{\bibfnamefont{E.}~\bibnamefont{Ponton}},
  \bibinfo{author}{\bibfnamefont{J.}~\bibnamefont{Santiago}}, \bibnamefont{and}
  \bibinfo{author}{\bibfnamefont{C.}~\bibnamefont{Wagner}},
  \bibinfo{journal}{Phys.Rev.} \textbf{\bibinfo{volume}{D76}},
  \bibinfo{pages}{035006} (\bibinfo{year}{2007}), \eprint{hep-ph/0701055}.

\bibitem[{\citenamefont{Carena et~al.}(2009)\citenamefont{Carena, Medina, Shah,
  and Wagner}}]{carena:2009yt}
\bibinfo{author}{\bibfnamefont{M.}~\bibnamefont{Carena}},
  \bibinfo{author}{\bibfnamefont{A.~D.} \bibnamefont{Medina}},
  \bibinfo{author}{\bibfnamefont{N.~R.} \bibnamefont{Shah}}, \bibnamefont{and}
  \bibinfo{author}{\bibfnamefont{C.~E.~M.} \bibnamefont{Wagner}},
  \bibinfo{journal}{Phys. Rev.} \textbf{\bibinfo{volume}{D79}},
  \bibinfo{pages}{096010} (\bibinfo{year}{2009}), \eprint{0901.0609}.

\bibitem[{\citenamefont{Albrecht et~al.}(2009)\citenamefont{Albrecht, Blanke,
  Buras, Duling, and Gemmler}}]{Albrecht:2009xr}
\bibinfo{author}{\bibfnamefont{M.~E.} \bibnamefont{Albrecht}},
  \bibinfo{author}{\bibfnamefont{M.}~\bibnamefont{Blanke}},
  \bibinfo{author}{\bibfnamefont{A.~J.} \bibnamefont{Buras}},
  \bibinfo{author}{\bibfnamefont{B.}~\bibnamefont{Duling}}, \bibnamefont{and}
  \bibinfo{author}{\bibfnamefont{K.}~\bibnamefont{Gemmler}},
  \bibinfo{journal}{JHEP} \textbf{\bibinfo{volume}{09}}, \bibinfo{pages}{064}
  (\bibinfo{year}{2009}), \eprint{0903.2415}.

\bibitem[{\citenamefont{del Aguila et~al.}(2010)\citenamefont{del Aguila,
  Carmona, and Santiago}}]{delAguila:2010vg}
\bibinfo{author}{\bibfnamefont{F.}~\bibnamefont{del Aguila}},
  \bibinfo{author}{\bibfnamefont{A.}~\bibnamefont{Carmona}}, \bibnamefont{and}
  \bibinfo{author}{\bibfnamefont{J.}~\bibnamefont{Santiago}},
  \bibinfo{journal}{JHEP} \textbf{\bibinfo{volume}{08}}, \bibinfo{pages}{127}
  (\bibinfo{year}{2010}), \eprint{1001.5151}.

\bibitem[{\citenamefont{Casagrande et~al.}(2010)\citenamefont{Casagrande,
  Goertz, Haisch, Neubert, and Pfoh}}]{Casagrande:2010si}
\bibinfo{author}{\bibfnamefont{S.}~\bibnamefont{Casagrande}},
  \bibinfo{author}{\bibfnamefont{F.}~\bibnamefont{Goertz}},
  \bibinfo{author}{\bibfnamefont{U.}~\bibnamefont{Haisch}},
  \bibinfo{author}{\bibfnamefont{M.}~\bibnamefont{Neubert}}, \bibnamefont{and}
  \bibinfo{author}{\bibfnamefont{T.}~\bibnamefont{Pfoh}}
  (\bibinfo{year}{2010}), \eprint{1005.4315}.

\bibitem[{\citenamefont{Carena et~al.}(2008)\citenamefont{Carena, Medina,
  Panes, Shah, and Wagner}}]{carena:2007tn}
\bibinfo{author}{\bibfnamefont{M.}~\bibnamefont{Carena}},
  \bibinfo{author}{\bibfnamefont{A.~D.} \bibnamefont{Medina}},
  \bibinfo{author}{\bibfnamefont{B.}~\bibnamefont{Panes}},
  \bibinfo{author}{\bibfnamefont{N.~R.} \bibnamefont{Shah}}, \bibnamefont{and}
  \bibinfo{author}{\bibfnamefont{C.~E.~M.} \bibnamefont{Wagner}},
  \bibinfo{journal}{Phys. Rev.} \textbf{\bibinfo{volume}{D77}},
  \bibinfo{pages}{076003} (\bibinfo{year}{2008}), \eprint{0712.0095}.

\bibitem[{\citenamefont{Contino and Servant}(2008)}]{Contino:2008hi}
\bibinfo{author}{\bibfnamefont{R.}~\bibnamefont{Contino}} \bibnamefont{and}
  \bibinfo{author}{\bibfnamefont{G.}~\bibnamefont{Servant}},
  \bibinfo{journal}{JHEP} \textbf{\bibinfo{volume}{06}}, \bibinfo{pages}{026}
  (\bibinfo{year}{2008}), \eprint{0801.1679}.

\bibitem[{\citenamefont{Aguilar-Saavedra}(2009)}]{AguilarSaavedra:2009es}
\bibinfo{author}{\bibfnamefont{J.~A.} \bibnamefont{Aguilar-Saavedra}},
  \bibinfo{journal}{JHEP} \textbf{\bibinfo{volume}{11}}, \bibinfo{pages}{030}
  (\bibinfo{year}{2009}), \eprint{0907.3155}.

\bibitem[{\citenamefont{Mrazek and Wulzer}(2010)}]{Mrazek:2009yu}
\bibinfo{author}{\bibfnamefont{J.}~\bibnamefont{Mrazek}} \bibnamefont{and}
  \bibinfo{author}{\bibfnamefont{A.}~\bibnamefont{Wulzer}},
  \bibinfo{journal}{Phys. Rev.} \textbf{\bibinfo{volume}{D81}},
  \bibinfo{pages}{075006} (\bibinfo{year}{2010}), \eprint{0909.3977}.

\bibitem[{\citenamefont{del Aguila et~al.}(2011)\citenamefont{del Aguila,
  Carmona, and Santiago}}]{delAguila:2010es}
\bibinfo{author}{\bibfnamefont{F.}~\bibnamefont{del Aguila}},
  \bibinfo{author}{\bibfnamefont{A.}~\bibnamefont{Carmona}}, \bibnamefont{and}
  \bibinfo{author}{\bibfnamefont{J.}~\bibnamefont{Santiago}},
  \bibinfo{journal}{Phys.Lett.} \textbf{\bibinfo{volume}{B695}},
  \bibinfo{pages}{449} (\bibinfo{year}{2011}), \eprint{1007.4206}.

\bibitem[{\citenamefont{Gursey and Serdaroglu}(1978)}]{Gursey:1978fu}
\bibinfo{author}{\bibfnamefont{F.}~\bibnamefont{Gursey}} \bibnamefont{and}
  \bibinfo{author}{\bibfnamefont{M.}~\bibnamefont{Serdaroglu}},
  \bibinfo{journal}{Nuovo Cim. Lett.} \textbf{\bibinfo{volume}{21}},
  \bibinfo{pages}{28} (\bibinfo{year}{1978}).

\bibitem[{\citenamefont{Gursey et~al.}(1976)\citenamefont{Gursey, Ramond, and
  Sikivie}}]{Gursey:1975ki}
\bibinfo{author}{\bibfnamefont{F.}~\bibnamefont{Gursey}},
  \bibinfo{author}{\bibfnamefont{P.}~\bibnamefont{Ramond}}, \bibnamefont{and}
  \bibinfo{author}{\bibfnamefont{P.}~\bibnamefont{Sikivie}},
  \bibinfo{journal}{Phys. Lett.} \textbf{\bibinfo{volume}{B60}},
  \bibinfo{pages}{177} (\bibinfo{year}{1976}).

\bibitem[{\citenamefont{Sultansoy and Unel}(2008)}]{Sultansoy:2006cw}
\bibinfo{author}{\bibfnamefont{S.}~\bibnamefont{Sultansoy}} \bibnamefont{and}
  \bibinfo{author}{\bibfnamefont{G.}~\bibnamefont{Unel}},
  \bibinfo{journal}{Phys. Lett.} \textbf{\bibinfo{volume}{B669}},
  \bibinfo{pages}{39} (\bibinfo{year}{2008}), \eprint{hep-ex/0610064}.

\bibitem[{\citenamefont{Mehdiyev et~al.}(2008)\citenamefont{Mehdiyev, Siodmok,
  Sultansoy, and Unel}}]{Mehdiyev:2007pf}
\bibinfo{author}{\bibfnamefont{R.}~\bibnamefont{Mehdiyev}},
  \bibinfo{author}{\bibfnamefont{A.}~\bibnamefont{Siodmok}},
  \bibinfo{author}{\bibfnamefont{S.}~\bibnamefont{Sultansoy}},
  \bibnamefont{and} \bibinfo{author}{\bibfnamefont{G.}~\bibnamefont{Unel}},
  \bibinfo{journal}{Eur. Phys. J.} \textbf{\bibinfo{volume}{C54}},
  \bibinfo{pages}{507} (\bibinfo{year}{2008}), \eprint{0711.1116}.

\bibitem[{\citenamefont{del Aguila
  et~al.}(2008{\natexlab{b}})}]{delAguila:2008iz}
\bibinfo{author}{\bibfnamefont{F.}~\bibnamefont{del Aguila}}
  \bibnamefont{et~al.}, \bibinfo{journal}{Eur. Phys. J.}
  \textbf{\bibinfo{volume}{C57}}, \bibinfo{pages}{183}
  (\bibinfo{year}{2008}{\natexlab{b}}), \eprint{0801.1800}.

\bibitem[{\citenamefont{Frampton et~al.}(2000)\citenamefont{Frampton, Hung, and
  Sher}}]{Frampton:1999xi}
\bibinfo{author}{\bibfnamefont{P.~H.} \bibnamefont{Frampton}},
  \bibinfo{author}{\bibfnamefont{P.~Q.} \bibnamefont{Hung}}, \bibnamefont{and}
  \bibinfo{author}{\bibfnamefont{M.}~\bibnamefont{Sher}},
  \bibinfo{journal}{Phys. Rept.} \textbf{\bibinfo{volume}{330}},
  \bibinfo{pages}{263} (\bibinfo{year}{2000}), \eprint{hep-ph/9903387}.

\bibitem[{\citenamefont{Kribs et~al.}(2007)\citenamefont{Kribs, Plehn,
  Spannowsky, and Tait}}]{Kribs:2007nz}
\bibinfo{author}{\bibfnamefont{G.~D.} \bibnamefont{Kribs}},
  \bibinfo{author}{\bibfnamefont{T.}~\bibnamefont{Plehn}},
  \bibinfo{author}{\bibfnamefont{M.}~\bibnamefont{Spannowsky}},
  \bibnamefont{and} \bibinfo{author}{\bibfnamefont{T.~M.~P.}
  \bibnamefont{Tait}}, \bibinfo{journal}{Phys. Rev.}
  \textbf{\bibinfo{volume}{D76}}, \bibinfo{pages}{075016}
  (\bibinfo{year}{2007}), \eprint{0706.3718}.

\bibitem[{\citenamefont{Erler and Langacker}(2010)}]{Erler:2010sk}
\bibinfo{author}{\bibfnamefont{J.}~\bibnamefont{Erler}} \bibnamefont{and}
  \bibinfo{author}{\bibfnamefont{P.}~\bibnamefont{Langacker}},
  \bibinfo{journal}{Phys. Rev. Lett.} \textbf{\bibinfo{volume}{105}},
  \bibinfo{pages}{031801} (\bibinfo{year}{2010}), \eprint{1003.3211}.

\bibitem[{\citenamefont{Martinez et~al.}(2011)\citenamefont{Martinez, Melfo,
  Nesti, and Senjanovic}}]{Martinez:2011ua}
\bibinfo{author}{\bibfnamefont{H.}~\bibnamefont{Martinez}},
  \bibinfo{author}{\bibfnamefont{A.}~\bibnamefont{Melfo}},
  \bibinfo{author}{\bibfnamefont{F.}~\bibnamefont{Nesti}}, \bibnamefont{and}
  \bibinfo{author}{\bibfnamefont{G.}~\bibnamefont{Senjanovic}}
  (\bibinfo{year}{2011}), \eprint{1101.3796}.

\bibitem[{\citenamefont{Holdom et~al.}(2009)}]{Holdom:2009rf}
\bibinfo{author}{\bibfnamefont{B.}~\bibnamefont{Holdom}} \bibnamefont{et~al.},
  \bibinfo{journal}{PMC Phys.} \textbf{\bibinfo{volume}{A3}},
  \bibinfo{pages}{4} (\bibinfo{year}{2009}), \eprint{0904.4698}.

\bibitem[{\citenamefont{del Aguila and Bowick}(1983)}]{delAguila:1982fs}
\bibinfo{author}{\bibfnamefont{F.}~\bibnamefont{del Aguila}} \bibnamefont{and}
  \bibinfo{author}{\bibfnamefont{M.~J.} \bibnamefont{Bowick}},
  \bibinfo{journal}{Nucl. Phys.} \textbf{\bibinfo{volume}{B224}},
  \bibinfo{pages}{107} (\bibinfo{year}{1983}).

\bibitem[{\citenamefont{Peskin and Takeuchi}(1992)}]{Peskin:1991sw}
\bibinfo{author}{\bibfnamefont{M.~E.} \bibnamefont{Peskin}} \bibnamefont{and}
  \bibinfo{author}{\bibfnamefont{T.}~\bibnamefont{Takeuchi}},
  \bibinfo{journal}{Phys. Rev.} \textbf{\bibinfo{volume}{D46}},
  \bibinfo{pages}{381} (\bibinfo{year}{1992}).

\bibitem[{\citenamefont{del Aguila and Bowick}(1982)}]{delAguila:1982yu}
\bibinfo{author}{\bibfnamefont{F.}~\bibnamefont{del Aguila}} \bibnamefont{and}
  \bibinfo{author}{\bibfnamefont{M.~J.} \bibnamefont{Bowick}},
  \bibinfo{journal}{Phys. Lett.} \textbf{\bibinfo{volume}{B119}},
  \bibinfo{pages}{144} (\bibinfo{year}{1982}).

\bibitem[{\citenamefont{Han and Skiba}(2005)}]{han:2004az}
\bibinfo{author}{\bibfnamefont{Z.}~\bibnamefont{Han}} \bibnamefont{and}
  \bibinfo{author}{\bibfnamefont{W.}~\bibnamefont{Skiba}},
  \bibinfo{journal}{Phys. Rev.} \textbf{\bibinfo{volume}{D71}},
  \bibinfo{pages}{075009} (\bibinfo{year}{2005}), \eprint{hep-ph/0412166}.

\bibitem[{\citenamefont{Han}(2006)}]{han:2005pr}
\bibinfo{author}{\bibfnamefont{Z.}~\bibnamefont{Han}}, \bibinfo{journal}{Phys.
  Rev.} \textbf{\bibinfo{volume}{D73}}, \bibinfo{pages}{015005}
  (\bibinfo{year}{2006}), \eprint{hep-ph/0510125}.

\bibitem[{\citenamefont{Brooijmans et~al.}(2008)}]{Brooijmans:2008se}
\bibinfo{author}{\bibfnamefont{G.~H.} \bibnamefont{Brooijmans}}
  \bibnamefont{et~al.} (\bibinfo{year}{2008}), \eprint{0802.3715}.

\bibitem[{\citenamefont{Cacciapaglia et~al.}(2010)\citenamefont{Cacciapaglia,
  Deandrea, Harada, and Okada}}]{Cacciapaglia:2010vn}
\bibinfo{author}{\bibfnamefont{G.}~\bibnamefont{Cacciapaglia}},
  \bibinfo{author}{\bibfnamefont{A.}~\bibnamefont{Deandrea}},
  \bibinfo{author}{\bibfnamefont{D.}~\bibnamefont{Harada}}, \bibnamefont{and}
  \bibinfo{author}{\bibfnamefont{Y.}~\bibnamefont{Okada}},
  \bibinfo{journal}{JHEP} \textbf{\bibinfo{volume}{11}}, \bibinfo{pages}{159}
  (\bibinfo{year}{2010}), \eprint{1007.2933}.

\bibitem[{\citenamefont{Pumplin et~al.}(2002)\citenamefont{Pumplin, Stump,
  Huston, Lai, Nadolsky et~al.}}]{Pumplin:2002vw}
\bibinfo{author}{\bibfnamefont{J.}~\bibnamefont{Pumplin}},
  \bibinfo{author}{\bibfnamefont{D.}~\bibnamefont{Stump}},
  \bibinfo{author}{\bibfnamefont{J.}~\bibnamefont{Huston}},
  \bibinfo{author}{\bibfnamefont{H.}~\bibnamefont{Lai}},
  \bibinfo{author}{\bibfnamefont{P.~M.} \bibnamefont{Nadolsky}},
  \bibnamefont{et~al.}, \bibinfo{journal}{JHEP}
  \textbf{\bibinfo{volume}{0207}}, \bibinfo{pages}{012} (\bibinfo{year}{2002}),
  \eprint{hep-ph/0201195}.

\bibitem[{\citenamefont{Alwall et~al.}(2007)}]{Alwall:2007st}
\bibinfo{author}{\bibfnamefont{J.}~\bibnamefont{Alwall}} \bibnamefont{et~al.},
  \bibinfo{journal}{JHEP} \textbf{\bibinfo{volume}{09}}, \bibinfo{pages}{028}
  (\bibinfo{year}{2007}), \eprint{0706.2334}.

\bibitem[{\citenamefont{Sjostrand et~al.}(2006)\citenamefont{Sjostrand, Mrenna,
  and Skands}}]{Sjostrand:2006za}
\bibinfo{author}{\bibfnamefont{T.}~\bibnamefont{Sjostrand}},
  \bibinfo{author}{\bibfnamefont{S.}~\bibnamefont{Mrenna}}, \bibnamefont{and}
  \bibinfo{author}{\bibfnamefont{P.~Z.} \bibnamefont{Skands}},
  \bibinfo{journal}{JHEP} \textbf{\bibinfo{volume}{0605}}, \bibinfo{pages}{026}
  (\bibinfo{year}{2006}), \eprint{hep-ph/0603175}.

\bibitem[{\citenamefont{Ovyn et~al.}(2009)\citenamefont{Ovyn, Rouby, and
  Lemaitre}}]{Ovyn:2009tx}
\bibinfo{author}{\bibfnamefont{S.}~\bibnamefont{Ovyn}},
  \bibinfo{author}{\bibfnamefont{X.}~\bibnamefont{Rouby}}, \bibnamefont{and}
  \bibinfo{author}{\bibfnamefont{V.}~\bibnamefont{Lemaitre}}
  (\bibinfo{year}{2009}), \eprint{0903.2225}.

\bibitem[{\citenamefont{Han et~al.}(1992)\citenamefont{Han, Valencia, and
  Willenbrock}}]{Han:1992hr}
\bibinfo{author}{\bibfnamefont{T.}~\bibnamefont{Han}},
  \bibinfo{author}{\bibfnamefont{G.}~\bibnamefont{Valencia}}, \bibnamefont{and}
  \bibinfo{author}{\bibfnamefont{S.}~\bibnamefont{Willenbrock}},
  \bibinfo{journal}{Phys. Rev. Lett.} \textbf{\bibinfo{volume}{69}},
  \bibinfo{pages}{3274} (\bibinfo{year}{1992}), \eprint{hep-ph/9206246}.

\bibitem[{\citenamefont{Figy et~al.}(2004)\citenamefont{Figy, Oleari, and
  Zeppenfeld}}]{Figy:2004ec}
\bibinfo{author}{\bibfnamefont{T.}~\bibnamefont{Figy}},
  \bibinfo{author}{\bibfnamefont{C.}~\bibnamefont{Oleari}}, \bibnamefont{and}
  \bibinfo{author}{\bibfnamefont{D.}~\bibnamefont{Zeppenfeld}},
  \bibinfo{journal}{Nucl. Phys. Proc. Suppl.} \textbf{\bibinfo{volume}{135}},
  \bibinfo{pages}{9} (\bibinfo{year}{2004}), \eprint{hep-ph/0407066}.

\bibitem[{\citenamefont{Alwall et~al.}(2008)}]{Alwall:2007fs}
\bibinfo{author}{\bibfnamefont{J.}~\bibnamefont{Alwall}} \bibnamefont{et~al.},
  \bibinfo{journal}{Eur. Phys. J.} \textbf{\bibinfo{volume}{C53}},
  \bibinfo{pages}{473} (\bibinfo{year}{2008}), \eprint{0706.2569}.

\bibitem[{\citenamefont{Alwall et~al.}(2009)}]{Alwall:2008pm}
\bibinfo{author}{\bibfnamefont{J.}~\bibnamefont{Alwall}} \bibnamefont{et~al.},
  \bibinfo{journal}{AIP Conf. Proc.} \textbf{\bibinfo{volume}{1078}},
  \bibinfo{pages}{84} (\bibinfo{year}{2009}), \eprint{0809.2410}.

\bibitem[{R-x()}]{R-xtal-ball}
\bibinfo{howpublished}{{The Crystal Ball function is a differentiable function
  that consists of a Gaussian term and a power-law tail. J. E. Gaiser, Ph. D.
  Thesis, SLAC-R-255 (1982)}}.

\bibitem[{atl()}]{atlas-csc}
\bibinfo{howpublished}{{ATLAS Collaboration, Expected Performance of the ATLAS
  Experiment, Detector, Trigger and Physics, CERN-OPEN-2008-020 (2009).}}

\bibitem[{mqb()}]{mqbound}
\emph{\bibinfo{title}{Search for heavy top $t^\prime \to$ w q in lepton plus
  jets events}}, \bibinfo{howpublished}{{CDF-note-10110}}.

\bibitem[{\citenamefont{Aaltonen et~al.}(2007)}]{Aaltonen:2007je}
\bibinfo{author}{\bibfnamefont{T.}~\bibnamefont{Aaltonen}} \bibnamefont{et~al.}
  (\bibinfo{collaboration}{CDF Collaboration}), \bibinfo{journal}{Phys.Rev.}
  \textbf{\bibinfo{volume}{D76}}, \bibinfo{pages}{072006}
  (\bibinfo{year}{2007}), \eprint{arXiv:0706.3264}.

\bibitem[{cdf()}]{cdf_hq}
\emph{\bibinfo{title}{Search for single production of heavy quarks}},
  \bibinfo{howpublished}{{CDF-note-10261}}.

\bibitem[{\citenamefont{Abazov et~al.}(2010)}]{Abazov:2010ku}
\bibinfo{author}{\bibfnamefont{V.~M.} \bibnamefont{Abazov}}
  \bibnamefont{et~al.} (\bibinfo{collaboration}{D0 Collaboration})
  (\bibinfo{year}{2010}), \eprint{1010.1466}.

\bibitem[{CMS()}]{CMS_significance}
\bibinfo{howpublished}{{CMS Collaboration, CMS Physics, Technical Design
  Report, CERN/LHCC 2006-001}}.

\end{thebibliography}

\end{document}